\documentclass[prd,nofootinbib,twocolumn]{revtex4-1}
\usepackage{mathrsfs,color}

\usepackage{amssymb,amsmath,enumerate}
\usepackage{graphicx}
\usepackage{subfigure}

\newcommand{\alm}{a_{\ell m}}
\newcommand{\Ylm}{Y_{\ell m}}

\newcommand{\Wb}{{\bf W}}
\newcommand{\vv}{{\bf v}}
\newcommand{\cc}{{\cal C}}
\newcommand{\degr}{\ensuremath{^\circ}} 

\newcommand{\Npix}{{N_{\rm pix}}}
\newcommand{\Omegapix}{{\Omega_{\rm pix}}}
\newcommand{\Nside}{{N_{\rm side}}}

\newcommand{\lmax}{{\ell_{\rm max}}}
\newcommand{\lmaxtot}{{\ell_{\rm max, tot}}}
\newcommand{\lmaxrec}{{\ell_{\rm max, rec}}}
\newcommand{\zpeak}{{z_{\rm peak}}}
\newcommand{\Cl}{C_\ell}
\newcommand{\fsky}{f_{\rm sky}}

\newcommand{\bra}[1]{\left(#1\right)}
\newcommand{\bras}[1]{\left[#1\right]}
\newcommand{\brac}[1]{\left\{#1\right\}}

\newcommand{\wn}{{\bf n}}
\newcommand{\wx}{{\bf x}}
\newcommand{\wy}{{\bf y}}
\newcommand{\wa}{{\bf a}}
\newcommand{\wc}{{\bf C}}
\newcommand{\wN}{{\bf N}}
\newcommand{\ws}{{\bf S}}

\newcommand{\we}{{\bf e}}
\newcommand{\wk}{{\bf k}}

\newcommand{\wS}{{\bf \Sigma}}

\newcommand{\apjl}{ApJL}
\newcommand{\apjs}{ApJS}
\newcommand{\mnras}{MNRAS}
\newcommand{\physrev}{Phys. Rev.}
\newcommand{\physrevlett}{Phys. Rev. Lett.}

\begin{document}
\title{Testing the statistical isotropy of large scale structure with
  multipole vectors}
\author{Caroline Zunckel}
\affiliation{Astrophysics Department, Princeton University, Peyton Hall, 4 Ivy
  Lane, NJ, 08544, USA}
\affiliation{Astrophysics and Cosmology Research Unit, University of
  Kwazulu-Natal, Westville, Durban, 4000, South Africa}

\author{Dragan Huterer}
\affiliation{Department of Physics, University of Michigan, 450 Church St, Ann Arbor, MI 48109-1040, USA}

\author{Glenn D. Starkman}
\affiliation{ISO/CERCA and Department of Physics, Case Western Reserve University, Cleveland,
Ohio, 44106-7079, USA }

\begin{abstract}
  A fundamental assumption in cosmology is that of statistical isotropy ---
  that the universe, on average, looks the same in every direction in the
  sky. Statistical isotropy has recently been tested stringently using Cosmic
  Microwave Background (CMB) data, leading to intriguing results on large
  angular scales. Here we apply some of the same techniques used in the CMB to
  the distribution of galaxies on the sky. Using the multipole vector
  approach, where each multipole in the harmonic decomposition of galaxy
  density field is described by unit vectors and an amplitude, we lay out the
  basic formalism of how to reconstruct the multipole vectors and their
  statistics out of galaxy survey catalogs.  We apply the algorithm to
  synthetic galaxy maps, and study the sensitivity of the multipole vector
  reconstruction accuracy to the density, depth, sky coverage, and
  pixelization of galaxy catalog maps.
\end{abstract}
\date{\today}
\pacs{PACS Numbers : }
\maketitle

\section{Introduction}

In the standard model of cosmology the primordial density perturbations in the
early Universe are generated by a Gaussian, statistically isotropic random
process. There are two reasons for this: the cosmological principle tells us
that the Universe is homogeneous and isotropic on large scales and the
standard (single-field, slow-roll) inflationary theory predicts near-perfect
Gaussianity and statistical isotropy of primordial fluctuations in the universe.

It is useful to differentiate the sometimes conflated concepts of statistical
isotropy (hereafter SI) and Gaussianity. Statistical isotropy means that the
expectation values of measurable quantities are invariant under rotations.
For example, the expected two-point correlation function of the Cosmic
Microwave Background (CMB) temperature (or galaxy overdensity) $\Delta$ in two
directions in the sky $\hat{e}_i$ and $\hat{e}_j$
\begin{equation} 
C(\hat{e}_i, \hat{e}_j) = \langle
  \Delta(\hat{e}_i) \Delta (\hat{e}_j) \rangle
\label{correlation}
\end{equation} 
(where $\langle \cdot \rangle$ represents the ensemble average) would, under
SI, depend only on the angle $\theta$ between $\hat{e}_i$ and $\hat{e}_j$,
i.e. $C(\hat{e}_i, \hat{e}_j) = C(\hat{e}_i \cdot \hat{e}_j)$.
Gaussianity, on the other hand, refers to the statistical distribution from
which the quantity $\Delta$ is drawn.  As a consequence of Gaussianity, all of
the statistical properties of the field are encapsulated in the two-point
correlation function $C(\hat{e}_i \cdot \hat{e}_j)$; all of the odd
higher-point correlation functions are zero, and the even-point correlation
functions can be related to the two-point function by Wick's theorem. In
general, a given field can be Gaussian but not SI, or SI but not Gaussian, or
neither. The standard cosmological theory predicts it to be both (except to the
extent that nonlinear evolution spoils the Gaussianity).

Much of the information used to construct the current concordance model has
been derived from examination of the statistical properties of the CMB
temperature anisotropies on the sky.  Following in the footsteps of the Cosmic
Background Explorer (COBE) \cite{COBE,COBE2}, experiments such as the
Wilkinson Microwave Anisotropy Probe (WMAP)
\cite{Spergel2003,Spergel2006,wmap7} have succeeded in measuring the
temperature anisotropies to high precision, engendering widespread confidence
that we have arrived at a convincing model, based on standard inflationary
cosmology, in which the perturbations are presumably Gaussian and
statistically isotropic.

However, certain anomalies at low $\ell$ have been pointed out and suggest
possible deviations from this paradigm.  Over a decade ago, the COBE
Differential Microwave Radiometer (COBE-DMR) first reported a lack of
large-angle correlations in the two-point angular-correlation function,
$\mathcal{C}(\theta)$, of the CMB \cite{DMR4}.  This was confirmed by the WMAP
team in their analysis of their first year of data~\cite{Spergel2003}, and by
some of us in the WMAP three, five and seven-year
data~\cite{wmap123,wmap12345,Sarkar}, and further confirmed by independent
analyses \cite{Hajian:2007pi,Bunn_Bourdon}.  The angular two-point function is
approximately zero at scales $\theta>60\degr$ in all wavebands, in contrast to
the theoretical prediction from the standard inflationary cosmology. Such a
result is expected in only $\sim 0.03$\% of the Gaussian random, isotropic
skies based on the standard inflationary model (and using a statistic
suggested in \cite{Spergel2003}).  This vanishing of $C(\theta)$ is unexpected
not only because of its low likelihood (which admittedly has been defined {\it
  a posteriori}), but for at least four other reasons. First, missing
correlations are inferred from cut-sky (i.e.\ masked) maps of the CMB, which
makes the results insensitive to assumptions about what lies behind the
cut. Second, what little large-angle correlation does appear in the full-sky
maps is associated with points inside the masked region, further casting into
doubt the full-sky reconstruction-based results \citep{wmap12345}. Third, the
vanishing power is not as clearly seen in multipole space where the quadrupole
is only moderately low, and it is really a range of low multipoles that
conspire to ``interfere'' in just such a way to make up the near-vanishing
$C(\theta)$ \citep{wmap12345}. Fourth, the missing power occurs on the largest
observable scales, where a cosmological origin is arguably most likely.

Moreover, some of us and others found that the two largest cosmologically
interesting modes of the CMB, the quadrupole and octopole ($\ell=2$ and $3$),
are correlated with the direction of motion and geometry of the solar system
\citep{Schwarz2004}. [Recall that each multipole $\ell$ corresponds to scales
  of about $180/\ell$ degrees on the sky]. In brief, the quadrupole and
octopole are unusually planar (as first pointed out by \cite{TOH}); their
plane is perpendicular to the ecliptic plane and pointed to the cosmic dipole;
and the ecliptic plane itself traces out a nodal line between the big hot and
cold spots in the quadrupole-octopole map. The alignments persist to smaller
scales (higher multipoles of the CMB), where it has been found that $\ell\leq
6$ multipoles have unusually large fraction of power in a preferred frame
\citep{Land2005a}.  Even at the first peak, it has been shown
\citep{Yoho:2010pb} that there is an ecliptically-associated anomaly -- the
first peak is significantly under-powered near the north ecliptic pole.  It
has also been found that the northern ecliptic hemisphere has significantly
less power than the southern hemisphere on scales larger than about 3 degrees
(multipoles $\ell\lesssim 60$)
\citep{Eriksen_asym,Hansen_asym,Eriksen:2007pc,Hansen:2008ym,Hoftuft}.  These
non-Gaussianities at large and small scales have been confirmed by other
analyses \citep{Raeth:2010kx}.  These alignments, being indicative of a real
effect whether it is cosmological or astrophysical, have caused wide interest,
and some of us followed them up by performing a comprehensive study of the
findings and comparing different statistics, considering the foreground
contamination, and studying the COBE data as well \citep{lowl2}. The most
recent WMAP paper on anomalies \cite{Bennett_anomalies}, while disagreeing
with some of the above findings and agreeing with others, does not appear to
offer convincing explanations of the observed anomalies.  For a brief review
of the anomalies, see \cite{Huterer_NewAst_review}; for a comprehensive
review, see \cite{CHSS_review}.

At this time there is no convincing explanation for alignments or the missing
large-angle correlations found in the CMB.  However, the consequences are
clear: if indeed the observed $\ell=2$ and $3$ CMB fluctuations are not
cosmological, one must reconsider all cosmological results that rely on low
$\ell$ of the CMB.  Even more importantly, a cosmological origin of the
violation of statistical isotropy would invalidate the basic assumptions used
in the standard analyses to extract cosmological parameters, requiring our
full understanding of the physics behind the anomalies.

In the past 15 years or so, galaxy surveys have revolutionized our
understanding of the universe. Most recently, the Sloan Digital Sky Survey
(SDSS) and the Two Degree-Field Survey (2dF) have measured the locations of
about a hundred million galaxies over $\sim\!10,000$ sq.\ deg.\ of the sky,
and measured about a million redshifts. The main product of these massive
efforts was precision measurement of the cosmological parameters, and also the
precise measurement of the matter power spectrum.  Perhaps surprisingly,
however, except for a few searches for modulations in power in the large-scale
structure (LSS) \cite{Hirata,Pullen_Hirata} and theoretical predictions for
clustering of halos in models that break the SI \cite{Ando:2008zza}, there
have been few explicit tests of statistical isotropy using the LSS. Instead,
most of the studies have been either theoretical or applied exclusively to the
CMB, and concerned with how the CMB anisotropy would look in inflationary (or
other) models that break SI
\cite{Gordon2005,ACW,Donoghue:2007ze,ArmendarizPicon07,Gumrukcuoglu,Rodrigues_Bianchi,Pullen_Kam,
  Pitrou08,Erickcek_hem,Erickcek_iso,Battye09}. Such models, where the
primordial power spectrum $P(\wk)$ depends on the magnitude {\it and}
direction $\hat{\wk}$ of the wavevector, may be detectable with WMAP or future
CMB experiments, and there has recently been a lot of effort searching for
signatures of broken SI in the CMB
\cite{Gordon2005,Groeneboom:2008fz,Hoftuft,Hanson_Lewis,Groeneboom_sys,Bennett_anomalies,Hanson_beams}. Given
that a set of robust statistical tools have been developed for such tests of
the CMB, the natural next step would be to adopt some of the same methods to
the study of LSS.

The CMB anomalies found using WMAP data have only whetted the appetite of
cosmologists to investigate the aforementioned anomalies further. While the
Planck CMB mission will --- like WMAP --- surely produce spectacular results
revolutionizing our understanding of the universe, it is generally expected
that Planck will confirm WMAP's findings on the largest scales as both
experiments are measuring the same physical phenomenon at scales where
Planck's better resolution makes no difference. Observations of large-scale
fluctuations are subject to sample variance (sometimes referred to as 
cosmic variance): our universe provides only a relatively small number of
independent samples of largest-scale structures, limiting the extent to which
the CMB alone can shed light on them. Therefore, it is imperative to extract
every last bit of information provided. In particular, {\it galaxy surveys}
complement the CMB in providing a picture of the largest scales with
different tracers of fluctuations than the CMB, emitting light at different
wavelengths, and whose analysis includes different systematic errors than that
of the CMB. Here we propose to stringently test the cosmological principle
using archival data from the upcoming large-scale structure surveys.

This is an excellent time to perform analyses of statistical isotropy on the
largest observable scales because  full-sky maps of the LSS, with tracers
at multiple wavelengths, are finally becoming available.  In this paper we
adapt the statistical tools used in tests of SI of the CMB to LSS measured by
galaxy surveys.  We investigate how the characteristics of LSS surveys impact
the accuracy of the extracted quantities and present one example of the
efficacy of detecting alignments in a specific, purely phenomenological, toy
model.

The structure of this paper is as follows.  In Sec.\ \ref{sec:prelim}, the
relevant cosmological quantities are defined and followed, in Sec.\
\ref{sec:tools}, by a brief overview of the statistical tools
available to conduct tests of SI. In Sec.\ \ref{sec:LSS} we construct a
framework in which the LSS observables are mapped to the selected
statistics. The reconstruction technique used to estimate these quantities and
how the accuracy of the reconstruction varies with the characteristics of the
galaxy survey are discussed in Sec.\ \ref{sec:reconstruction}.  We then
proceed to test how this accuracy translates into detection of possible violations
of SI in Sec.\ \ref{sec:alignments}.  In Sec.\ \ref{sec:discussion} we
discuss our findings and future work.

\section{Preliminaries}
\label{sec:prelim}
Consider a cosmological dataset which can be characterized by the function
$f(\theta, \phi)$ on the celestial sphere. It can be decomposed into multipole
moments as follows:
\begin{equation}
  f(\theta, \phi) = \sum_{\ell} f_{\ell}(\theta, \phi) = 
\sum_{\ell=0}^{\ell=\infty} \sum_{m=-\ell}^{\ell} a_{\ell m} Y_{\ell m}\bra{\theta, \phi},
\label{eq:f}
\end{equation}
where $0 \leq \theta \leq \pi$ and $0 \leq \phi \leq 2\pi $ and the $\alm$
are the multipole coefficients and the complex spherical harmonic functions
are given by
\begin{equation}
Y_{\ell m}\bra{\theta, \phi}=\sqrt{\frac{(2\ell+1)(\ell-m)!}{4\pi (\ell +
    m)!}} P_{\ell m}\bra{\cos \theta}e^{i m \phi},
\label{harmonic}
\end{equation}
where $P_{\ell m}$ are the associated Legendre polynomials.  If the
cosmological data are indeed produced by a statistically isotropic and Gaussian
process, then the $\alm$ are realizations of Gaussian random variables
of zero mean, characterized fully by their variances. The added property of
statistical isotropy (SI) further implies that their variances depend only on
$\ell$ and means that we can write
\begin{equation}
  C_{\ell m  \ell' m'} \equiv \langle \alm a^{*}_{\ell', m'}\rangle
=\Cl \delta_{\ell \ell'} \delta_{m m'} 
\label{ensemble}
\end{equation} 
where $\Cl$ is the expected power in the $\ell$-th multipole. Note that the theoretically predicted
coefficients $\alm$ and the power spectrum $C_\ell$ correspond to averages
over an ensemble of universes.  While we unfortunately have only a single
sample of $\alm$ for each $\ell$ and $m$, corresponding to values measured in
our universe, the power spectrum $C_\ell$ can be estimated with a finite
sample variance by averaging the power in $\alm$ for each $m$
\begin{equation}
\tilde{\cc}_\ell  \equiv \frac{1}{2\ell +1}\sum^{\ell}_{m=-\ell} |\alm|^2. 
\label{C}
\end{equation} 
If SI holds, then $\tilde{\cc}_\ell$ is an \emph{unbiased}
estimator of $\Cl$.  If Gaussianity additionally holds, then it is the best
estimator, with cosmic variance $2 \tilde{\cc}^2_{\ell}/(2 \ell +1)$.

Since the power spectrum can be readily calculated from theory, we can compare
predictions of our cosmological models to the observationally determined
$\Cl$, placing precise constraints on the parameters.

\section{Statistical Tools}
\label{sec:tools}

In this section we consider the various quantities related to the above which
can be used to test the isotropic nature of cosmological data which is
characterized by the function $f(\theta, \phi)$ on the sky given in
Eq.~(\ref{eq:f}).

\subsection{Multipole coefficients}\label{sec:alm}

A caveat that comes with using the power spectrum as a tool for searches of
statistical anisotropies is that it is sensitive to only specific types of
departures from SI. It is possible for the distribution of power in $\Cl$
throughout the $m$-modes to violate SI with no bearing on the $\Cl$ spectrum.

It is therefore important to measure quantities that contain information
about Gaussianity and SI such as the multipole coefficients $\alm$. They are
another representation of the information in $f(\hat\Omega)$, where
$\hat\Omega = (\theta, \phi)$, related by
\begin{equation} 
\alm = \int  f(\hat\Omega)   Y^{*}_{\ell m}\bra{\hat\Omega} d \Omega.
\label{inversion}
\end{equation}

If $f(\hat\Omega)$ is a realization of a Gaussian and isotropic process, then the
equality in Eq.~(\ref{ensemble}) holds and the $\alm$ are independent, random
variables with Gaussian distributions and variances that depend only on
$\ell$. This implies that the distribution of the overall power throughout the
$\alm$ (i.e.\ their magnitudes) should be a function of $\ell$ only and the
distribution of the power in a particular scale (i.e.\ $\Cl$) through the
$m$-modes should depend only the selected coordinate system.

In \cite{deOliveira2004}, a statistic was introduced which associates an axis with
each $\ell$ around which the angular dispersion is maximized
\begin{equation}
  S_{\ell} = \text{max}_{\wn} \sum_{m} m^2 |\alm|^2.  
\end{equation} 
This statistic finds the frame of reference with its z-axis in the
$\hat{\wn}_{\ell}$ direction which maximizes the angular dispersion, with the
extent of this preference gauged by the magnitude of $_{\ell}$. As mentioned
previously, when applied to the WMAP1 data \cite{Spergel2003}, this statistic
indicated that $\hat{\wn}_{2}$ and $\hat{\wn}_{3}$ were unexpectedly aligned
in a direction in which the power $\cc_{2}$ is significantly suppressed.
Another such statistic introduced in \cite{Land2005a} is
\begin{equation}
r_{\ell} = \text{max}_{m \wn}\bras{\frac{\cc_{\ell m}}{\sum_{\tilde m} |{a_{\ell \tilde m}}| ^2}}
\label{eq:aoe}
\end{equation} where $\cc_{\ell 0} =|a_{\ell 0}|^2$ and $\cc_{\ell m} = 2|\alm|^2$
for $m>0$.  Here $r_{\ell}$ is the ratio of power of the $\ell$-th multipole
that lies in the $m$ mode in the direction $\wn$. This statistic explicitly
returns the axis and direction in which the power distribution is most uneven
(i.e.\ $\wn$) and the extent to which it is uneven (i.e.\ magnitude of
$r_{\ell}$). When applied to the WMAP1 data, this statistic returned the same
preferred axis as in \cite{deOliveira2004}.  These features of the CMB sky may
be suggesting inter-$m$ correlations between the $\alm$ and a break down of
SI.

\subsection{Multipole Vectors}
\label{sec:MV}

While the multipole vector formalism was first introduced by
\cite{Copi:2003kt} into the analysis of the CMB, its full history is much
longer. More than 100 years ago, Maxwell \cite{Maxwell} pointed out that for
any real function $f_{\ell}(x, y, z)$, which is an eigenfunction of the
Laplacian on the unit sphere with eigenvalue $-\ell(\ell+1)$, there exist
$\ell$ unit vectors $\bra{\vv_1, \vv_2, ...\vv_{\ell}}$ such that
\begin{equation} 
f(x,y,z)_{\ell} = \nabla_{\vv_1}...\nabla_{\vv_{\ell}} \frac{1}{r}\,,
\label{fxyz}
\end{equation} 
where 
$(x, y, z) = (\cos\theta\sin\theta, \sin\theta\sin\phi,\cos\phi)$, 
$\nabla_{\vv_{\ell}}\equiv\vv_{\ell}\cdot\nabla$ is the directional derivative operator, 
and $r = \sqrt{x^2+y^2+z^2}$.
A multipole
can then be represented in terms of $\ell$ unit vectors
$\brac{\vv_{\ell,i}\mid i=.1...\ell}$, termed the multipole vectors (MVs) and
an invariant scalar $A_{\ell}$. Heuristically, the $\ell$-th multipole of
the CMB can be written as a product of $\ell$ unit vectors and an overall
normalization  so that we can write 
\begin{equation} 
f_{\ell} \sim A_{\ell} \Pi_{i = 1}^{\ell} \bra{\vv_{\ell,i} \cdot \hat{\we}}
\label{eq:MV}
\end{equation} 
where $\hat{\we} = \bra{\sin \theta \cos\phi, \sin \theta \sin \phi, \cos
  \theta}$ is the unit radial vector. Note that the signs of all the vectors
can be absorbed into the sign of $A^{(\ell)}$, so one is free to choose the
hemisphere of each vector.  These multipole vectors encode all the information
about the phase relationships of the $a_{\ell m}$.  The MVs can be understood
in the context of harmonic polynomials \cite{Katz2004} and have many
interesting properties (e.g.\ \cite{Dennis2005}). An efficient algorithm to
compute the multipole vectors for low-$\ell$ has been presented in
\cite{Copi:2003kt} and is publicly available \cite{MV_code}; other algorithms
have been proposed as well \cite{Katz2004,Weeks04,Helling}.

Note that {\it multipole vectors are defined in exactly the same way for the
  galaxy surveys} provided one makes the obvious identification
\begin{equation}
{\delta T\over T}(\hat {\bf n}) \longleftrightarrow
{\delta n\over n}(\hat {\bf n})
\end{equation}
where $n$ is the number of galaxies (or other tracers of the LSS) per unit
area of the sky.

\begin{figure*}[t]
\begin{center}
  \includegraphics[scale=0.4]{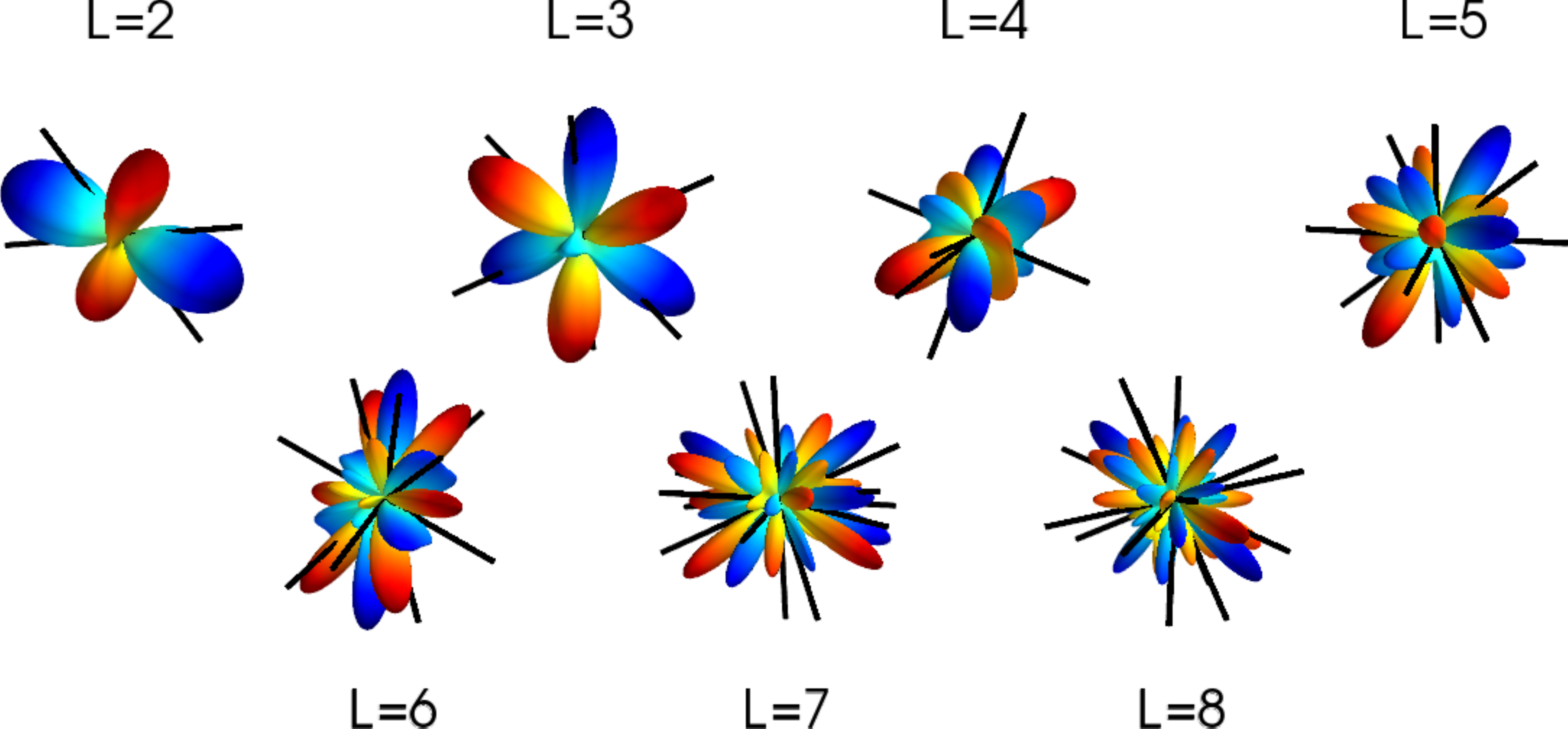} 
\end{center}
\caption{Multipole vectors of our sky, with the corresponding multipoles
  $\ell=2-8$ computed from WMAP's 3-year Internal Linear Combination (ILC) map
  \cite{Hinshaw:2006ia}. The lobes represent the CMB temperature
    pattern seen at each multipole, where the observer is at the center and
    the observed sky anisotropy can be projected to a sphere of a fixed
    radius.  The sticks are the multipole vectors, each pointing in a fixed
    direction (or its opposite) on this sphere. Figure kindly provided by
  Craig Copi.  }
\label{fig:MV}
\end{figure*}

Figure~\ref{fig:MV} shows the multipole vectors of our sky, with the
corresponding multipoles $\ell=2-8$ computed from WMAP's 3-year Internal
Linear Combination (ILC) map \cite{Hinshaw:2006ia}.  Multipole vectors still
contain the full information about the map, but are often more sensitive to
different aspects of the temperature pattern than the usual spherical harmonic
representation.

Mutual cross products of $\ell$ vectors in the $\ell$-th multipole define
$\ell(\ell-1)/2$ planes, and these planes are also useful for testing the SI.
For example, in \cite{CHSS_review}, the three octopole planes of the CMB were
found to be nearly parallel and aligned with the single plane of the
quadrupole, and this alignment is statistically significant at the $99.9\%$
level.

To illustrate the advantage of decomposing a multipole in this fashion, we
consider MVs of the real part of a pure harmonic mode; $Re Y_{\ell m}(\theta,
\phi)$, so that all the power $\Cl$ lies in that particular $m$-mode.  In this
case, $\ell-|m|$ of the $\ell$ MVs are aligned with the $z$-axis (which is the
frame of the $Y_{\ell m}$), while the remaining $|m|$ MVs line in the $x-y$
plane. Since the configuration of MVs rotates with the function
$f_{\ell}(\theta, \phi)$, the pure harmonic modes are readily identified in
any frame of reference. This is true of any function $f_{\ell}(\theta, \phi)$
which makes the MVs very useful for investigation issues such as SI
\cite{Dennis2004}.

For our purposes the MVs are the quantities of interest and represent all
information contained in the data regarding the phase relationships between
the $\alm$.

\section{Large scale structure: mathematical description}
\label{sec:LSS}

Galaxy surveys measure positions of galaxies either in three dimensions (as
redshift surveys) or as a 2D projection on the sky (angular surveys).
However, most surveys contain information that is somewhere between 2D and
3D, since galaxies have photometric redshifts that enable {\it approximate} 
rendering of radial distance to galaxies (given good knowledge
of the underlying cosmological parameters).

In this work we consider projected (i.e.\ two-dimensional) large-scale
structure surveys.  We wish to reconstruct the underlying density
distribution, $\sigma(\hat\Omega)$, given counts of galaxies on the sky. When
multiplied by the bias parameter $b$, the density field gives an angular
number density distribution function of the catalog on the sky
$\nu(\hat\Omega)$.

We can split the number density of objects on the sky, $\nu(\hat\Omega)$, into
its mean and relative variation across the sky
\begin{equation} 
  \nu(\hat\Omega) = {\bar \nu}\left(1 + \delta(\hat\Omega)\right),
\end{equation} 
where the ${\bar \nu}$ is the average density over the sky, given by
${\bar \nu} = \int d\Omega ~\nu(\hat\Omega) / \int d\Omega $  and
$\delta(\hat\Omega)$ are the fluctuations around the mean at position
$\hat\Omega$.

To enable connection with observable counts of galaxies, we bin the sky into
$\Npix$ equal-area pixels and define
\begin{equation}
n_i  = S \int_{{\rm ith~pixel}} d\Omega~\nu(\hat\Omega),
\label{ni}
\end{equation} 
where $n_i$ is the expected number of objects in the $i$-th pixel centered
at $\Omega_i$ and $S$ is a selection function which accounts for the physical
attributes of the survey construction, such as the exposure time and the
sensitivity of the instruments.  For simplicity, we assume that the selection
function is independent of direction on the sky; while clearly simplistic,
this assumption is straightforwardly relaxed provided that the full selection
function is known. Effects of the uncertainties in the selection function,
however, may be important and certainly warrant further investigation, but are
outside of scope of the present foundational work.

The mean number of expected objects per pixel is then given by
\begin{equation} 
{\bar n} \equiv \frac{1}{\Npix}\sum_{i=1}^{\Npix} n_i.
\end{equation}
We now express the expected fluctuations around the mean $\bar{n}$ by
\begin{equation}
\Delta_i \equiv \Delta(\Omega_i) = \frac{n_i - {\bar n}}{{\bar n}}.
\label{eq:Delta_i}
\end{equation}
We see that the \emph{binned} fluctuation $\Delta_i$ in the i$^{th}$
pixel relates to the true underlying fluctuation $\delta$ via
\begin{equation}
\Delta_i = \frac{1}{\Omegapix}\int_{\rm  i^{th}~pixel} \delta(\Omega)d\Omega,
\end{equation}
where $ \Omegapix$ is the area of a pixel, so that the $\Delta_i$ is the
average fluctuation around the mean in the i$^{th}$ pixel. Hence the
disparity between $\Delta_i$ at a point $\Omega_i$ on the sky and the true
underlying $\delta(\Omega_i)$ depends on the level of pixelization of the sky,
so that $\Delta_i\to\delta(\Omega_i)$ in the limit of perfect resolution
($\Npix\to \infty$). 

The function $\Delta(\Omega)$ has a constant value $\Delta_i$ within the
i$^{th}$ pixel, but otherwise varies across the sky. We expand it into
spherical harmonics
\begin{equation}
\Delta(\hat\Omega) = \sum_{\ell=1}^\infty\sum_{m=-\ell}^\ell \alm Y_{\ell m}(\hat\Omega)
\end{equation}
or 
\begin{equation}
\label{eqn:Deltaialms}
\Delta_i = \sum_{\ell=1}^\infty\sum_{m=-\ell}^\ell \alm Y_{\ell m}(\Omega_i) .
\end{equation}
We are now able to apply the same treatment of the CMB temperature
anisotropies to the case of LSS.

\section{Multipole Vector reconstruction}
\label{sec:reconstruction}

\subsection{The Reconstruction Methodology }\label{sec:method}

In the last section, we described the transformation of a galaxy catalog
into a set of measurements $\Delta\bra{\Omega_i}$ of object numbers in a set
of pixels, centered at $\Omega_i$ where $i=1...\Npix$ on the full celestial
sphere.  The $\alm$ can be determined from these observations by inverting
Eq.~(\ref{eqn:Deltaialms})
\begin{equation}
\alm = \int Y^*_{\ell m} (\hat\Omega) \Delta(\hat\Omega) d \Omega = 
\Omegapix \sum_{\hat{\Omega}} Y^*_{\ell m} (\hat{\Omega}) \Delta(\hat{\Omega}), 
\label{full_alm}
\end{equation} 
where $\hat{\Omega}$ is the direction on the sky.

\begin{figure*}
\centering
\includegraphics[scale=0.3, angle=0, trim = 60mm 60mm 30mm 10mm ]{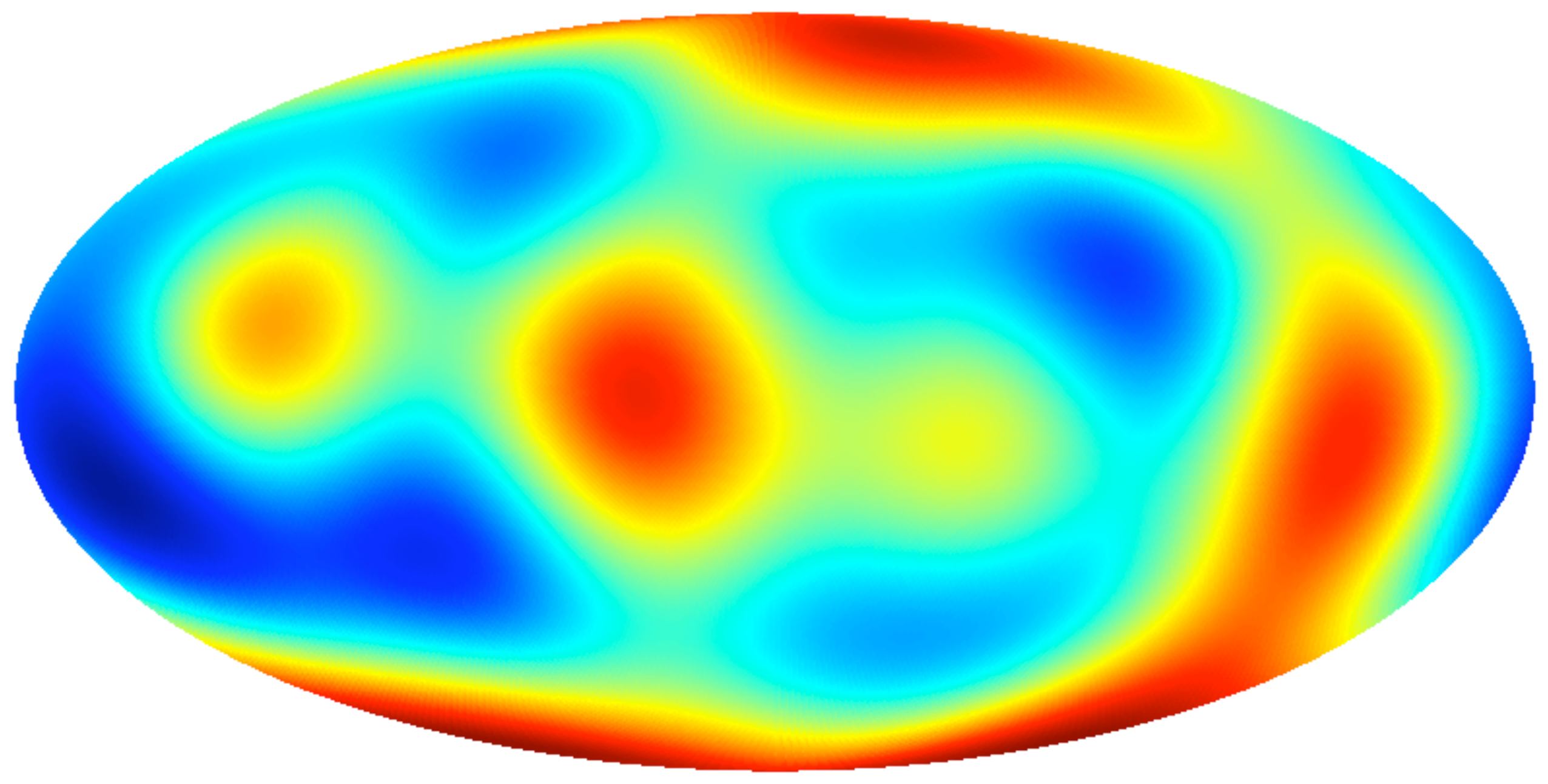}\hspace{2.5cm}
\includegraphics[scale=0.3, angle=0,  trim = 40mm 60mm 40mm 10mm]{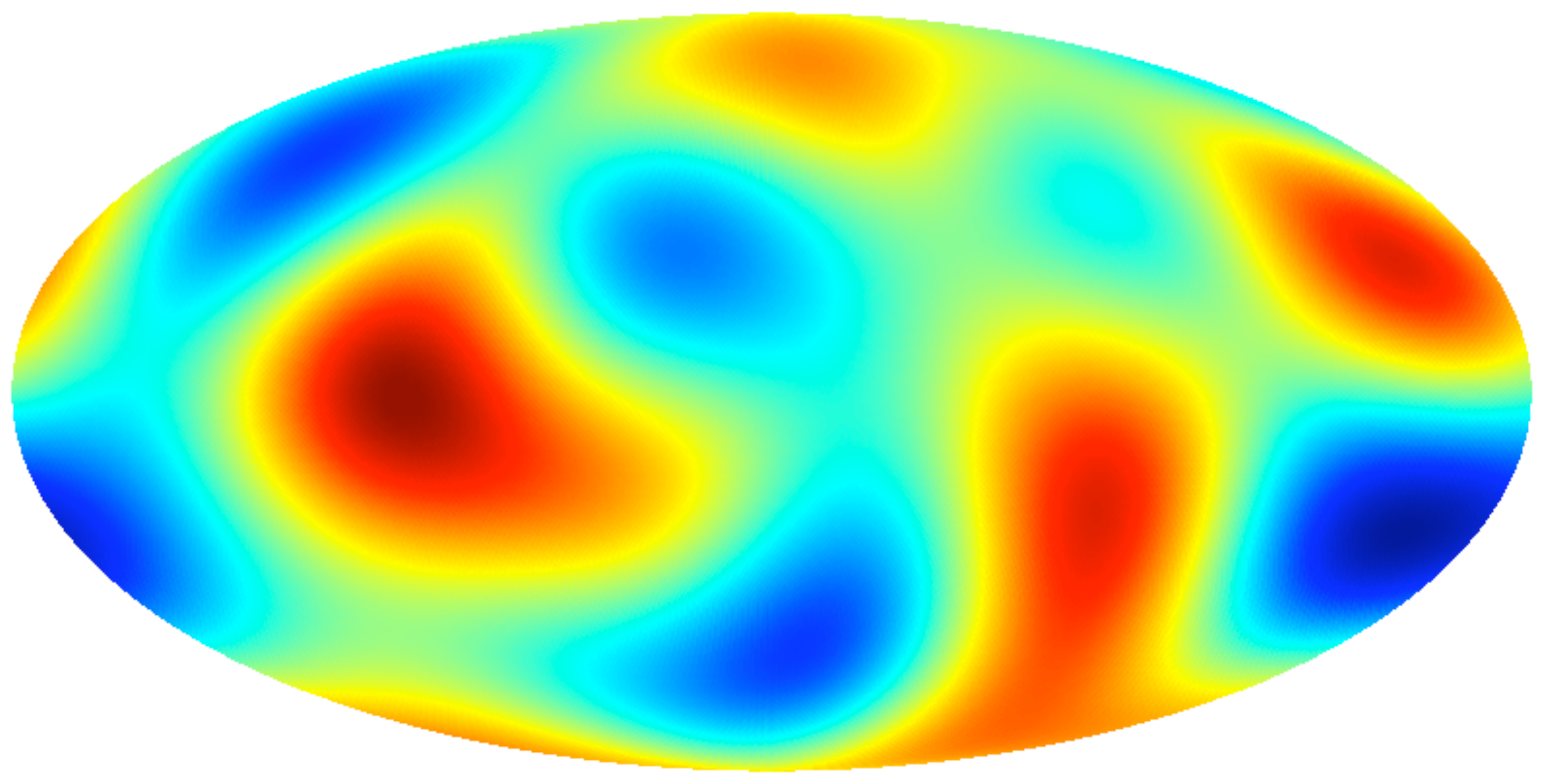}\\[2.0cm]
\hspace{-0.6cm}Starting map \hspace{6.4cm} Starting map \\[0.8cm]
\includegraphics[scale=0.3, angle=0, trim = 60mm 60mm 30mm 10mm]{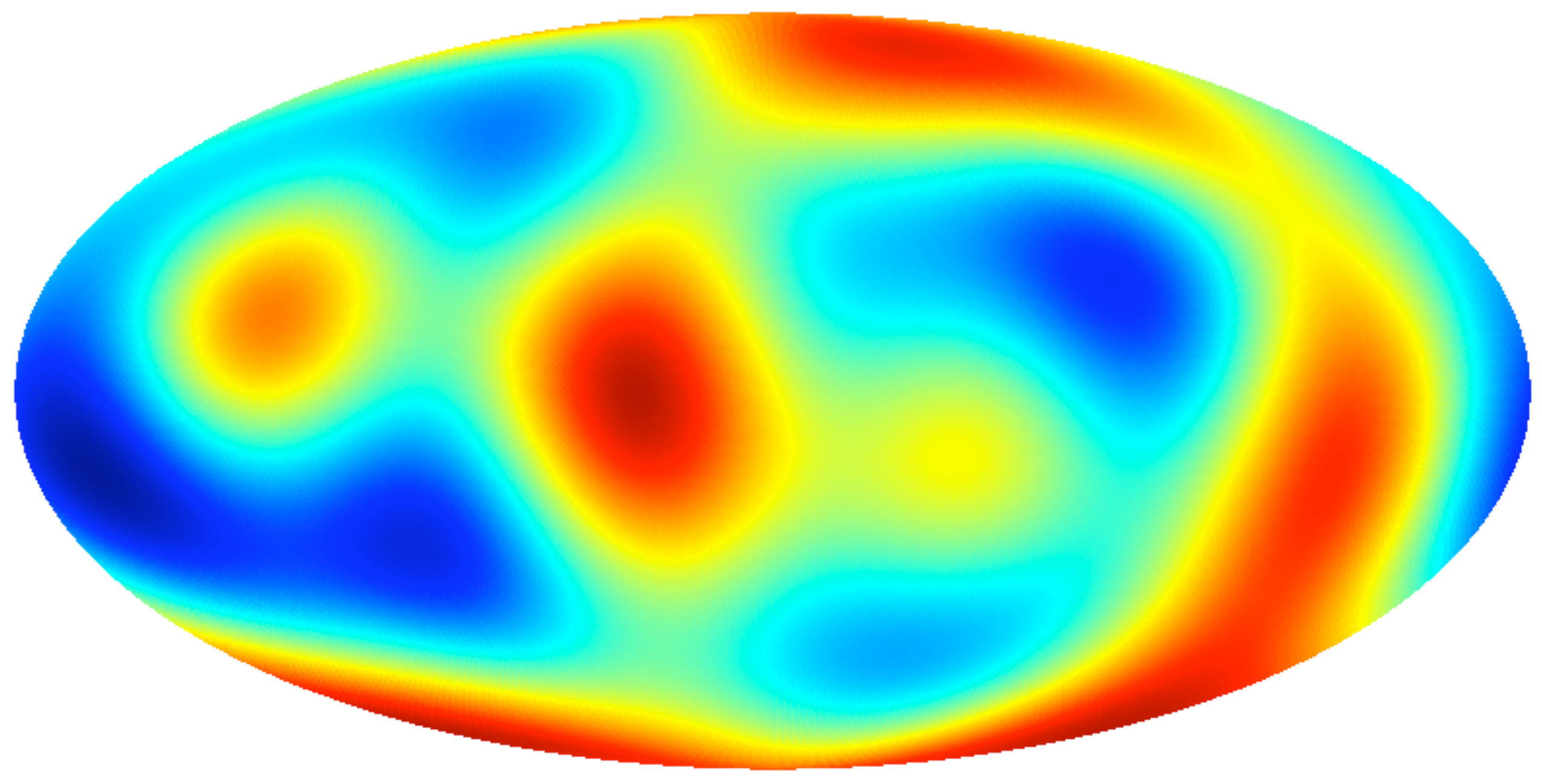}\hspace{2.5cm}
\includegraphics[scale=0.3, angle=0,  trim = 40mm 60mm 40mm 10mm]{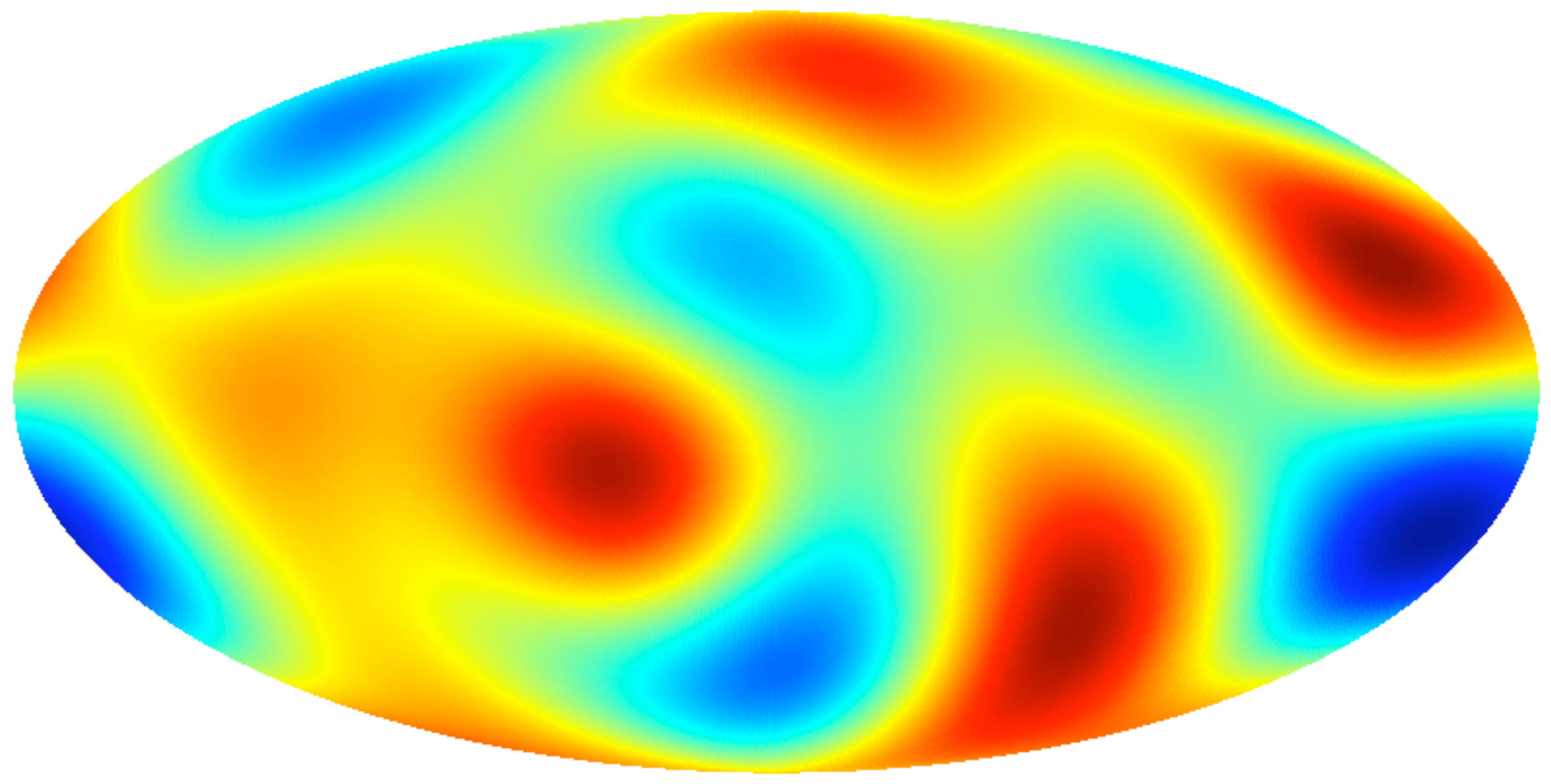}\\[2.0cm]
\hspace{-0.4cm}From cut-sky $\alm$ \hspace{6.0cm} From cut-sky $\alm$  \\[0.8cm]
\includegraphics[scale=0.3, angle=0, trim = 60mm 60mm 30mm 10mm]{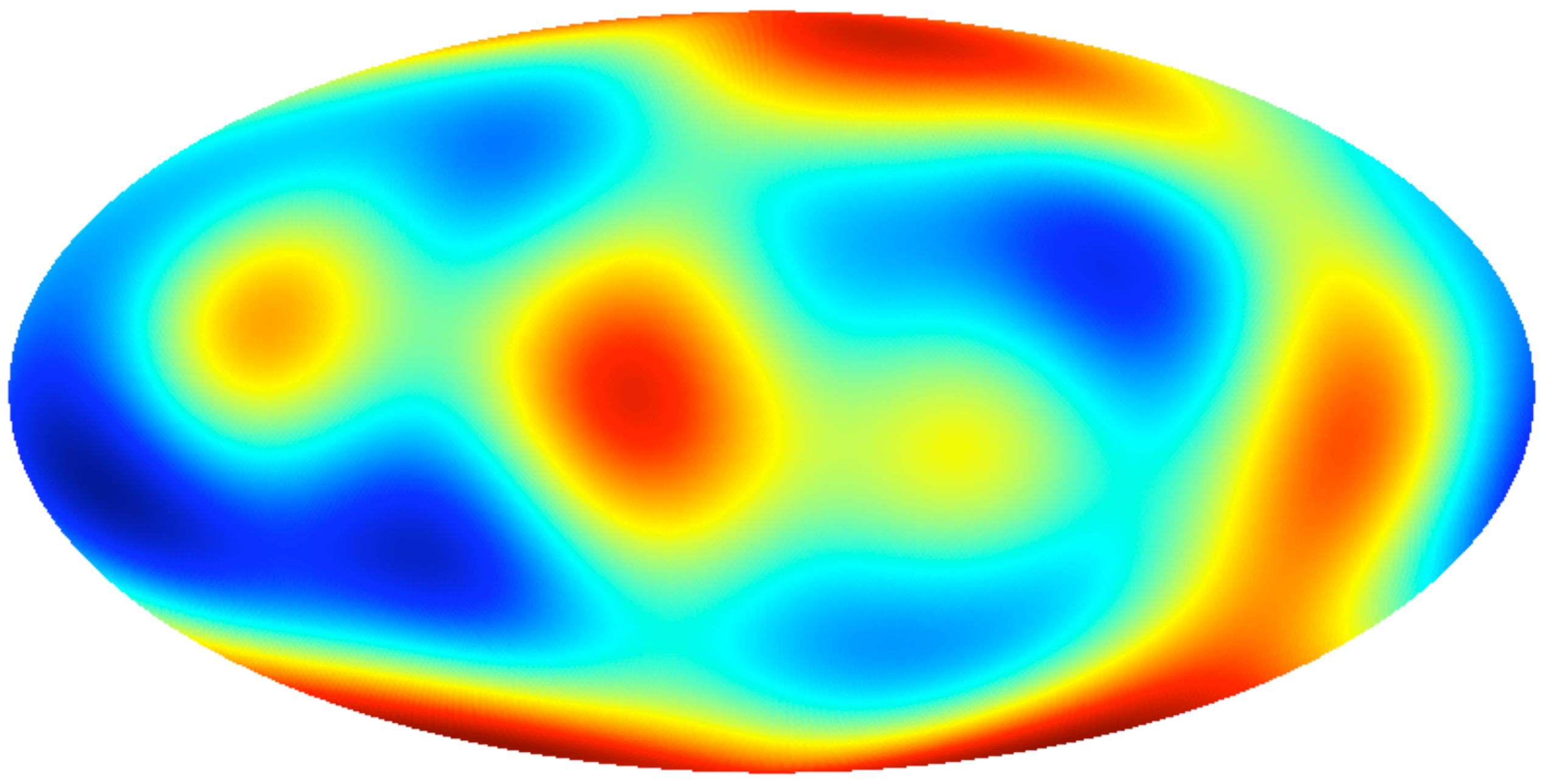}\hspace{2.5cm}
\includegraphics[scale=0.3, angle=0,  trim = 40mm 60mm 40mm 10mm]{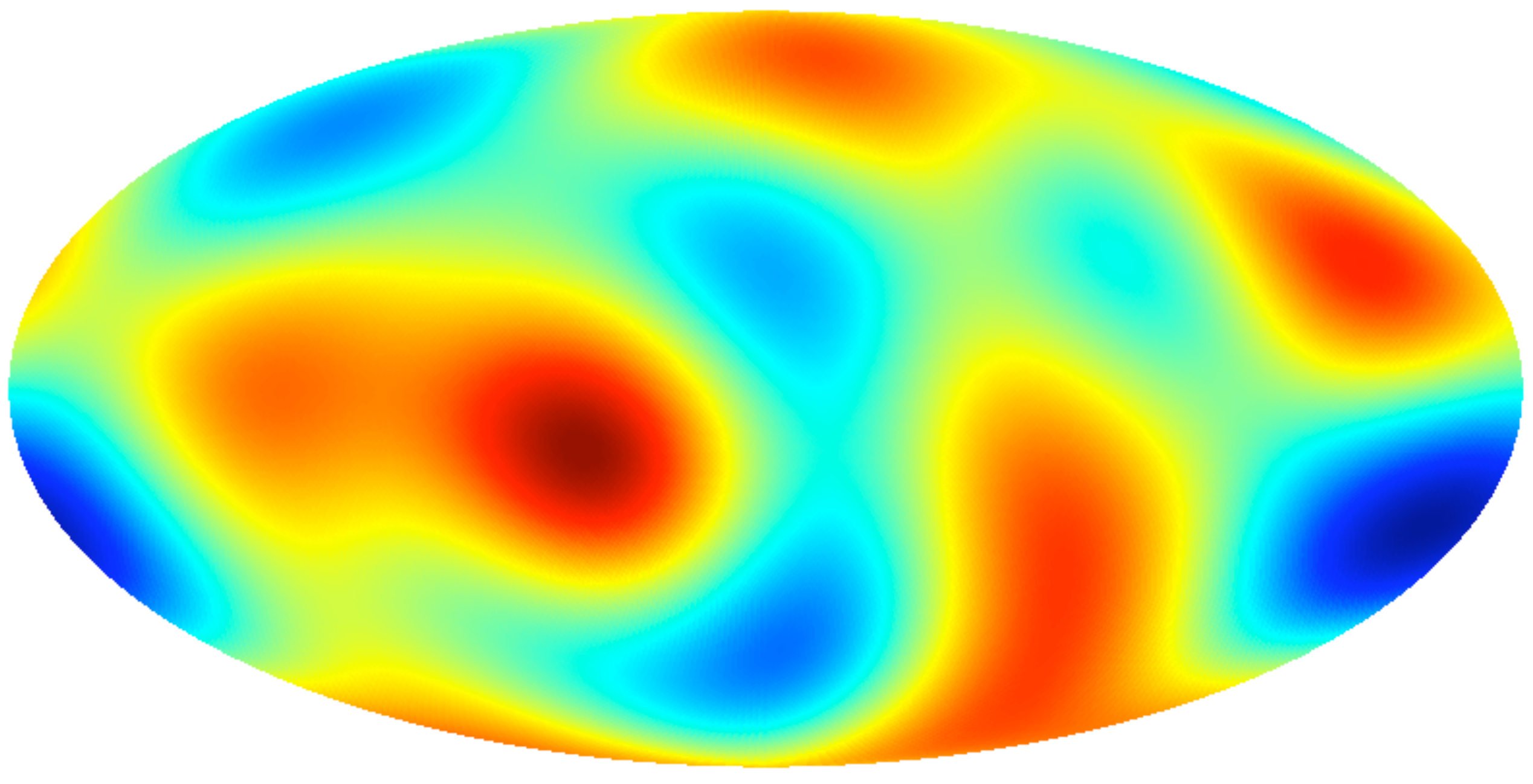}\\[2.0cm]
\hspace{-0.2cm}Full-sky reconstruction \hspace{5.0cm} Full-sky reconstruction  \\
\caption{Illustration of the efficacy of our reconstruction scheme for a mock
  galaxy survey with $N_g=10^6$. The top panel shows our starting map. The
  middle panels show the map made up from the cut-sky coefficients
  (i.e.\ using Eq.~\ref{full_alm}), while the bottom row shows the full-sky
  reconstruction that we adopted. The left columns show the full-sky case,
  while the right columns show the case where $\pm 4.5^\circ$ galactic cut
  (removing $\simeq 8\%$ of pixels) have been applied.  }
\label{reconstruct_method}
\end{figure*}

Depending on which tracer objects we are considering for our tests, a fraction
of the sky in the direction of the Galactic center may be obscured by stars
and dust, as well as point sources. These contaminated regions must typically
be avoided in all cosmological analyses of the large-scale structure, just
like for the case of the CMB.  In the CMB, for example, cosmological signal
from the contaminated regions can be recovered using multiwavelength
information \cite{Bennett_foregr,LILC}, though such cleaning may be risky and
prone to biases \cite{Tegmark:1995pn,Hansen:2006rj}.  For the case of LSS, data is given
by the object positions given in (e.g.\ galaxy) catalogs; thus inevitably we
are forced to deal with data that sample only parts of the sky.

The presence of the sky mask and measurement noise imply that
Eq.~(\ref{full_alm}) may be inaccurate in reconstructing the $\alm$. Instead, one
can implement a weighting scheme on the unmasked part of the sky. Such an
approach was advocated in \cite{deOliveiraCosta:2006zj} and applied to the CMB
and has been shown to optimally estimate the low-$\ell$ multipoles for cut
skies (under certain assumptions about the statistical properties of the sky).
 We now review this method and apply the reconstruction technique to
galaxy catalogs.

Let $x_i = \Delta_i$ represent the number of objects measured in a pixel
centered at the points $\Omega_i \equiv (\theta_i, \phi_i)$.  The information
in the catalog can then be represented by the vector $\wx = \bra{x_1,
  x_2....x_{\Npix}}$.  We wish to measure a set of multipole coefficients
$\alm$ which are reassigned for convenience as the vector $\wa=(a_1, a_2,
....a_M)$.  We choose to reconstruct only those coefficients with $\ell\leq
\lmaxrec$ which means that $M = \sum_{\ell=0}^{\lmaxrec} (2\ell+1)$.  We can
then write
\begin{equation}
\wx=\wy \wa+ \wn \,,
\end{equation}
where $\wy$ is a $\Npix \times M$ matrix containing the spherical harmonics --
$\wy_{ij}
\equiv Y_{\ell_j m_j }(\theta_i, \phi_i)$.  
Our conventions for casting the coefficients $\alm$ and spherical
harmonics $\Ylm$ in terms of purely real numbers, suitable for numerical
calculations, are given in Appendix \ref{app:conventions}.

The matrix $\wn$ has two contributions: the detector noise with
covariance matrix $\wN$ and the sky signal $\ws$ from multipole coefficients
that have not been included in the vector $\wa$, i.e.\ contamination from
$\alm$ with $\ell>\lmaxrec$. Assuming isotropic noise with zero mean, $\langle
\wn \rangle =0$, the covariance matrix can be written as
\begin{equation}
\wc \equiv \langle \wn \wn^{T}\rangle =\ws + \wN.
\end{equation}
The noise matrix $\wN$ is dominated by the shot noise, encoding the fact that
the number of sources in a pixel is only a statistical sample of the underlying
density field.

The covariance matrix
of the remaining contribution to the map $\ws$, from the uncertainty in the
multipoles that will not be reconstructed, is given by
\cite{deOliveiraCosta:2006zj}
\begin{equation}
  \ws_{ij}=\sum_{\ell=\lmaxrec+1}^{\lmaxtot}\frac{2\ell+1}{4\pi}
  P_{\ell}(\hat{\Omega_i} \cdot \hat{\Omega_j}) \Cl,
\label{eq:ws}
\end{equation}
where $\Cl$ is an estimate of the angular power spectrum of the galaxy survey
(see next subsection and Appendix \ref{app:power_spectrum} on how it is
calculated, and see Fig.~\ref{fig:Cl}). Note that the $\ell$ included in the
summation correspond to those $\alm$ that are {\it not} included in the
vector $\wa$. Heuristically, the structures with $\ell>\ell_{\rm rec, max}$
serve as noise for the reconstructed signal at $\ell\leq\ell_{\rm rec,
  max}$. Here we adopt $\lmaxtot=50$, which is more than sufficient for the
reconstruction of multipoles out to $\lmaxrec=4$.

The aim is then to find an approximation $\hat{\wa}$ to the true $\wa$ that is
unbiased and has minimum variance.  For problems such as this where there are
far more pixels than parameters for which we need to solve, the optimal
solution to the above system of equations is 
\cite{Tegmark:1997vs}
\begin{equation}
\hat{\wa}=\Wb \wx, \quad \Wb \equiv [\wy^T \wc^{-1} \wy]^{-1} \wy^T \wc^{-1}
\label{wa}
\end{equation} with a covariance matrix 
\begin{equation}
\wS \equiv \langle \hat{\wa} \hat{\wa}^{T}\rangle -\langle \hat{\wa} 
\rangle \langle \hat{\wa} \rangle^T = [\wy^T \wc^{-1} \wy]^{-1}.
\label{sigma}
\end{equation}
Here $\wS$ is the covariance matrix of the reconstructed $\alm$.  With
full-sky coverage, the covariance matrix {\boldmath $\Sigma$} is diagonal;
with the sky cut, it is not. In the latter case the algorithm corrects for the
mixing of the different $(\ell, m)$ at the cost of larger error bars
\cite{deOliveiraCosta:2006zj}.

In Fig.~\ref{reconstruct_method} we illustrate the effectiveness of the above
reconstruction method to estimate the $\alm$, and contrasted to the alternative
approach of merely using Eq.~(\ref{full_alm}). Using a subset of known
$\alm^{\rm true}$ for $\ell=2 - 4$, we generated a mock dataset $\wx$
representing a catalog of $10^6$ objects with noise $\wN$; the details of
the computation of $\wN$ are shown in Appendix \ref{app:covmatr}. The middle
panels show the map made up from the cut-sky coefficients (i.e.\ using
Eq.~\ref{full_alm}), which is clearly biased.  The bottom panels of
Fig.~\ref{reconstruct_method} show the reconstructed density maps using our
algorithm. Left panels show the case when full-sky information is available,
while right panels show the case when $\pm 4.5\degr$ galactic cut has been
applied (i.e.\ when about $\simeq 8\%$ of the area has been removed).  The
improved accuracy with which the multipoles are reconstructed using our
selected method is clearly seen.

\subsection{Generating mock galaxy catalogs}
\label{sec:mock}

We now describe the technology to generate synthetic, pixelated maps of galaxy
counts.  We wish to create a field with the number density given by
\begin{equation}
\Delta\bra{\theta, \phi}=\sum^{\lmaxtot}_{\ell=0} \sum_{m} \alm Y_{\ell m}\bra{\theta, \phi},
\end{equation}
so that it is consistent with the density field $\nu(\Omega)$. Since we are
mainly interested in testing statistical isotropy on large scales, generating
maps out to $\lmaxtot=50$ is sufficient. 

\begin{figure}[t]
\centering
\includegraphics[width=\linewidth]{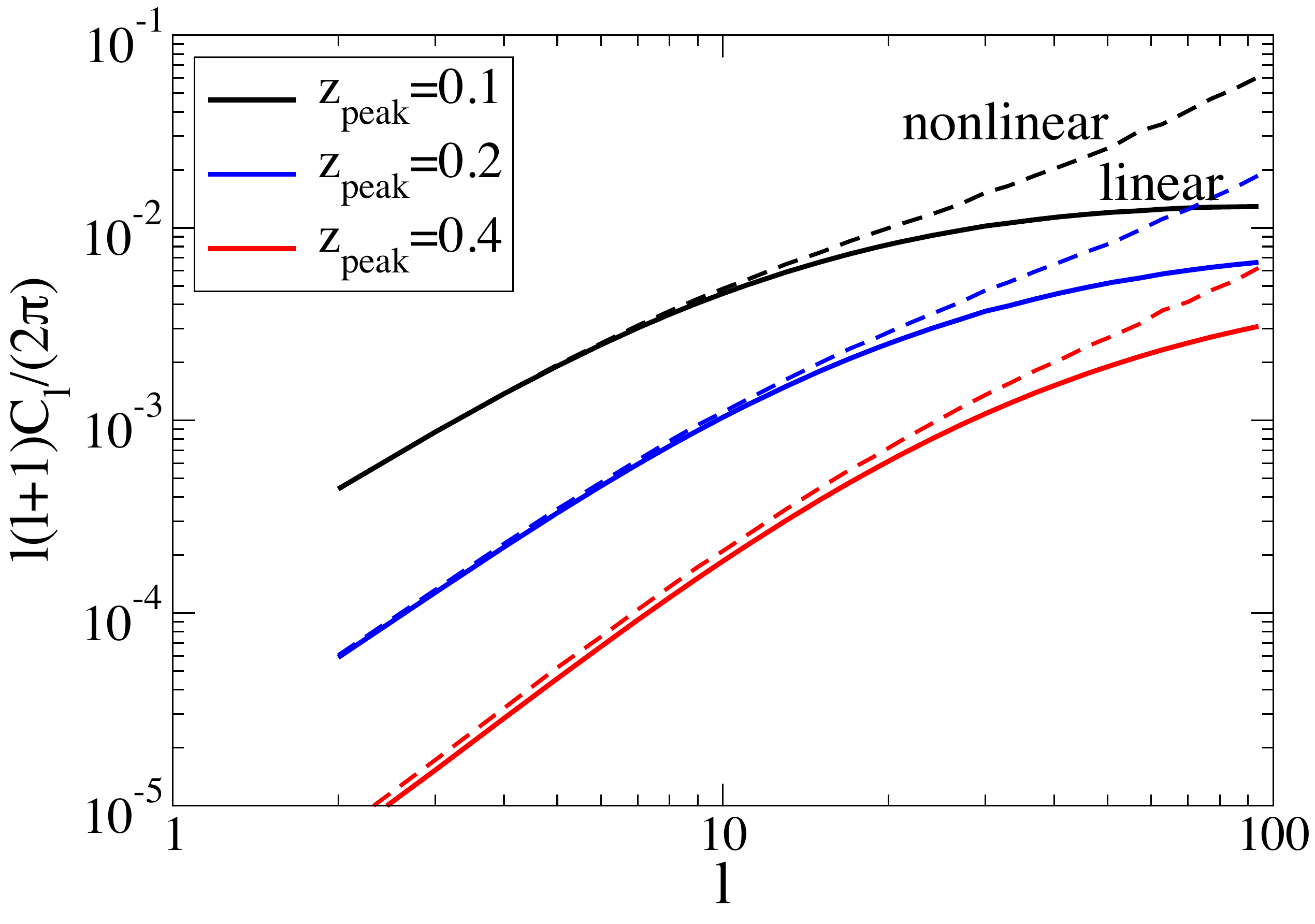} 
\caption{The theoretical angular power spectra calculated using the radial
  number density function $n(z)$ from the SDSS for different redshifts at
  which the radial number density of objects peaks. See Appendix
  \ref{app:power_spectrum} for details of the calculation. }
\label{fig:Cl}
\end{figure}

The starting ingredient for mapmaking is the theoretical angular power
spectrum of dark matter, $\Cl$, which we calculate according to the
prescription given in Appendix \ref{app:power_spectrum}. Notice that the
number density of galaxies, $dN/dz$, is necessary for calculation of the
theoretical angular power spectrum (see Appendix
\ref{app:power_spectrum}). Here we assume a number density of the form
\cite{HTBJ}
\begin{equation}
n(z) = \frac{z^2 e^{-z/z_0}}{2 z_0^3}.
\label{eq:m=nz}
\end{equation}
that peaks at $\zpeak = 2z_0$. In Fig.~\ref{fig:Cl} we show the angular power
spectra for $\zpeak=0.1, 0.2$ and $0.4$; the angular spectra are of course
smooth because they correspond to matter overdensity projected along the line
of sight. This figure also shows that nonlinearities enter at $\ell\gtrsim
20$; in our analysis, we are interested in reconstructing $\ell$ of a few, and
thus it is sufficient to use the linear angular power spectra.

Details of how we first generate a smooth projected matter density map, and
from it the distribution of galaxies on the sky, are spelled out in Appendix
\ref{app:mock}.  In brief, starting with the choice of the form of the galaxy
density $dN/dz$ and its peak value $\zpeak$, we use the calculated theoretical
$\Cl$ at $\ell\leq 50$ to generate a set of random $\alm$ with zero mean and
variance $\Cl$. We then use the HEALPix \cite{healpix} routine {\tt alm2map} to generate a
smooth density map.

Next, we generate a galaxy catalog with $N_g$ galaxies consistent with the
smooth map; details are described in Appendix \ref{app:mock}.  Starting with
the coefficients $\Cl$, we generate 100 random sets of $\alm$ coefficients,
and from each we produce 3 realizations of the corresponding galaxy catalog.
This gives us a total of 300 realizations of galaxies on which we base the
statistics. This number was smaller than we might have liked, 
because the galaxy generation step is time consuming for large $N_g$ ($\gtrsim 10^8$). 
We found, however, that this number of realizations produced sufficiently accurate results.

\subsection{Testing the reconstruction accuracy}
\label{sec:testing}

We now investigate how the accuracy of the estimated quantities of interest
(i.e.\ the $\alm$ and the multipole vectors) depends on the characteristics of
the survey -- its depth, and the sky density of tracer objects.  We follow the
procedure outlined in \cite{deOliveiraCosta:2006zj} and optimally reconstruct
the full-sky $\alm$ from each mock catalog using the method described in
Sec.\ \ref{sec:reconstruction}.  The corresponding MVs are subsequently
extracted from the $\alm$ using the publicly available code \cite{MV_code}.

\begin{figure*}[t]
\begin{center}
\includegraphics[width=6in]{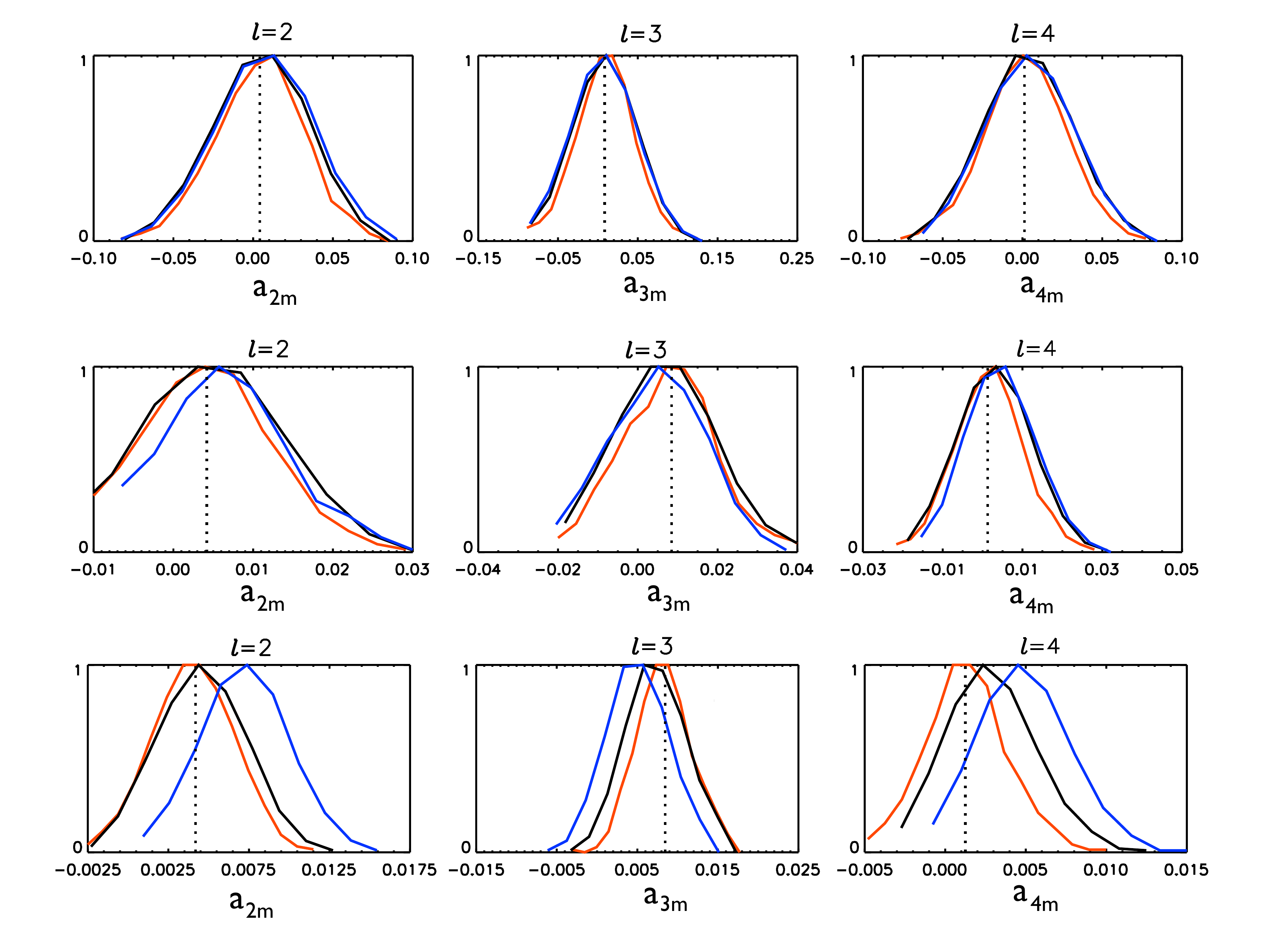} 
\end{center}
\caption{Reconstruction of the coefficients $\alm$ for $\ell=2 - 4$ for 300
  realizations with $N_g=10^4$ (top row), $N_g=10^5$ (middle row) and $10^6$
  (bottom row).  We show results for three HEALPix map resolutions:
  pixelizations of $\Nside$ = 4 (blue), 8 (black) and 16 (red). The total
  number of pixels on the full sky is $\Npix = 12\times \Nside^2$. The true
  underlying $\alm$ are shown by the dotted line.  An increase in resolution
  (i.e.\ higher $\Nside$) improves the accuracy of the reconstruction only for
  mock catalogs of size $N_g = 10^6$ and higher.}
\label{fig:nside}
\end{figure*}

\medskip
{\it Sufficiently fine pixelization.}  In our approach, one performs counts-in-cells of
galaxies on the sky.  To test effects of finite resolution imposed by
pixelization, we consider a single realization of a galaxy survey with $N_g$
objects and reconstruct the $\alm$ using different values of the HEALPix
parameter $\Nside$, where the number of pixels is $\Npix = 12 \Nside^2$ (the
angular size of a pixel is roughly $\theta_{\rm pix} \approx 60\degr/\Nside$).

Figure \ref{fig:nside} shows the reconstructed $\alm$ for three choices of
$\Nside$ and for 300 realizations of mock catalogs with $N_g=10^5$, $10^6$ and
$10^7$ objects.  The width of each distribution encapsulates the variance on
the measurement of the multipole coefficient and remains relatively unchanged
as the pixelization varies. Clearly, for catalogs with smaller galaxy
density (i.e.\ larger shot noise), an increase from $\Nside=4$ to $\Nside=8$
improves the accuracy of the reconstruction only marginally, rendering
$\Nside=8$ sufficient to guarantee that the contribution to noise is dominated
by the shot noise for a survey with $10^5$ objects (which is reduced with
increased resolution).  For larger number density catalogs ($N_g=10^6$ in
the Figure), a higher pixelization of $\Nside=16$ does make a slight
improvement in the $\alm$ estimation but not enough to warrant the additional
computation time. For the rest of the analysis, $\Nside=8$ will be used.

\begin{figure*}[t]
\begin{center}
\includegraphics[width=\linewidth]{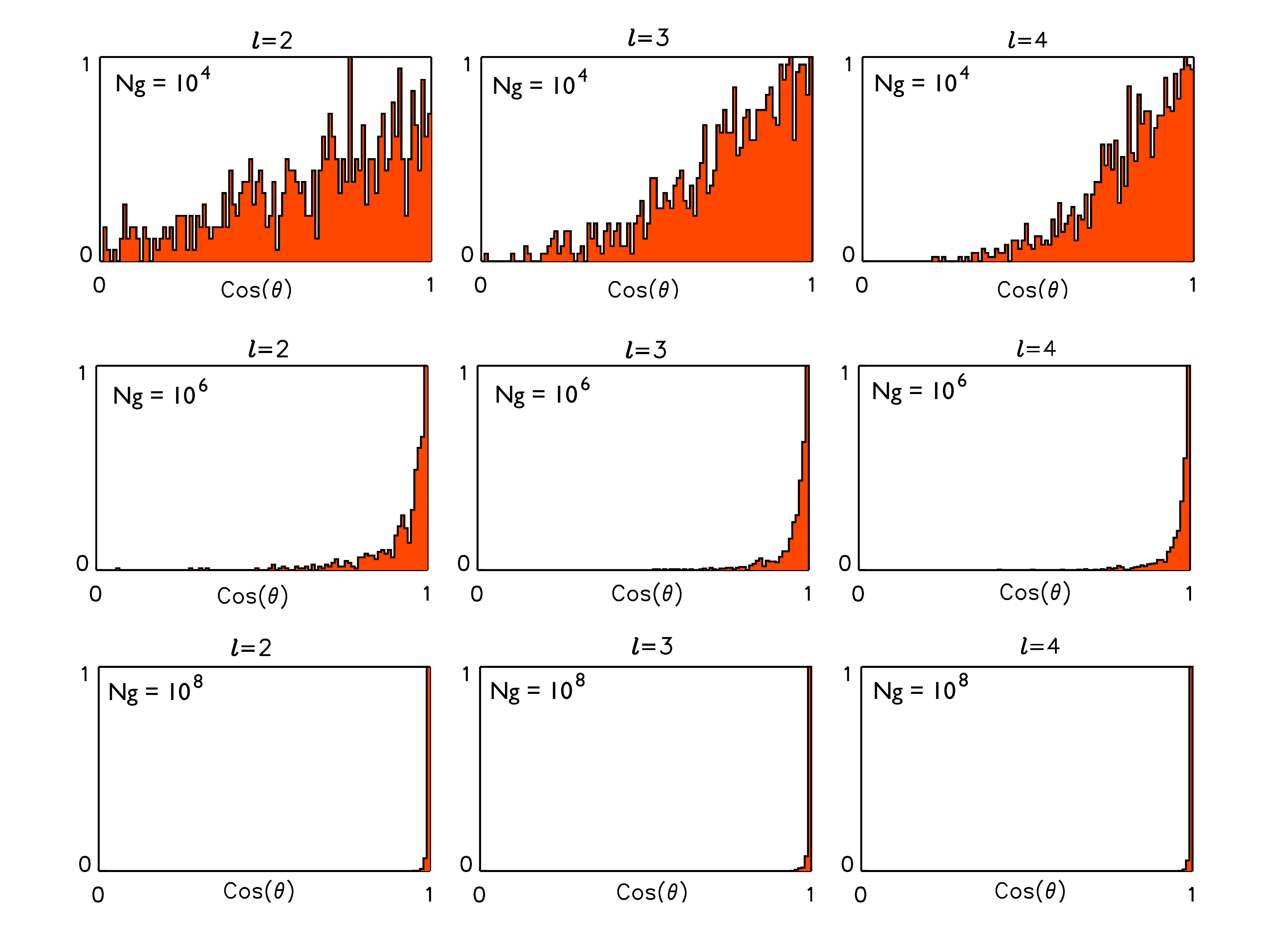}
\end{center}
\caption{Effects of the number density of LSS
  tracers. Histograms of the dot product of the true and reconstructed MVs,
  $\cos(\Theta^{(\ell, i)}) = v^{(\ell,  i)}\cdot v_{\rm true}^{(\ell, i)}$, from $300$
  realizations for surveys with $N_g=10^4$ (top row), $N_g=10^6$ (middle row),
  and $N_g=10^8$ (bottom row). We assume a fixed pixelization level of
  $\Nside=8$, and the radial distribution of objects $\zpeak=0.2$.  An
  improvement in accuracy is indicated by a closer proximity to $1$, at which
  the MVs are reconstructed perfectly. The narrowing of the histograms
  suggests a considerably better recovery of the MVs as the survey size is
  increased.}
\label{fig:hist}
\end{figure*}

\medskip
{\it Sky density of objects.}  The projected sky density of objects will vary
dramatically between different classes of objects. For example, using all
galaxies as tracers will provide higher counts than using only the luminous
red galaxies, and those in turn have a much higher density than quasars or
gamma-ray bursts. More accurate reconstruction of the underlying density field
is expected to be revealed from catalogs with a larger numbers of objects.
Therefore, the number of tracer objects in the survey is likely to play an
important role in the precision of our tests.

Let us examine the effect of the available number of sources in the
reconstruction accuracy of multipole vectors $v^{({\ell}, i)}$. To do that, we
compare the MVs $v^{(\ell,i)}$ obtained from the reconstructed $\alm$ to those
$v_{\rm true}^{(\ell,i)}$ which
corresponds to the $\alm$ used to generate the density map of the mock
catalog.  The results are quantified by the angles $\Theta^{(\ell, i)}$
\begin{equation}
\cos\left (\Theta^{(\ell, i)}\right ) = v_{\rm true}^{(\ell,i)}\cdot v^{(\ell, i)}
\end{equation}
from $300$ realizations as a function of the total number of galaxies $N_g$.
Fig.~\ref{fig:hist} shows the histograms for catalogs increasing with $N_g =
10^4, 10^6$ and $10^8$.  The loss of accuracy is gauged by how much
$\cos(\Theta^{(\ell, i)})$ deviates from perfect reconstruction where its
value is unity. The widths of the one-sided distributions decrease
dramatically as the number of objects in the survey $N_g$ increases,
indicating substantial increase in the ability of a galaxy catalog to
represent the underlying density field. The rapid degradation in the accuracy
of estimated MVs for $N_g\ll 10^6$ already hints that large catalogs may be
required to test SI reliably.

\begin{figure*}[t]
\begin{center}
  \includegraphics[width=\linewidth]{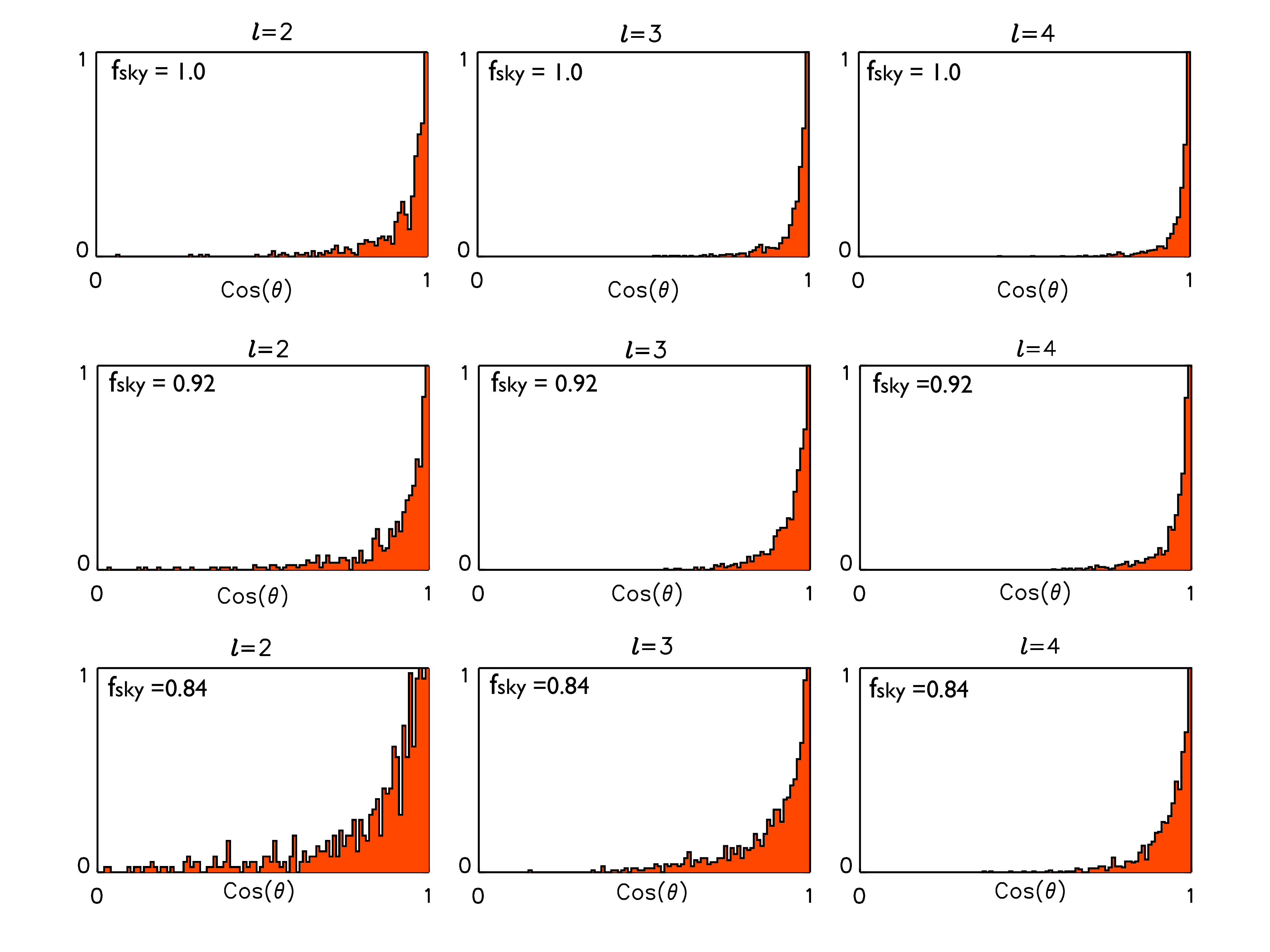}
\end{center}
\caption{Effects of the sky cut. Histogram of the dot products of the true and
  reconstructed MVs $\cos(\Theta^{(\ell, i)}) = v_{\rm true}^{(\ell,i)}\cdot
  v^{(\ell, i)}$, from $300$ realizations when the following areas of the sky
  are removed: $0$ (top row), $\pm 4.5\degr$ (middle row), and $\pm 9\degr$
  (bottom row). The second and third case correspond to $\fsky\simeq 0.92$ and
  $0.84$ respectively.  The pixelization level is fixed at $\Nside=8$ and we
  assume a survey with $N_g = 10^6$ objects which radial distribution of
  tracers that peaks at $\zpeak=0.2$.}
\label{fig:hist_cut}
\end{figure*}

\begin{figure*}[t]
\begin{center}
\includegraphics[width=\linewidth]{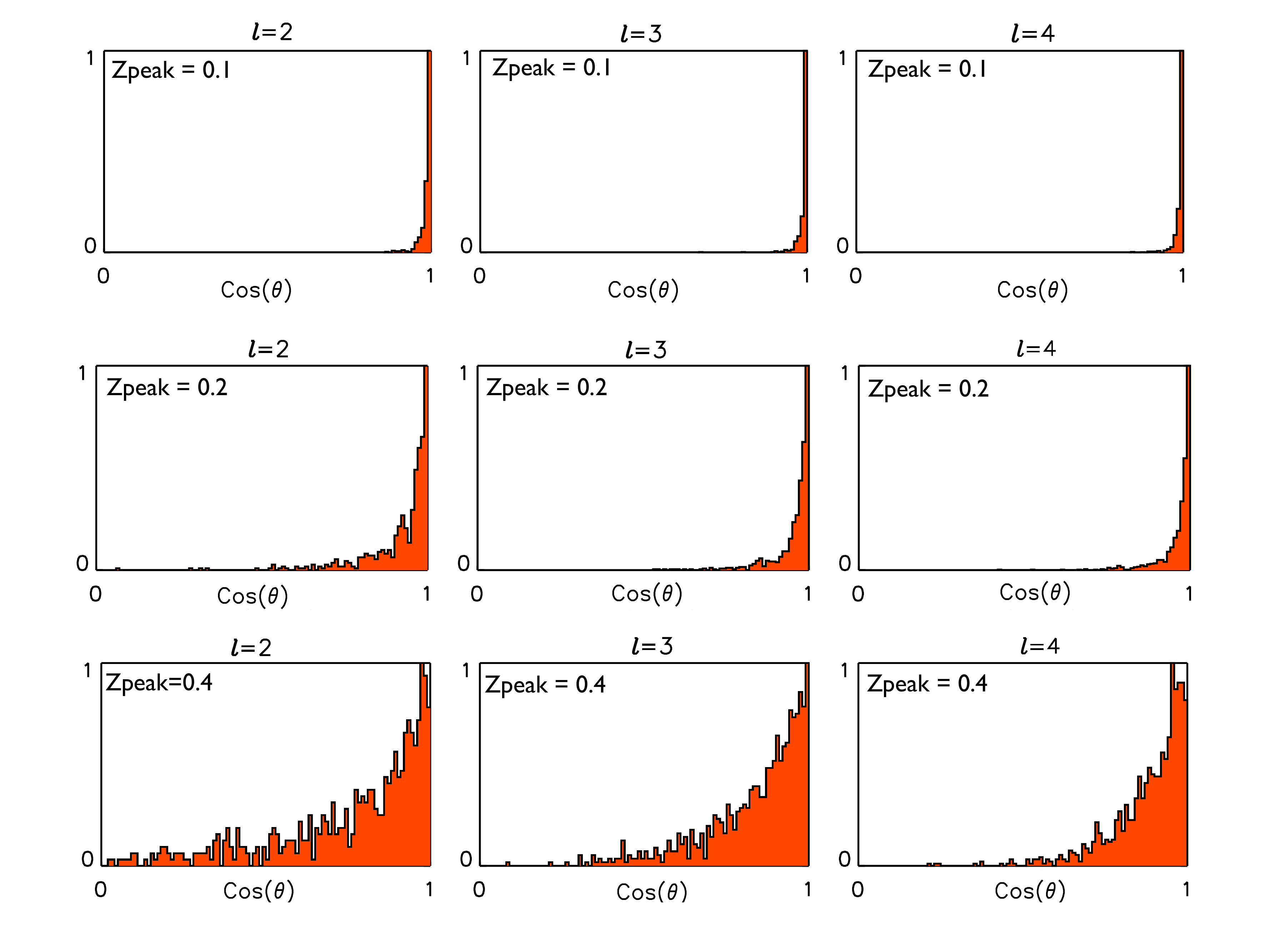}
\end{center}
\caption{Effects of the survey depth. Histogram of the dot products of the
  true and reconstructed MVs $\cos(\Theta^{(\ell, i)}) =v_{\rm
    true}^{(\ell,i)}\cdot v^{(\ell, i)}$ from $300$ realizations of a full sky
  for a surveys with $\zpeak=0.1, 0.2$ and $0.4$ (top to bottom rows). The
  adopted pixelization is $\Nside=8$ and the total number of sources is $N_g =
  10^6$.}
 \label{hist_zpeak04}
\end{figure*}

\begin{figure*}[t]
\begin{center}
\begin{tabular}{ccc} 
\includegraphics[scale=0.5]{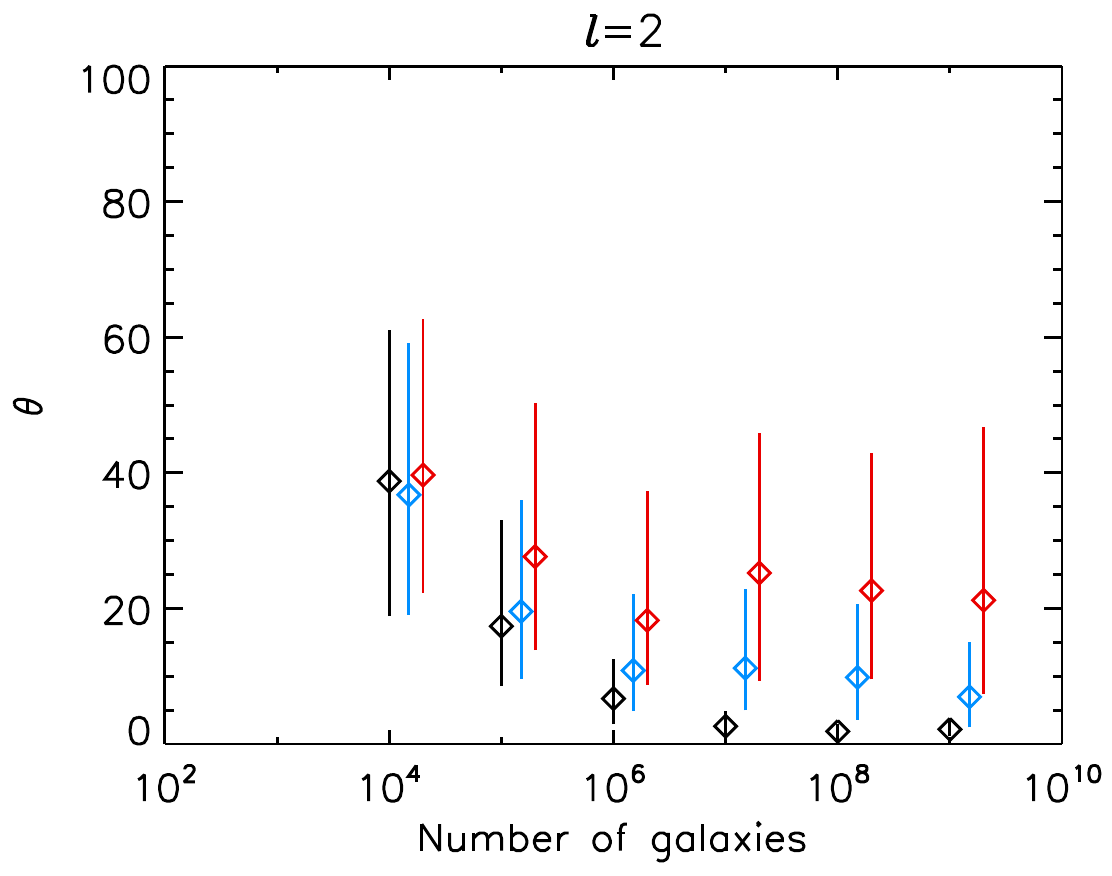}  
\includegraphics[scale=0.5]{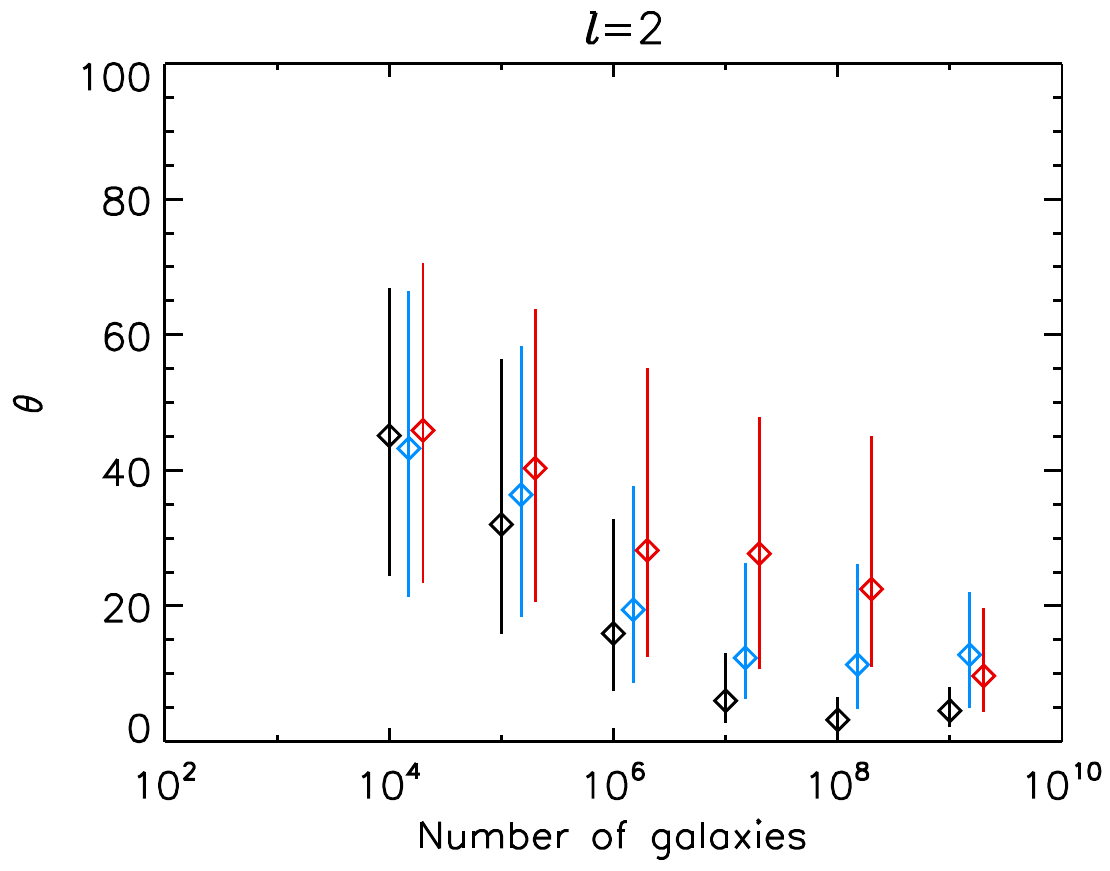} 
\includegraphics[scale=0.5]{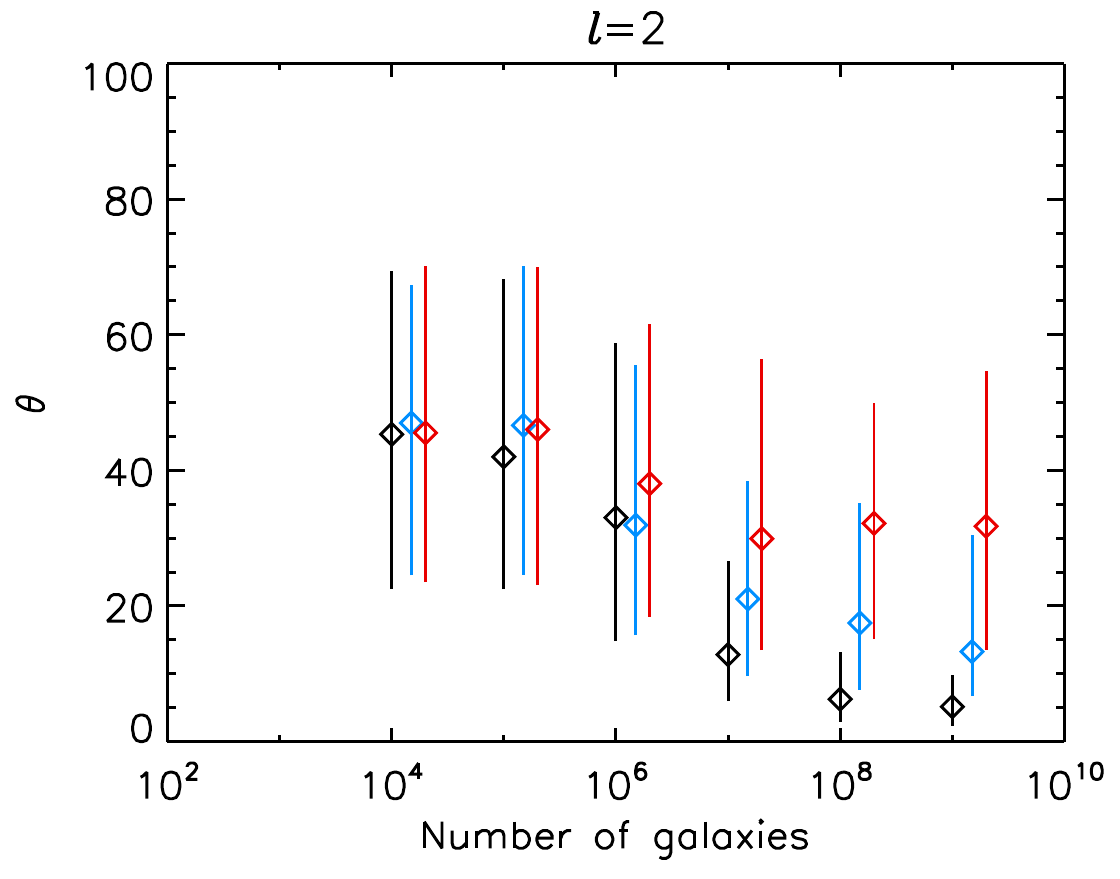}\\ 
\includegraphics[scale=0.5]{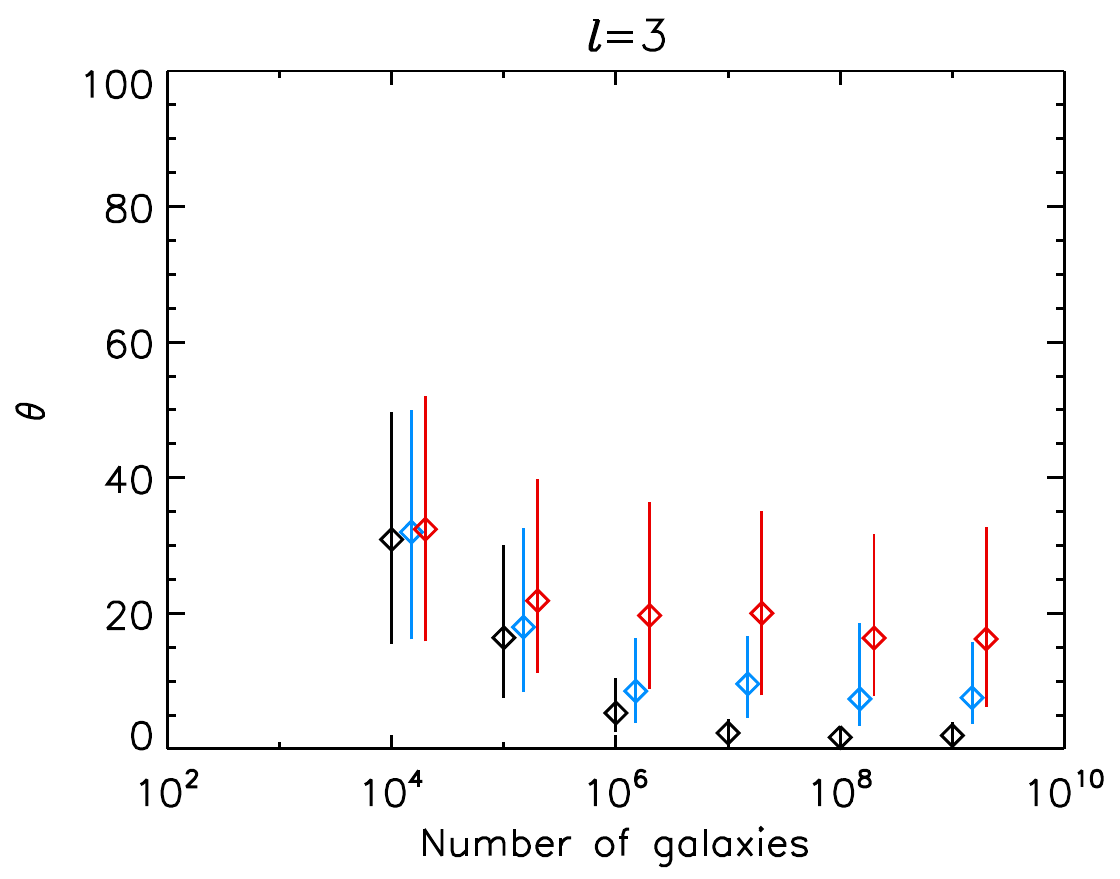}
\includegraphics[scale=0.5]{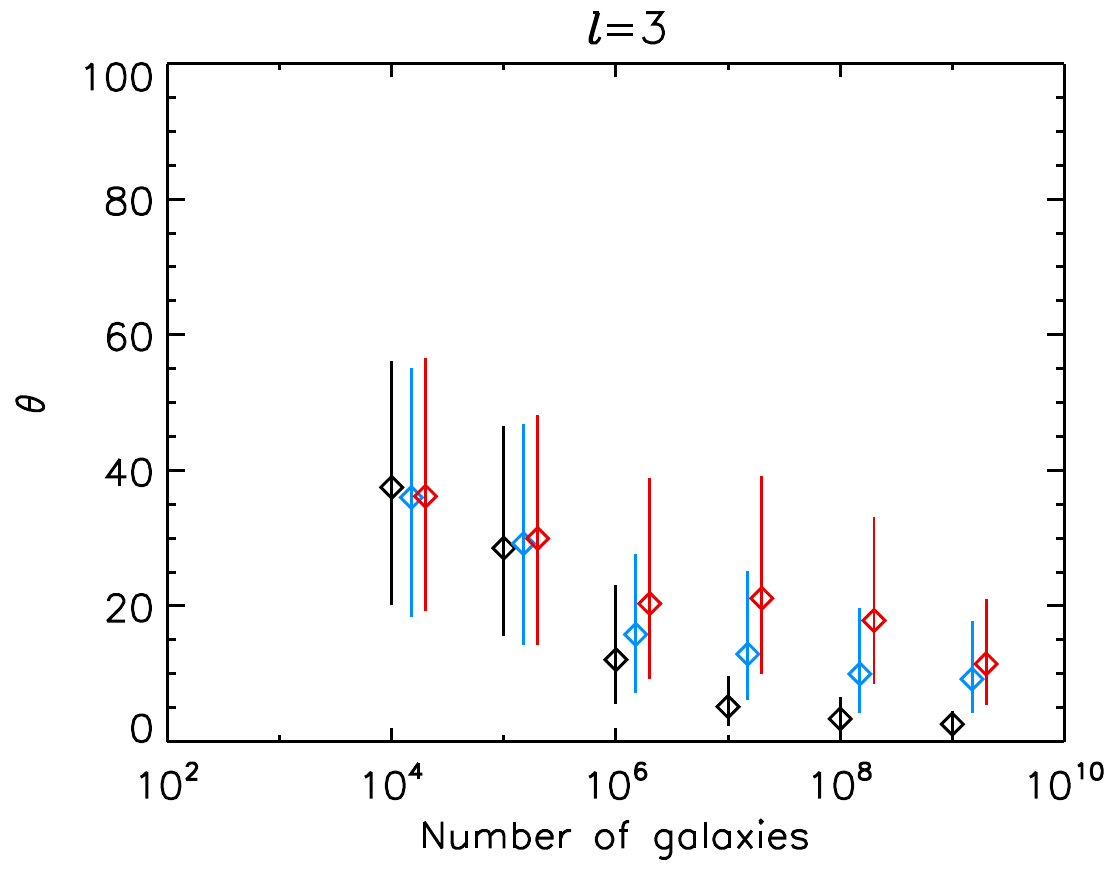} 
\includegraphics[scale=0.5]{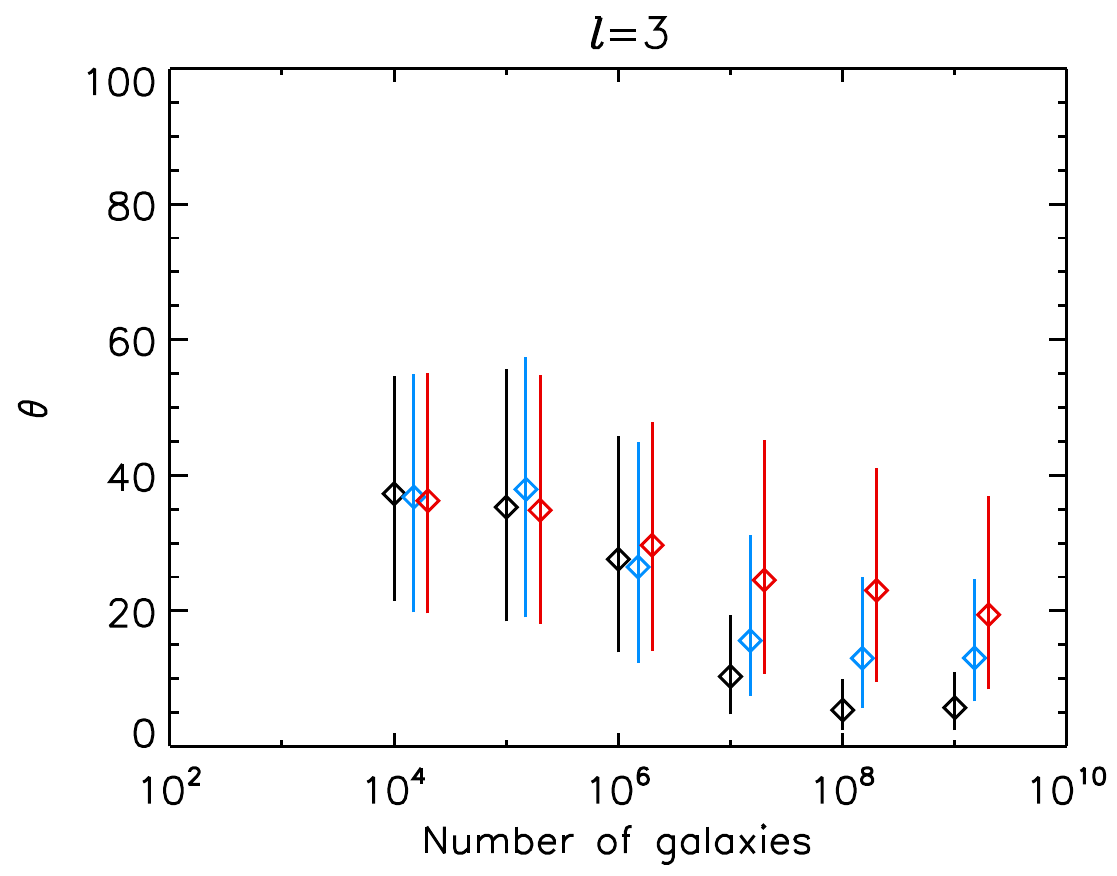}\\
\includegraphics[scale=0.5]{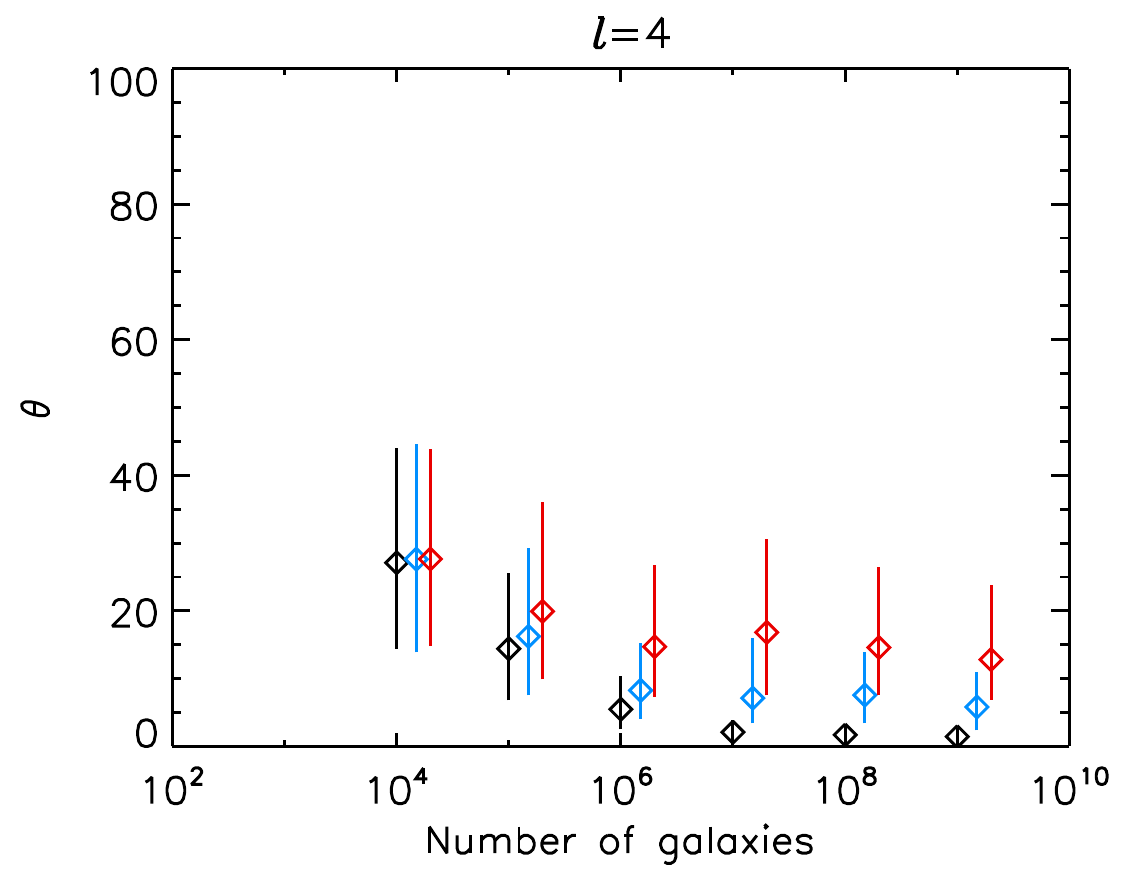}
\includegraphics[scale=0.5]{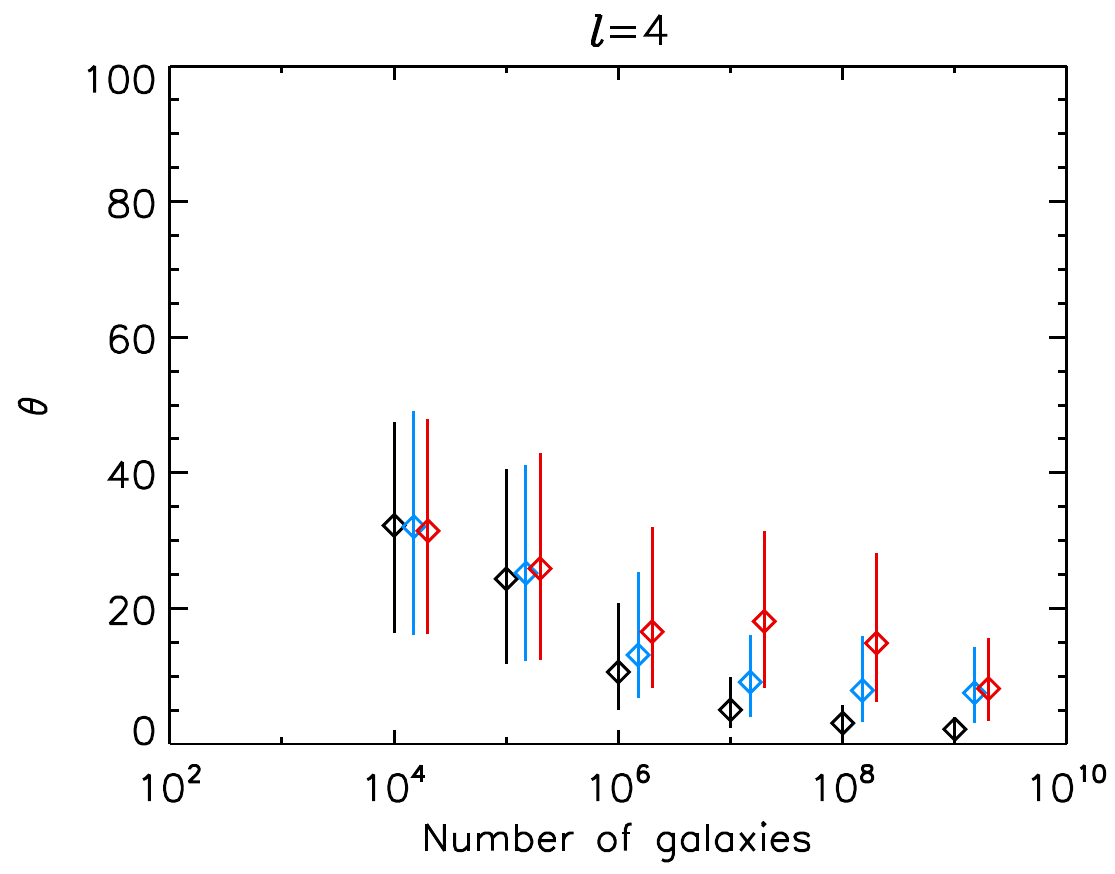} 
\includegraphics[scale=0.5]{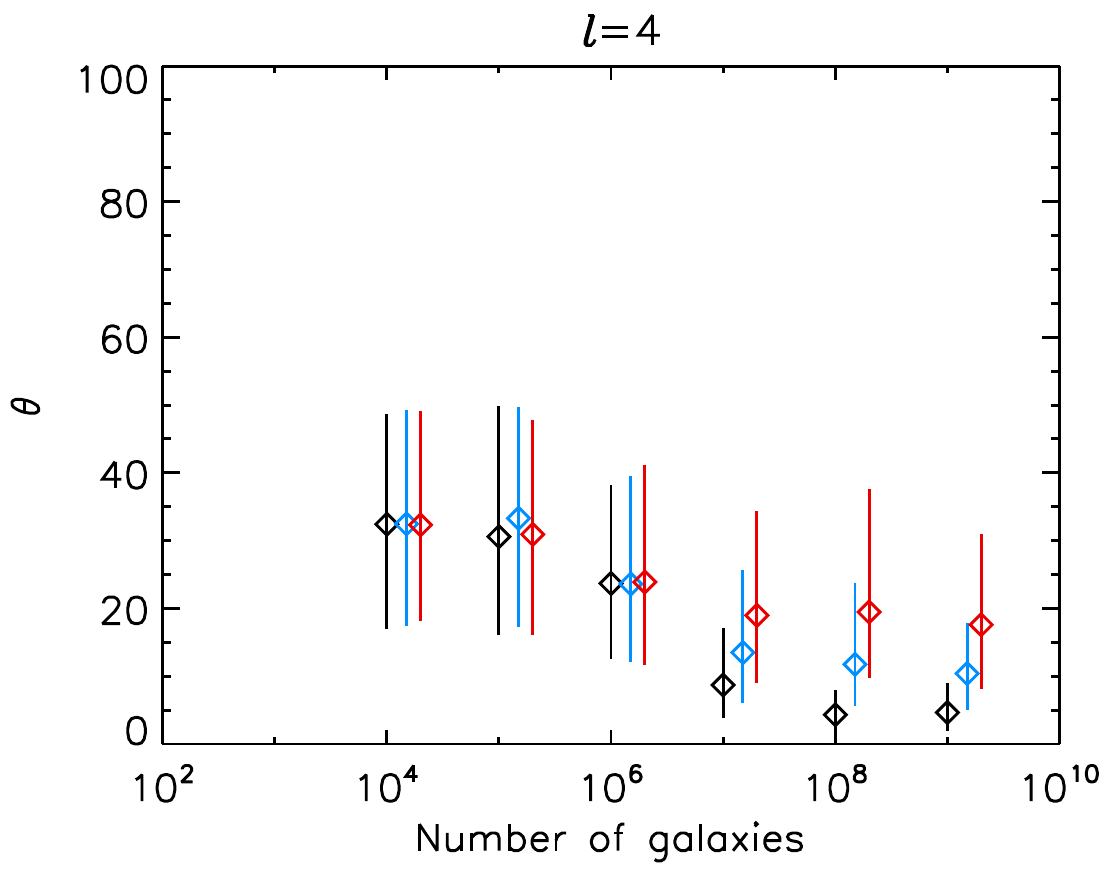}
\end{tabular}
  \end{center}
  \caption{Summary of all effects. Plot of the average angle between the
    reconstructed and input MVs, $\Theta^{(\ell, i)} =\arccos\bra{v_{\rm
        true}^{(\ell,i)}\cdot v^{(\ell, i)}}$ as a function of $N_g$, with
    error bars indicating the $16-84$ percentile ranges for different choices
    of $\zpeak$; $\zpeak=0.1$ (left column), $\zpeak=0.2$ (middle column) and
    $\zpeak=0.4$ (right column) for $\ell=2$ (top row), $\ell=3$ (middle row)
    and $\ell=4$ (bottom row). The different colors indicate different sky
    masks: $0^{\circ}$ (black), $\pm 4.5^{\circ}$ (blue) and $\pm 9^{\circ}$
    (red).  }
\label{fig:MVplot}
\end{figure*} 

\medskip
{\it Sky cut.}  It is likely that, for most tracer objects of the large-scale
structure, parts of the sky will have to be masked either to incomplete
observations, or to the presence of point sources\footnote{Gamma-ray bursts
  may be an exception here, but tests of SI might prove challenging given that
  the density of the bursts will be orders of magnitude lower than that of
  galaxies.}.  The removal of data from part of the sky will inevitably
degrade the accuracy of the reconstruction of the $\alm$, multipole vectors,
and any other statistics. In Ref.~\cite{Copi:2003kt}, it was shown that accurate
reconstruction of the MVs of the CMB temperature anisotropy (to about a degree
or better) requires a galaxy cut no larger than a few degrees. Here we perform
a similar analysis for the MVs of the large-scale structure.

We assume the following isolatitude cuts: $0\degr$, $\pm 4.5\degr$ and $\pm
9\degr$, corresponding respectively to the full sky, 8\%, and 16\% of the sky
area masked.  Given that our test skies are statistically isotropic, the
fiducial orientation of the cuts is irrelevant. And while the fact that
isolatitude cuts are assumed is certainly a simplifying assumption, we do not
expect that the azimuthally uneven cut with roughly the same area will lead to
very different results. We leave the analysis with cuts with more general
geometries for future work when cuts motivated by specific surveys will be
used.

Fig.~\ref{fig:hist_cut} shows histograms of the dot products of the true input
MVs and the reconstructed MVs $\cos(\Theta^{(\ell i)}) = v_{\rm
  true}^{(\ell,i)}\cdot v^{(\ell, i)}$ for 300 realizations of a galaxy survey
with $N_{g}=10^6$ and the three different cuts.  When only part of the sky is
observed, mixing of the higher multipoles, $\ell\gtrsim 1/\theta_{\rm cut}$,
with those describing the reconstructed sky ($\wa$) is introduced. The
reconstruction method implemented here accounts for this mode-mixing in the
reconstructed multipoles $\wa$ at the cost of larger error bars,
indicated by the increase in the widths of the histograms as $\fsky$ decreases.

\medskip
{\it Survey depth.}  Reconstruction also depends on the depth of the survey,
which we here parametrize with the peak of the redshift distribution of
sources $\zpeak$. While a deeper survey enables a larger effective
representative volume of the universe from which to test statistical isotropy,
it turns out that the angular power spectrum has a lower amplitude for a
deeper survey; see Fig.~\ref{fig:Cl}. This is why deeper surveys lead to
worsening in the reconstruction of the multipole
vectors. Fig.~\ref{hist_zpeak04} shows a marked increase in the error of the
reconstructions with increasing redshifts of the source distribution.

This analysis illustrates the role of the additional factors which must be
taken into account when adapting CMB tests of SI to the case of LSS. The full
set of results are summarized in Fig.~\ref{fig:MVplot}.  One interesting
observation is that the accuracy of the reconstruction is comparable for all
$\ell$ when the entire sky is observed (black lines) but deteriorates from
high to low $\ell$ (bottom to top panel) when part of the sky is surveyed
(blue and red lines).  This trend becomes more apparent as $\fsky$ decreases
from $0.92$ (blue) to $0.84$ (red). Furthermore, we find that the
reconstruction accuracy plateaus at around $N_g = 10^6$-$10^8$ in almost all
cases considered, with little improvement at higher source densities.
Overall, and perhaps as expected, we find the primary limiting factor to be
incomplete sky coverage and not the density of the sources.

\section{Recovering Evidence of Alignments}
\label{sec:alignments}

The robustness tests from the previous section imply a certain accuracy in
reconstructing the multipole vectors out of noisy data. We now test how this
accuracy translates into detection of the violations of SI.

For the sake of definitiveness, let us assume a purely phenomenological model
where the sky has a quadrupole and octopole that are perfectly planar. That
is, we assume that the quadrupole and octopole $\alm$ coefficients are pure
$a_{22}$ and $a_{33}$. [Any mix of $a_{22}^{RE}$, $a_{22}^{IM}$, $a_{33}^{RE}$
  and $a_{33}^{IM}$ will do, since the real/imaginary mixing only affects the
  azimuthal structure in the plane.] We first create Monte Carlo realizations
of skies that have this type of perfect alignment at $\ell=2, 3$ while having
other $\alm$ drawn from the usual Gaussian distributions.  We then apply our
reconstruction of the sky temperature, and thus the multipole vectors, and
study whether the alignment is observable.

If the aligned model has planar structures --- as observed on our sky by WMAP
--- then it is advantageous to study the directions and magnitudes of the
mutual cross products of multipole vectors, which are referred to as the
``oriented area'' vectors: \cite{CHSS_review}
\begin{equation}
w^{(\ell,i,j)}\equiv v^{(\ell, i)}\times v^{(\ell, j)}.
\end{equation}

Let us illustrate how one could search for planar alignments
  represented by the near-collinear oriented area vectors that we use as an
  example. Let us define a new statistic 
\begin{equation}
B_{\rm signal} = 
\underset{\hat{d}}{\min}
\left [
\frac{1}{N_{\rm pairs}} 
\sum_{\ell=2}^{\lmax}\sum_{j=2}^{\ell} \sum_{\substack{ i=1 }}^{j-1} 
\left ( 
1- \frac{\mid w^{(\ell,i,j)}\cdot \hat{d}\mid}{\mid w^{(\ell,i,j)}\mid}
\right )^2
\right ]^{1/2}
\label{eq:Bsignal}
\end{equation}
%
where $\lmax=3$ and the minimization is over all possible directions
$\hat{d}$.  For our alignment model, a perfect reconstruction of multipole
vectors would imply that all oriented area vectors are collinear, so that
$B_{\rm signal}=0$. In the presence of the uncertainty in the reconstruction,
however, the oriented area vectors $w^{(\ell,i,j)}$ will generally not be
aligned, and $B_{\rm signal}$ will be greater than zero but presumably small.
Finally, for a statistically isotropic sky, we expect that the oriented areas
do not preferentially lie close to any single direction $\hat{d}$, so that
$B_{\rm signal}^{\rm unaligned}\gg B_{\rm signal}^{\rm aligned}$.

We generate 50,000 Monte Carlo realizations of the perfectly aligned skies
with purely planar quadrupole and octopole as described above and higher
multipoles consistent with statistically isotropy. We also generate 50,000
statistically isotropic skies. In each case, we reconstruct the coefficients
$\alm$, and the corresponding multipole vectors, as described in
Sec.~\ref{sec:reconstruction}. We consider one case where the survey has $10^6$ galaxies
whose distribution peaks at $\zpeak=0.1$, and another case with $10^9$ galaxies
with $\zpeak=0.4$, representing examples of a shallow and
a deep survey respectively. For the reconstruction, we use $\Nside=8$, and a sky cut of
either $0\degr$ (i.e.\ $\fsky=1$) or $\pm 9\degr$ (i.e.\ $\fsky\simeq 0.84$).

\begin{figure}[!ht]
\begin{center}
\includegraphics[width=\linewidth]{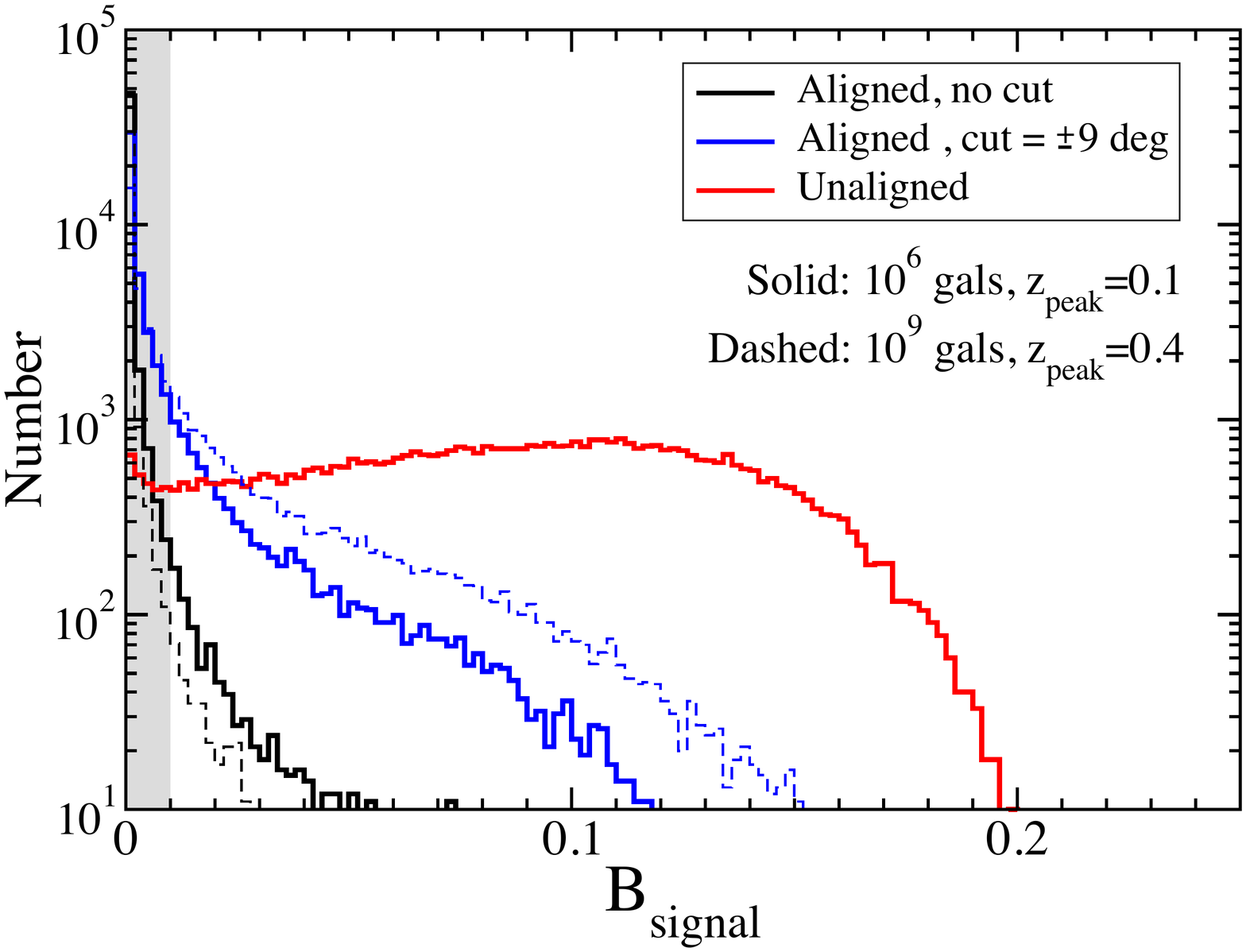} 
\end{center}
\caption{Detectability of perfectly aligned quadrupole and octopole in a mock
  survey using the $B_{\rm signal}$ statistic (see Eq.~\ref{eq:Bsignal}).
  Each histogram is based on 50,000 Monte Carlo realizations.  Solid lines
  shows survey with $N_g = 10^6$ objects and the radial distribution that
  peaks at $\zpeak=0.1$, while the dashed lines show a survey with $N_g =
  10^9$ and $\zpeak=0.4$ (the 'unaligned' case, shown with the red solid line,
  is independent of the presence of the cut and the values of $\zpeak$ and
  $N_g$). The grey region covers values of $B_{\rm signal}$ which correspond
  to the bottom 5\% of the isotropic (i.e.\ unaligned) sky cases.  We find
  that 98-99\% of the aligned skies without the galactic cut (and for either
  of the two $\zpeak$ cases) lie {\it below} this value --- in other words, it
  is roughly at the 20:1 odds that the value of $B_{\rm signal}$ found below
  this value favors the aligned model.  }
\label{fig:Bsignal}
\end{figure}


The histogram of the statistics $B_{\rm signal}$ is shown in
Fig.~\ref{fig:Bsignal}. As expected, the values of $B_{\rm signal}$ for the
aligned skies are preferentially smaller than for the unaligned
(i.e.\ isotropic) realizations. The shaded region covers values of $B_{\rm
  signal}$ which correspond to the bottom 5\% of the isotropic
(i.e.\ unaligned) sky cases; therefore, finding $B_{\rm signal}$ below this
value would indicate a $\sim 2\sigma$ evidence for this particular alignment.
We find that 98-99\% of the aligned sky realizations without the galactic cut
(and for either of the two $\zpeak$ cases) lie below this value of $B_{\rm
  signal}$, and so it is with this probability that one would find a $\sim
2\sigma$ evidence for the alignment. With the $\pm9\degr$ sky cut, evidence
for alignments will be weaker, and the $2\sigma$ evidence can be made in 65\%
($\zpeak=0.4$, $N_g=10^9$) or 85\% ($\zpeak=0.1$, $N_g=10^6$) percent of the
realizations of the aligned model.

These results are encouraging, given that we did not optimize over the choice
of the statistic to detect the assumed alignment. In this exploratory paper we
do not study the issue of detectability any further, perform a complete
likelihood analysis, or study more specific models for the alignment; this is
left for future work.

\section{Discussion and future work}
\label{sec:discussion}

In this paper we have proposed to apply the statistical tools developed for
studies of the CMB to conduct tests of the statistical isotropy (SI) of
large-scale structure. We considered the projected (i.e.\ two-dimensional)
density field provided by a galaxy catalog, and expanded it into multipole
moments analogously to how the CMB temperature field is conventionally
analyzed.  Each multipole can be decomposed into a set of $\ell$ multipole
vectors $\brac{\hat{v}^{(\ell,i)} i=1...,\ell}$ and a scalar
$A_{(\ell)}$. These vectors represent all phase information contained in the
projected density field, and enable a variety of tests of directionality in
the galaxy distribution. We developed an algorithm to reconstruct the full-sky
multipole vectors out the the cut-sky galaxy catalog, while carefully
accounting for the signal and noise specific to the galaxy maps. Note that
galaxies are not the only feasible tracers of the LSS; clusters of galaxies,
gamma-ray bursts, X-ray and radio sources, and other tracers could also be
potentially very useful in testing the SI.

In this work we have concentrated on the large scales, in particular only
considering the multipoles $\ell=2-4$; extension to smaller scales is in
principle straightforward.  Because LSS surveys typically do not typically
cover the full sky, we have implemented the reconstruction of the full-sky
(recently applied to CMB temperature maps in \cite{deOliveiraCosta:2006zj}).
Exactly to what extent this reconstruction effectively {\it assumes} SI has
recently been debated
\cite{CHSS_review,Efstathiou2009,Pontzen_Peiris,Aurich_Lustig}. The issue of
how to test SI with reconstructed full-sky information that explicitly does
not assume SI on relevant scales is an important problem in its own right, and
we leave it for future work.

Unlike the CMB temperature anisotropy field, which comes from a single,
well-defined redshift, galaxy surveys mapping the local universe are diverse
in their source density and redshift range and, like the CMB maps, can also
cover different areas of the sky.  We explored the impact of each of these
survey properties and found the primary limiting factor to be incomplete sky
coverage. Even a modest Galactic plane cut increases the
noise in the reconstruction due to mode mixing.  We find that if a significant
fraction ($\sim 16\%$) of the sky is not surveyed, the accuracy quickly
becomes limited by the uncertainty in the reconstructed full-sky properties
due to the cut, with little improvement in the errors achieved by increasing
the number of objects beyond $10^6$.

We also find that the accuracy of the reconstruction is comparable for all
$\ell$ when the entire sky is observed, but deteriorates from high to low
$\ell$ when part of the sky is surveyed; see Fig.~\ref{fig:MVplot}.  The
reconstruction accuracy typically plateaus at around $N_g = 10^6$-$10^8$,
suggesting that there is an intrinsic limit on how well the multipole vectors
can be recovered.  Furthermore, the recovery of the multipole vectors is more
accurate in a catalogs with sources at lower redshifts due to a higher power
in those cases (see Fig.~\ref{fig:Cl}).

Using a statistic constructed to detect planar alignments, we tested for
violations of SI in Monte Carlo simulations of isotropic skies, and of skies
in which the quadrupole and octopole are perfectly aligned.  We found a $98\%$
chance of making a $2\sigma$ detection of this particular alignment using a
galaxy catalog with $10^6$ sources of mean redshift $z=0.1$, detected over the
entire sky.  This likelihood drops to $85\%$ when $16\%$ of the sky is masked
out.  Similarly, for the $\zpeak=0.4$, $N_g=10^9$ survey, we find the
probabilities of 99\% ($\fsky=1$) and 65\% ($\fsky=0.84$) of finding a
$2\sigma$ detection of this particular alignment. Note, however, that we have
not optimized over the choice of the detection statistic, nor considered any
physical models for the alignment, so actual success in detecting such
anomalies may well be different from these numbers.

The next decade or two will see a dramatic improvement in the galaxy data on
largest observable scales.  For example, the Wide-field Infrared Survey
Explorer (WISE), currently observing, will provide an all-sky survey from 3.5
to 23 $\mu m$ about a thousand times more sensitive than IRAS, and should
produce a very large number of objects out to redshift of $z \sim 3$. Clearly,
data provided by surveys such as WISE in the infrared, and perhaps other
radio, X-ray and optical surveys, would be perfect targets to test the SI with
the multipole vectors. Such wide and deep surveys could even start to probe
the scales probed by the large-angle CMB; for example, it is possible (though
somewhat unlikely) that LSS can confirm or refute the missing large-angle
primordial power favored by the CMB in this scenario \cite{Gibelyou:2010qe}.

Quite possibly the biggest challenge in studying realistic surveys may be
understanding the details of any given survey, and culling out a
representative subsample of objects that can be used for tests of
isotropy. Fortunately, since we are primarily interested in large scale
information, we do not need to worry as much about other commonly found
systematic effects in galaxy surveys due to nonlinear clustering.  However, it
is clear that details of the selection function for each survey will need to
be known fairly accurately, as spatial or temporal variations in depth of
observations can masquerade as evidence for violations of SI.

In conclusion, we hope that multipole vectors will do the same for the LSS
maps that they did for the CMB: provide a novel and useful way to quantify
anisotropies on the sky. In the case of the CMB, this has led to a variety of
new tests of the SI with interesting results. We hope that the applications
to real LSS surveys will be equally fruitful.

\acknowledgments

CZ is funded by a NRF/DST (SA) Innovation Fellowship and a National Science
Foundation (USA) fellowship under grant PIRE-0507768.  DH is supported by DOE
OJI grant under contract DE-FG02-95ER40899, NSF under contract AST-0807564,
and NASA under contract NNX09AC89G. GDS is supported by a grant from the US
Department of Energy and by NASA under cooperative agreement NNX07AG89G.

\appendix
\section{Conventions}
\label{app:conventions}

The temperature on the sky can be decomposed in terms of spherical harmonics
\begin{equation}
{\delta T\over T}(\theta, \phi) = \sum_{\ell, m}\alm \Ylm(\theta, \phi).
\end{equation}
Spherical harmonics $\Ylm$ can be defined in terms of the associated Legendre
polynomials $P_{\ell m}$
\begin{equation}
  Y_{\ell m}\bra{\theta, \phi}=
  \sqrt{\frac{(2\ell+1)(\ell-m)!}{4\pi (\ell + m)!}}P_{\ell m}\bra{\cos
    \theta}e^{i m \phi}.
\end{equation}
For computing convenience, we wish to to work with real numbers only.
Breaking up the spherical harmonics $\Ylm$ and the coefficients $\alm$ into
real and imaginary parts
\begin{eqnarray}
\alm &=& \alm^{Re}+i\alm^{Im}\\[0.2cm]
Y_{\ell m} &=& Y_{\ell m}^{Re}+i Y_{\ell m}^{Im}
\end{eqnarray}

For negative $m$
\begin{eqnarray}
a_{\ell -m} &=& (-1)^m \alm^{*} = (-1)^m\bra{\alm^{Re}-i\alm^{Im}}\\[0.2cm]
Y_{\ell -m} &=& (-1)^m Y_{\ell m}^{*} = (-1)^m\bra{Y_{\ell m}^{Re}-i Y_{\ell m}^{Im}}.
\end{eqnarray}

The contribution to the sum  $\sum_m \alm Y_{\ell m}$ from a single value of $|m|$ is
\begin{eqnarray}
&&\alm Y_{\ell m} + a_{\ell -m} Y_{\ell -m} =\\[0.2cm]
&&\left \{  
\begin{array}{cl}
2\bra{\alm^{Re} Y_{\ell m}^{Re} - \alm^{Im} Y_{\ell m}^{Im}}  & 
(m \neq 0) \\[0.2cm]
a_{\ell 0} Y_{\ell 0}   & (m=0 ) \\[0.2cm]
\end{array} 
\right .
\label{eq:alm_sum}
\end{eqnarray}

We define the following: $Y_{\ell m} \equiv |Y_{\ell m}|\cos\bra{m\phi}+
i |Y_{\ell m}|\sin\bra{m\phi}$. Following \cite{deOliveiraCosta:2006zj}, we define
\begin{enumerate}
\item $Y^1_{\ell m}\equiv\sqrt{2}\, |Y_{\ell m}|\,\cos\bra{m\phi} \hfill~\bra{\text{for}~ m > 0} $ 
\item $Y^2_{\ell m}\equiv\sqrt{2}\, |Y_{\ell m}|\,\sin\bra{m\phi} \hfill \bra{\text{for}~ m < 0}$
\item $Y^3_{\ell m}\equiv |Y_{\ell m}|\,                       \hfill \bra{\text{for}~ m = 0}$
\end{enumerate}

We then define the following parameters:
\begin{enumerate}
\item $b^1_{\ell m} \equiv \sqrt{2} \alm^{RE}  \hfill\bra{\text{for}~ m > 0} $ 
\item $b^2_{\ell m} \equiv -\sqrt{2} \alm^{IM} \hfill\bra{\text{for}~ m < 0}$
\item $b^3_{\ell m} \equiv \alm                \hfill\bra{ \text{for}~ m = 0}$
\end{enumerate}
Hence, we can obtain the right-hand side of Eq.~(\ref{eq:alm_sum}) using the
following summation over real quantities, $b_{\ell m}^1 Y_{\ell m}^1 +
b^2_{\ell m} Y^2_{\ell m}$ (for $m\neq 0$) or $b_{\ell m}^3 Y_{\ell m}^3$ (for
$m = 0$).

\section{The covariance matrix}
\label{app:covmatr}

The reconstruction method described in Sec.\ \ref{sec:method} requires the
calculation of the covariance matrix $\wc$.  We discuss this in detail given
the various subtleties which require attention.

Firstly, we consider the sources of detector noise encapsulated in $\wN$.
The reconstruction of the underlying function $\nu(\Omega)$ from a galaxy
survey introduces two types of noise.  The nature of the sampling process
means that in an actual catalog, the number of objects in the i$^{th}$ pixel
will not be $n_i$ defined in Eq.~(\ref{ni}), but rather an integer
$\tilde{n}_i$.  This difference is due to \emph{shot noise}, encompassed in
the parameter $\nu_i$, given by
\begin{equation}
\nu_i = \frac{\tilde n_i - n_i}{\bar n}.
\end{equation}
In the same way, the average number of objects per pixel will not be $\bar n$
but rather $\tilde n$, given by
\begin{equation}
\tilde n \equiv \frac{1}{\Npix}\sum_{i=1}^{N_p} \tilde n_i .
\end{equation}
The above $\tilde{n}$ is the \emph{survey mean} and is taken to be our best
estimate of the ensemble mean $\bar{n}$.  We estimate the density contrast
$\Delta_i$ using the mean number density of the survey on its largest scales
\cite{Tegmark:1997cx}
\begin{equation}
\tilde \Delta_i = \frac{\tilde n_i - \tilde n}{\tilde n}.
\end{equation}
This procedure forces our estimates of the fluctuations on the largest scale
of the survey to zero, an effect sometimes called the 'integral
  constraint'. Following \cite{EDSGC}, a parameter $\epsilon$ is introduced
to account for the fractional difference between the survey mean and the
ensemble mean
\begin{equation} 
\epsilon \equiv \frac{\tilde n - \bar n}{\bar n}.
\end{equation}
Using the fact that $\Delta_i\equiv (n_i - {\bar n})/{\bar n}$ -- see
  Eq.~(\ref{eq:Delta_i}) -- we can relate our estimate $\tilde \Delta_i$ to the
    true value $\Delta_i$ in terms of $\epsilon$ and $\nu_i$ as
\begin{equation}
\tilde\Delta_i = \frac{\Delta_i + \nu_i - \epsilon}{1+\epsilon}.
\end{equation}
This equation relates the measured density contrast $\tilde\Delta_i$ to the
theoretically predicted density contrast $\Delta_i$.

We now wish to calculate the statistical properties of catalog density
contrast $\tilde\Delta_i$, in particular, its mean and covariance.
We need to express these in terms of statistical properties of
  the ensemble density contrast $\Delta_i$.

It will be useful to rewrite
\begin{equation}
\tilde\Delta_i \simeq \left( \Delta_i + \nu_i - \epsilon\right)
	\left(1 - \epsilon + \epsilon^2 + {\cal O}(\epsilon^3)\right)
\end{equation}
where the following hold
\begin{eqnarray}
\langle \nu_i \nu_j\rangle &=& \delta_{ij}\frac{(1+\Delta_i)}{\bar n} + {\cal O}(\bar N_g^{-2}) \nonumber \\
\langle \epsilon \rangle &=& 0 \nonumber \\
\langle \epsilon^2 \rangle &=& \frac{1}{\bar N_g} \nonumber \\
\langle \nu_i  \rangle &=& 0\nonumber \\
\langle \nu_i \epsilon \rangle &=& \frac{1}{\bar N_g} \nonumber \\
\langle \nu_i \epsilon^2 \rangle &=& 0 \nonumber \\
 \langle \nu_i \nu_j \epsilon^2\rangle &=& \frac{\delta_{ij}(1+\Delta_i)}{\bar N_g \bar n} + {\cal O}(\bar N_g^{-2})
\end{eqnarray}
Note that the expectation value is $\langle \Delta_i \rangle= \Delta_i$ and
not zero, as in the case of the ensemble. Putting this together we find
\begin{equation}
\langle \tilde\Delta_i\rangle = \Delta_i\left(1 + \frac{1}{\bar N_g}\right). 
\end{equation}
Furthermore, we find that 
\begin{eqnarray}
\langle \tilde\Delta_i \tilde\Delta_j \rangle &=&
\Delta_i\Delta_j \langle \left(1+3\epsilon^2\right)\rangle -
 2 \left(\Delta_i  \langle \nu_j \epsilon\rangle +  
 \Delta_j \langle \nu_i \epsilon\rangle \right) \nonumber \\[0.2cm]
&+& \langle \nu_i\nu_j (1+3\epsilon^2)\rangle + 
 2\left(\Delta_i+\Delta_j\right)\langle \epsilon^2\rangle - 
 \langle\left(\nu_i+\nu_j\right)\epsilon\rangle \nonumber \\[0.2cm]
&+& \langle \epsilon^2\rangle + {\cal O}(\bar N_g^{-2})\nonumber \\
&=& \Delta_i\Delta_j  \left(1+\frac{3}{\bar N_g}\right) + 
\frac{\delta_{ij}(1+\Delta_i)}{ \bar n}\left(1+\frac{3}{\bar N_g}\right) \nonumber \\
&-&\frac{1}{\bar N_g} + {\cal O}(\bar N_g^{-2}).
\end{eqnarray}

The covariance matrix of $\tilde\Delta_i$ is therefore:
\begin{eqnarray} 
\label{eqn:Cij}
C_{ij} &\equiv& \langle \tilde\Delta_i \tilde\Delta_j \rangle  - 
 \langle \tilde\Delta_i\rangle \langle \tilde\Delta_j\rangle \nonumber \\
&=&  \frac{1}{\bar{N}_g} \bra{\Delta_i \Delta_j - 1} + 
\frac{\delta_{ij}(1+\Delta_i)}{\bar n} \nonumber\\
&+& {\cal O}(\bar N_g^{-2}).
\end{eqnarray}

We need to write both $\langle\tilde\Delta_i\rangle$ and $C_{ij}$ in terms of
the $\alm$.  Using Eq.~(\ref{eqn:Deltaialms}), we can write
\begin{equation}
\langle \tilde\Delta_i\rangle = \sum_{\ell=2}^{\lmaxrec}\sum_m \alm Y_{\ell m}(\Omega_i)
\left(1 + \frac{1}{\bar N_g}\right).
\end{equation}
Notice that we have truncated the sum over $\alm$ at $\lmaxrec$
which is the last multipole that we reconstruct.  The $\alm$ at higher
$\ell$ are replaced by their expectation values in the ensemble of universes,
in which $\langle \alm \rangle = 0.$ We treat the covariance matrix
$C_{ij}$ in a similar fashion and replace $\alm a_{\ell' m'}$ by its
expectation value in the ensemble of universes:
\begin{equation}
\langle \alm a_{\ell' m'} \rangle = \delta_{\ell\ell'}\delta_{mm'}\cc_\ell \, .
\end{equation}

We follow \cite{deOliveiraCosta:2006zj} in their reconstruction of the
$\alm$, and reconstruct a limited range of multipoles, $2 \leq \ell\leq
\lmaxrec$.
This procedure treats the higher multipoles $\lmaxrec+1\leq \ell\leq
\lmaxtot$ as ``noise'' to the reconstructed multipoles'``signal''.
Following this logic, we split the pixel density fluctuations into the suitably
chosen signal and noise parts
\begin{eqnarray}
\tilde\Delta_i &=& \frac{\Delta_i + \left(\nu_i-\epsilon\right)}{1+\epsilon}\nonumber  \\
&\simeq& \Delta_i\left(1-\epsilon+\epsilon^2\right) + \frac{\nu_i-\epsilon}{1+\epsilon}\nonumber \\
&=& \sum_{\ell=2}^{ \lmaxrec}b_{\ell m} Y_{\ell m} (\Omega_i) \nonumber\\
&+& \left[\sum_{\ell= \lmaxrec+1}^{\ell_{max}} \sum_m b_{\ell m} Y_{\ell m}
  (\Omega_i)+
  \left(\epsilon^2-\epsilon\right)\Delta_i + \frac{\nu_i-\epsilon}{1+\epsilon}\right] \nonumber \\
&\equiv& \sum_{\ell=2}^{ \lmaxrec} \sum_m b_{\ell m} Y_{\ell m} (\Omega_i) + {\cal N}_i .
\end{eqnarray}
where
\begin{eqnarray*}
{\cal N}_i = \sum_{\ell= \lmaxrec+1}^{\lmaxtot} \sum_m b_{\ell m} Y_{\ell m}
(\Omega_i)+
\left(\epsilon^2-\epsilon\right)\Delta_i + \frac{\nu_i-\epsilon}{1+\epsilon} 
\label{eq:Ni}
\end{eqnarray*}
and where we take $\lmaxtot=50$. Note that the $\alm$ have been recast in new
variables denoted $b_{\ell m}$ defined in Appendix \ref{app:conventions} in
order to simplify the calculation. In the above, ${\cal N}_i$ is the noise in
the i$^{th}$ pixel. The first term in Eq.~(\ref{eq:Ni}) is the contribution
from leakage from multipoles which are not reconstructed, while the next two
terms are due to shot noise arising from the sampling process.  Taking
expectation value of Eq.~(\ref{eq:Ni}) we get 
\begin{eqnarray}
\langle {\cal N}_i\rangle &=& \sum_{\ell= \lmaxrec+1}^{\lmaxtot} \sum_m b_{\ell m} Y_{\ell m} (\Omega_i)  
+ \langle\epsilon^2\rangle\Delta_i  \nonumber \\
  &+& \Big\langle\frac{\nu_i}{1+\epsilon}\Big\rangle - 
  \Big\langle\frac{\epsilon}{1+\epsilon}\Big\rangle \\
&=& \frac{\Delta_i}{\bar N_g} +  \sum_{\ell= \lmaxrec+1}^{\lmaxtot} \sum_m b_{\ell m} Y_{\ell m} (\Omega_i) .
\end{eqnarray}
As usual, $\alm$ terms with $\ell > \lmaxrec$ are neglected as they are
unknown and will not be reconstructed.  Our treatment of the unknown true
underlying perturbation $\Delta_i$ is limited and we merely replace it with
our current best estimate in an iterative process: 
\begin{equation}
\langle {\cal N}_i\rangle^{(p)} \simeq \frac{1}{\bar N_g}
\sum_{\ell=2}^{ \lmaxrec}b_{\ell m} Y_{\ell m} (\Omega_i) +
	\sum_{\ell=2}^{ \lmaxrec}\sum_m b_{\ell m}^{(p)}Y_{\ell m}(\Omega_i)
\end{equation}
where $(p)$ numbers the iterative step. At the $0^{th}$ iteration we use
$b_{\ell m}^{(0)}=0$, which is then replaced by
estimates of $b_{\ell m}$ in successive iterations until convergence is achieved.

The covariance of the noise is given by
\begin{eqnarray}
&&\langle {\cal N}_i {\cal N}_j\rangle^{(p)} = \nonumber \\
&& \Bigg\langle \Bigg(\sum_{\ell= \lmaxrec+1}^{\ell_{max}} \sum_m b_{\ell m}
  Y_{\ell m} (\Omega_i) + 
  \left(\epsilon^2-\epsilon\right)\Delta_i \nonumber\\ 
&+&\frac{\nu_i}{1+\epsilon} - \frac{\epsilon}{1+\epsilon} \Bigg)\nonumber\\
&&\times\Bigg(\sum_{\ell'= \lmaxrec+1}^{\ell_{max}}\sum_{m'} b_{\ell' m'}
  Y_{\ell' m'} (\Omega_j) +
  \left(\epsilon^2-\epsilon\right)\Delta_j \nonumber\\ 
&+&\frac{\nu_j}{1+\epsilon}-\frac{\epsilon}{1+\epsilon}\Bigg)\Bigg\rangle .
\end{eqnarray}
Replacing $\langle b_{\ell m} b_{\ell' m'}\rangle$ by its expectation value in
the ensemble of Universes (for $\ell \geq \lmaxrec$, $\cc_\ell\delta_{\ell
  \ell'}\delta_{m m'}$), and $\langle b_{\ell m}\rangle$ by its expectation
value (i.e.\ zero) we find
\begin{eqnarray*}
\langle {\cal N}_i {\cal N}_j\rangle^{(p)} &=& 
\sum_{\ell= \lmaxrec+1}^{\ell_{max}}\frac{2\ell+1}{4\pi}\cc_\ell P_\ell(\cos\theta_{ij})\nonumber \\
&+&[\langle\epsilon^2\rangle\left(1+\Delta_i^{(p)}+\Delta_j^{(p)}+\Delta_i^{(p)}\Delta_j^{(p)} \right) \nonumber \\
&-&\langle\epsilon\nu_j\rangle (\Delta_i^{(p)}+1) -
  \langle\epsilon\nu_i\rangle(\Delta_j^{(p)} +1)\\[0.2cm]
  &+& \langle\nu_i\nu_j\rangle +3\langle\epsilon^2\nu_i\nu_j\rangle ]\nonumber\\
  &=& \sum_{\ell= \lmaxrec+1}^{\ell_{max}}\frac{2\ell+1}{4\pi}\cc_\ell P_\ell(\cos\theta_{ij})\nonumber \\
&+& \left[ \frac{\delta_{ij}(1+\Delta_i^{(p)})}{\bar n  }\left(1 + \frac{3}{\bar N_g}\right) + 
\frac{1}{\bar N_g}(\Delta_i^{(p)}\Delta_j^{(p)} -1) \right] \nonumber \\
&+& {\cal O}\left(\frac{1}{\bar N_g^2}\right) .
\end{eqnarray*}

Since $\langle {\cal N}_i\rangle\langle{\cal N}_j\rangle^{(p)} = 
{\cal O}(1/\bar N_g^2)$, $C_{ij}^{(p)} = \langle{\cal N}_i {\cal N}_j \rangle^{(p)}$.
For clarity, we separate the covariance matrix out into its two contributions;
\begin{eqnarray}
C^{(p)}_{ij} &=&  S_{ij}^{(p)} + N_{ij}^{(p)} 
\nonumber
\end{eqnarray}
where
\begin{eqnarray}
S_{ij}^{(p)} &=& \sum_{\ell= \lmaxrec+1}^{\ell_{max}}\frac{2\ell+1}{4\pi}
\cc_\ell P_\ell(\cos\theta_{ij}) \\
 N_{ij}^{(p)} &=& \left[ \frac{\delta_{ij}(1+\Delta_i^{(p)}) }{\bar n }
   \left(1 + \frac{3}{\bar N_g}\right) + 
\frac{1}{\bar N_g}(\Delta_i^{(p)}\Delta_j^{(p)} -1) \right]  \nonumber .
\label{eq:nfinal}
\end{eqnarray}
In the first evaluation we use $\Delta_i^{(p)}=\tilde\Delta_i$.  Once the
first set of reconstructed $b_{\ell m}$ are extracted, they will be used to
update $\Delta_i^{(p)}= \sum_{\ell=2}^{ \lmaxrec}b_{\ell m} Y_{\ell m}
(\Omega_i)$ for the subsequent iterations.  Note that the value of the $\Cl$
used in the computation of the signal matrix $S$ is not crucial: error in the
estimation of the angular power spectrum will merely mean that more iterations
will be required for convergence.

As discussed above, the true average number of galaxies per pixel is unknown
and can only be estimated by the mean calculated from the survey. This
assumption $\tilde N_g = \bar{N_g}$ however artificially suppresses the
estimates of the power on large scales and is accounted for by the factor of
$1/\bar{N}_{g}$ in the last term of Eq.~(\ref{eq:nfinal}).  Comparing the
expression in Eq.~(\ref{eq:nfinal}) with the covariance matrix calculated for
the CMB in \cite{deOliveiraCosta:2006zj}, we find that they are in agreement
if we bear in mind that the case of the CMB effectively corresponds to the
case where $N_{g} \rightarrow \infty$ 

\section{The theoretical angular power spectrum $\Cl$}
\label{app:power_spectrum}

Equation (\ref{eq:ws}) shows that an estimate of the angular power spectrum
$\Cl$ is required for our reconstruction. We now show how to calculate the
angular power spectrum of a large-scale structure survey (for pioneering
works on this, see \cite{Baugh_Efstathiou,EDSGC,Efstathiou_Moody}).
We only consider a single, vanilla best-fit $\Lambda$CDM cosmological model,
as the cosmological model dependence of the $\Cl$ is not expected to affect
the results. 

The angular power spectrum in harmonic space can be related to its counterpart
in Fourier space via
\begin{equation}
\Cl= \int^{\infty}_{0}K_{\ell}(k) P(k) k^2 d k
\end{equation}
where, as shown in \cite{EDSGC}, $K_{\ell}$ is an integral kernel given by
$\frac{2}{\pi}f_{\ell}^2(k)$ where $f_{\ell}$ is the Bessel transform
\footnote{A Bessel transform is equivalent to a two-dimensional Fourier
  transform but with a radially symmetric integral kernel. They arise from
  solving Laplace's equation in spherical coordinates and are related to
  ordinary Bessel function of the same kind $J$ by
  $j_{n}(x)=\sqrt{\frac{\pi}{2x}}J_{n+1/2}(x)$. } of the radial selection
function $f(r)=g(r) h(r)$.  Here $g(r)$ is the radial probability distribution
of galaxies
\begin{equation}
g(r) \propto \frac{dN}{dr}= \frac{dN/dz}{dr/dz}=H(z)\,\frac{dN}{dz}
\end{equation}
where $dN/dz$ is the radial redshift distribution of objects in the survey.
The objects which constitute potential catalogs are biased tracers of dark
matter; while this bias primarily depends on the object's mass, for
definitiveness we assume $b=1$.  The function $h(r)$ which accounts for this
galaxy bias as well as clustering, is therefore assumed to be unity.  This
means that the power spectrum above is measured at a radial distance of $r\sim
\ell/k$. Hence,
\begin{equation}
f_{\ell}(k) \equiv \int^{\infty}_0 j_{\ell}(kr) f(r) dr = 
\int^{\infty}_{0}j_{\ell}(kr(z))\frac{dN}{dz} dz,
\end{equation}
where $j_{\ell}(kr)$ is the spherical Bessel function of order $\ell$.  As
mentioned in the text, we assume the distribution of objects of the form
$dN/dz\equiv n(z)\propto z^2\exp(-z/z_0)$ that peaks at $\zpeak = 2z_0$. The
power spectrum $P(k)$ is approximated to be scale-invariant with $P(k)\propto
k^{n_s}$ where we adopt $n_s = 0.96$ and normalization consistent with WMAP data.

So far we have assumed the linear clustering regime, which will dominate on
the large scales that we are interested in. Nevertheless, it is important to
check what the role of nonlinearities will be. To that effect, we adopt the
following simple correction formula proposed in \cite{Cole:2005sx} relating
the linear and nonlinear power spectra
\begin{equation} 
P_{\rm nl}(k) = b^2 \frac{1+Q_{\rm nl} k^2}{1+A_{\rm nl}k} P(k)
\end{equation} 
where $A_{\rm nl} = 1.4$.  The factor $Q_{\rm nl}$ is determined from the
galaxy catalog itself, and we adopt the value obtained by the Sloan Digital
Sky Survey Luminous Red Galaxies of $Q_{\rm nl} = 31$ \cite{Tegmark:2006az}.
The linear and nonlinear angular power spectra of surveys with $\zpeak=0.1$,
$0.2$ and $0.4$ are shown in Fig.~\ref{fig:Cl}.

\section{Mock catalog generation}\label{app:mock} 

A density map is constructed in the following way:
\begin{enumerate}
\item The theoretical power spectrum (based on the SDSS power spectrum) is
  calculated for a $\Lambda$CDM Universe for a given set of cosmological
  parameters. The amplitude of the spectrum is determined by the redshift
  distribution of sources, $d N /d z$, which is assumed to be a Gaussian
  peaking at $\zpeak$. The theoretical power spectra for the three cases
  considered ($\zpeak=0.1, 0.2$ and $0.4$) are shown in Fig.~\ref{fig:Cl}.
\item A set of $\alm$ are drawn randomly from a distribution centered at zero
  with variance $\Cl^{th}$, so that $\alm\in N(0, \Cl)$. The
  corresponding power spectrum is denoted $\Cl^{\rm realiz} \equiv
  \sum_m |\alm|^2$.
\item The HEALpix routine {\tt alm2map} is used to generate a density map
  of $12 \Nside^2 $ pixels from the input $\alm$.  Initially we use a high
  pixelization of $\Nside=64$ to produce a smoother density field.
  \newcounter{enumsaved}
  \setcounter{enumsaved}{\theenumi} 
\end{enumerate}
The density map generated in the above manner is used as the basis for
constructing each realization of a galaxy survey as follows;
\begin{enumerate}
\setcounter{enumi}{\theenumsaved} 
\item The density map is populated with $N_g$ ``galaxies'' (i.e.\ points) so
  that the fraction of sources allocated to each pixel represents the
  underlying average fluctuation in density around the mean.  Given that we
  would like to investigate the impact of the number of galaxies in the survey
  and sky coverage of the survey separately, regardless of the sky cut we
  first create {\it full-maps} with the number of galaxies of $N_g/\fsky$, so that
  the total number of galaxies on the {\it cut sky} will be a fixed $N_g$.
\item In order to speed up the computation (which requires inversion of
  matrices of size $\Npix\times \Npix$ where $\Npix=12\Nside^2$), we downgrade
  the maps to a lower resolution using the HEALPix routine {\tt udgrade}. The
  cost of the reduced accuracy in the reconstruction due to the downgrading
  process is considered in Sec.~\ref{sec:testing}.
\item In cases where we are simulating a masked sky, we remove (i.e.\ set to
  zero counts) galaxies in the isolatitude cut of $\pm 4.5\degr$ or $\pm
  9\degr$.
\item Elements of the noise matrix ${\cal N}$ are initially estimated using
  the measured map.  In the subsequent iterations, the elements are computed
  using the reconstructed $\alm$. We perform three such iterations of the
  reconstruction and update the $\alm$ at each step. Convergence is tested.
\end{enumerate}
The above process is repeated 300 times to produce a set of realizations from
which the necessary statistics can be calculated.

\bibliography{lssmv}

\begin{thebibliography}{10}%
\makeatletter
\providecommand \@ifxundefined [1]{%
 \ifx #1\undefined \expandafter \@firstoftwo
 \else \expandafter \@secondoftwo
\fi
}%
\providecommand \@ifnum [1]{%
 \ifnum #1\expandafter \@firstoftwo
 \else \expandafter \@secondoftwo
\fi
}%
\providecommand \enquote [1]{``#1''}%
\providecommand \bibnamefont  [1]{#1}%
\providecommand \bibfnamefont [1]{#1}%
\providecommand \citenamefont [1]{#1}%
\providecommand\href[0]{\@sanitize\@href}%
\providecommand\@href[1]{\endgroup\@@startlink{#1}\endgroup\@@href}%
\providecommand\@@href[1]{#1\@@endlink}%
\providecommand \@sanitize [0]{\begingroup\catcode`\&12\catcode`\#12\relax}%
\@ifxundefined \pdfoutput {\@firstoftwo}{%
 \@ifnum{\z@=\pdfoutput}{\@firstoftwo}{\@secondoftwo}%
}{%
 \providecommand\@@startlink[1]{\leavevmode\special{html:<a href="#1">}}%
 \providecommand\@@endlink[0]{\special{html:</a>}}%
}{%
 \providecommand\@@startlink[1]{%
  \leavevmode
  \pdfstartlink
   attr{/Border[0 0 1 ]/H/I/C[0 1 1]}%
   user{/Subtype/Link/A<</Type/Action/S/URI/URI(#1)>>}%
  \relax
 }%
 \providecommand\@@endlink[0]{\pdfendlink}%
}%
\providecommand \url  [0]{\begingroup\@sanitize \@url }%
\providecommand \@url [1]{\endgroup\@href {#1}{\urlprefix}}%
\providecommand \urlprefix [0]{URL }%
\providecommand \Eprint[0]{\href }%
\@ifxundefined \urlstyle {%
  \providecommand \doi [1]{doi:\discretionary{}{}{}#1}%
}{%
  \providecommand \doi [0]{doi:\discretionary{}{}{}\begingroup
  \urlstyle{rm}\Url }%
}%
\providecommand \doibase [0]{http://dx.doi.org/}%
\providecommand \Doi[1]{\href{\doibase#1}}%
\providecommand \bibAnnote [3]{%
  \BibitemShut{#1}%
  \begin{quotation}\noindent
    \textsc{Key:}\ #2\\\textsc{Annotation:}\ #3%
  \end{quotation}%
}%
\providecommand \bibAnnoteFile [2]{%
  \IfFileExists{#2}{\bibAnnote {#1} {#2} {\input{#2}}}{}%
}%
\providecommand \typeout [0]{\immediate \write \m@ne }%
\providecommand \selectlanguage [0]{\@gobble}%
\providecommand \bibinfo [0]{\@secondoftwo}%
\providecommand \bibfield [0]{\@secondoftwo}%
\providecommand \translation [1]{[#1]}%
\providecommand \BibitemOpen[0]{}%
\providecommand \bibitemStop [0]{}%
\providecommand \bibitemNoStop [0]{.\EOS\space}%
\providecommand \EOS [0]{\spacefactor3000\relax}%
\providecommand \BibitemShut [1]{\csname bibitem#1\endcsname}%
\bibitem{COBE}%
  \BibitemOpen
  \bibfield{author}{%
  \bibinfo {author} {\bibfnamefont{J.~C.}\ \bibnamefont{{Mather}}}, \bibinfo
  {author} {\bibfnamefont{E.~S.}\ \bibnamefont{{Cheng}}}, \bibinfo {author}
  {\bibfnamefont{R.~E.}\ \bibnamefont{{Eplee}}, \bibfnamefont{Jr.}}, \bibinfo
  {author} {\bibfnamefont{R.~B.}\ \bibnamefont{{Isaacman}}}, \bibinfo {author}
  {\bibfnamefont{S.~S.}\ \bibnamefont{{Meyer}}}, \bibinfo {author}
  {\bibfnamefont{R.~A.}\ \bibnamefont{{Shafer}}}, \bibinfo {author}
  {\bibfnamefont{R.}~\bibnamefont{{Weiss}}}, \bibinfo {author}
  {\bibfnamefont{E.~L.}\ \bibnamefont{{Wright}}}, \bibinfo {author}
  {\bibfnamefont{C.~L.}\ \bibnamefont{{Bennett}}}, \bibinfo {author}
  {\bibfnamefont{N.~W.}\ \bibnamefont{{Boggess}}}, \bibinfo {author}
  {\bibfnamefont{E.}~\bibnamefont{{Dwek}}}, \bibinfo {author}
  {\bibfnamefont{S.}~\bibnamefont{{Gulkis}}}, \bibinfo {author}
  {\bibfnamefont{M.~G.}\ \bibnamefont{{Hauser}}}, \bibinfo {author}
  {\bibfnamefont{M.}~\bibnamefont{{Janssen}}}, \bibinfo {author}
  {\bibfnamefont{T.}~\bibnamefont{{Kelsall}}}, \bibinfo {author}
  {\bibfnamefont{P.~M.}\ \bibnamefont{{Lubin}}}, \bibinfo {author}
  {\bibfnamefont{S.~H.}\ \bibnamefont{{Moseley}}, \bibfnamefont{Jr.}}, \bibinfo
  {author} {\bibfnamefont{T.~L.}\ \bibnamefont{{Murdock}}}, \bibinfo {author}
  {\bibfnamefont{R.~F.}\ \bibnamefont{{Silverberg}}}, \bibinfo {author}
  {\bibfnamefont{G.~F.}\ \bibnamefont{{Smoot}}},\ and\ \bibinfo {author}
  {\bibfnamefont{D.~T.}\ \bibnamefont{{Wilkinson}}},\ }%
  \bibfield{journal}{%
  \Doi{10.1086/185717}{\bibinfo {journal} {\apjl}}\ }%
  \textbf{\bibinfo {volume} {354}},\ \bibinfo {pages} {L37} (\bibinfo {month}
  {May}\ \bibinfo {year} {1990})%
  \bibAnnoteFile{NoStop}{COBE}%
\bibitem{COBE2}%
  \BibitemOpen
  \bibfield{author}{%
  \bibinfo {author} {\bibfnamefont{J.~C.}\ \bibnamefont{{Mather}}}, \bibinfo
  {author} {\bibfnamefont{E.~S.}\ \bibnamefont{{Cheng}}}, \bibinfo {author}
  {\bibfnamefont{D.~A.}\ \bibnamefont{{Cottingham}}}, \bibinfo {author}
  {\bibfnamefont{R.~E.}\ \bibnamefont{{Eplee}}, \bibfnamefont{Jr.}}, \bibinfo
  {author} {\bibfnamefont{D.~J.}\ \bibnamefont{{Fixsen}}}, \bibinfo {author}
  {\bibfnamefont{T.}~\bibnamefont{{Hewagama}}}, \bibinfo {author}
  {\bibfnamefont{R.~B.}\ \bibnamefont{{Isaacman}}}, \bibinfo {author}
  {\bibfnamefont{K.~A.}\ \bibnamefont{{Jensen}}}, \bibinfo {author}
  {\bibfnamefont{S.~S.}\ \bibnamefont{{Meyer}}}, \bibinfo {author}
  {\bibfnamefont{P.~D.}\ \bibnamefont{{Noerdlinger}}}, \bibinfo {author}
  {\bibfnamefont{S.~M.}\ \bibnamefont{{Read}}}, \bibinfo {author}
  {\bibfnamefont{L.~P.}\ \bibnamefont{{Rosen}}}, \bibinfo {author}
  {\bibfnamefont{R.~A.}\ \bibnamefont{{Shafer}}}, \bibinfo {author}
  {\bibfnamefont{E.~L.}\ \bibnamefont{{Wright}}}, \bibinfo {author}
  {\bibfnamefont{C.~L.}\ \bibnamefont{{Bennett}}}, \bibinfo {author}
  {\bibfnamefont{N.~W.}\ \bibnamefont{{Boggess}}}, \bibinfo {author}
  {\bibfnamefont{M.~G.}\ \bibnamefont{{Hauser}}}, \bibinfo {author}
  {\bibfnamefont{T.}~\bibnamefont{{Kelsall}}}, \bibinfo {author}
  {\bibfnamefont{S.~H.}\ \bibnamefont{{Moseley}}, \bibfnamefont{Jr.}}, \bibinfo
  {author} {\bibfnamefont{R.~F.}\ \bibnamefont{{Silverberg}}}, \bibinfo
  {author} {\bibfnamefont{G.~F.}\ \bibnamefont{{Smoot}}}, \bibinfo {author}
  {\bibfnamefont{R.}~\bibnamefont{{Weiss}}},\ and\ \bibinfo {author}
  {\bibfnamefont{D.~T.}\ \bibnamefont{{Wilkinson}}},\ }%
  \bibfield{journal}{%
  \Doi{10.1086/173574}{\bibinfo {journal} {\apj}}\ }%
  \textbf{\bibinfo {volume} {420}},\ \bibinfo {pages} {439} (\bibinfo {month}
  {Jan.}\ \bibinfo {year} {1994})%
  \bibAnnoteFile{NoStop}{COBE2}%
\bibitem{Spergel2003}%
  \BibitemOpen
  \bibfield{author}{%
  \bibinfo {author} {\bibfnamefont{D.~N.}\ \bibnamefont{{Spergel}}}
  \emph{et~al.} (\bibinfo {collaboration} {WMAP}),\ }%
  \bibfield{journal}{%
  \bibinfo {journal} {\apjs}\ }%
  \textbf{\bibinfo {volume} {148}},\ \bibinfo {pages} {175} (\bibinfo {year}
  {2003}),\ \Eprint{http://arxiv.org/abs/astro-ph/0302209}{astro-ph/0302209}%
  \bibAnnoteFile{NoStop}{Spergel2003}%
\bibitem{Spergel2006}%
  \BibitemOpen
  \bibfield{author}{%
  \bibinfo {author} {\bibfnamefont{D.}~\bibnamefont{{Spergel}}} \emph{et~al.},\
  }%
  \bibfield{journal}{%
  \bibinfo {journal} {\apjs}\ }%
  \textbf{\bibinfo {volume} {170}} (\bibinfo {year} {2007}),\
  \Eprint{http://arxiv.org/abs/astro-ph/0603449}{astro-ph/0603449}%
  \bibAnnoteFile{NoStop}{Spergel2006}%
\bibitem{wmap7}%
  \BibitemOpen
  \bibfield{author}{%
  \bibinfo {author} {\bibfnamefont{E.}~\bibnamefont{Komatsu}} \emph{et~al.}
  (\bibinfo {collaboration} {WMAP}),\ }%
  \bibfield{journal}{%
  \Doi{10.1088/0067-0049/192/2/18}{\bibinfo {journal} {Astrophys. J. Suppl.}}\
  }%
  \textbf{\bibinfo {volume} {192}},\ \bibinfo {pages} {18} (\bibinfo {year}
  {2011}),\ \Eprint{http://arxiv.org/abs/1001.4538}{arXiv:1001.4538
  [astro-ph.CO]}%
  \bibAnnoteFile{NoStop}{wmap7}%
\bibitem{DMR4}%
  \BibitemOpen
  \bibfield{author}{%
  \bibinfo {author} {\bibfnamefont{G.}~\bibnamefont{{Hinshaw}}}, \bibinfo
  {author} {\bibfnamefont{A.~J.}\ \bibnamefont{{Branday}}}, \bibinfo {author}
  {\bibfnamefont{C.~L.}\ \bibnamefont{{Bennett}}}, \bibinfo {author}
  {\bibfnamefont{K.~M.}\ \bibnamefont{{Gorski}}}, \bibinfo {author}
  {\bibfnamefont{A.}~\bibnamefont{{Kogut}}}, \bibinfo {author}
  {\bibfnamefont{C.~H.}\ \bibnamefont{{Lineweaver}}}, \bibinfo {author}
  {\bibfnamefont{G.~F.}\ \bibnamefont{{Smoot}}},\ and\ \bibinfo {author}
  {\bibfnamefont{E.~L.}\ \bibnamefont{{Wright}}},\ }%
  \bibfield{journal}{%
  \Doi{10.1086/310076}{\bibinfo {journal} {\apjl}}\ }%
  \textbf{\bibinfo {volume} {464}},\ \bibinfo {pages} {L25+} (\bibinfo {month}
  {Jun.}\ \bibinfo {year} {1996}),\
  \Eprint{http://arxiv.org/abs/astro-ph/9601061}{astro-ph/9601061}%
  \bibAnnoteFile{NoStop}{DMR4}%
\bibitem{wmap123}%
  \BibitemOpen
  \bibfield{author}{%
  \bibinfo {author} {\bibfnamefont{C.~J.}\ \bibnamefont{{Copi}}}, \bibinfo
  {author} {\bibfnamefont{D.}~\bibnamefont{{Huterer}}}, \bibinfo {author}
  {\bibfnamefont{D.~J.}\ \bibnamefont{{Schwarz}}},\ and\ \bibinfo {author}
  {\bibfnamefont{G.~D.}\ \bibnamefont{{Starkman}}},\ }%
  \bibfield{journal}{%
  \Doi{10.1103/PhysRevD.75.023507}{\bibinfo {journal} {\prd}}\ }%
  \textbf{\bibinfo {volume} {75}},\ \bibinfo {pages} {023507} (\bibinfo {month}
  {Jan.}\ \bibinfo {year} {2007}),\
  \Eprint{http://arxiv.org/abs/astro-ph/0605135}{astro-ph/0605135}%
  \bibAnnoteFile{NoStop}{wmap123}%
\bibitem{wmap12345}%
  \BibitemOpen
  \bibfield{author}{%
  \bibinfo {author} {\bibfnamefont{C.~J.}\ \bibnamefont{{Copi}}}, \bibinfo
  {author} {\bibfnamefont{D.}~\bibnamefont{{Huterer}}}, \bibinfo {author}
  {\bibfnamefont{D.~J.}\ \bibnamefont{{Schwarz}}},\ and\ \bibinfo {author}
  {\bibfnamefont{G.~D.}\ \bibnamefont{{Starkman}}},\ }%
  \bibfield{journal}{%
  \Doi{10.1111/j.1365-2966.2009.15270.x}{\bibinfo {journal} {\mnras}}\ }%
  \textbf{\bibinfo {volume} {399}},\ \bibinfo {pages} {295} (\bibinfo {month}
  {Oct.}\ \bibinfo {year} {2009}),\
  \Eprint{http://arxiv.org/abs/0808.3767}{arXiv:0808.3767}%
  \bibAnnoteFile{NoStop}{wmap12345}%
\bibitem{Sarkar}%
  \BibitemOpen
  \bibfield{author}{%
  \bibinfo {author} {\bibfnamefont{D.}~\bibnamefont{Sarkar}}, \bibinfo {author}
  {\bibfnamefont{D.}~\bibnamefont{Huterer}}, \bibinfo {author}
  {\bibfnamefont{C.~J.}\ \bibnamefont{Copi}}, \bibinfo {author}
  {\bibfnamefont{G.~D.}\ \bibnamefont{Starkman}},\ and\ \bibinfo {author}
  {\bibfnamefont{D.~J.}\ \bibnamefont{Schwarz}},\ }%
  \bibfield{journal}{%
  \Doi{10.1016/j.astropartphys.2010.12.009}{\bibinfo {journal}
  {Astropart.Phys.}}\ }%
  \textbf{\bibinfo {volume} {34}},\ \bibinfo {pages} {591} (\bibinfo {year}
  {2011}),\ \Eprint{http://arxiv.org/abs/1004.3784}{arXiv:1004.3784
  [astro-ph.CO]}%
  \bibAnnoteFile{NoStop}{Sarkar}%
\bibitem{Hajian:2007pi}%
  \BibitemOpen
  \bibfield{author}{%
  \bibinfo {author} {\bibfnamefont{A.}~\bibnamefont{{Hajian}}},\ }%
  \bibfield{journal}{%
  \bibinfo {journal} {ArXiv Astrophysics e-prints}\ }%
  \textbf{\bibinfo {volume} {astro-ph/0702723}} (\bibinfo {month} {Feb.}\
  \bibinfo {year} {2007}),\
  \Eprint{http://arxiv.org/abs/astro-ph/0702723}{astro-ph/0702723}%
  \bibAnnoteFile{NoStop}{Hajian:2007pi}%
\bibitem{Bunn_Bourdon}%
  \BibitemOpen
  \bibfield{author}{%
  \bibinfo {author} {\bibfnamefont{E.~F.}\ \bibnamefont{Bunn}}\ and\ \bibinfo
  {author} {\bibfnamefont{A.}~\bibnamefont{Bourdon}},\ }%
  \bibfield{journal}{%
  \bibinfo {journal} {Phys. Rev. D}\ }%
  \textbf{\bibinfo {volume} {78}},\ \bibinfo {pages} {123509} (\bibinfo {year}
  {2008}),\
  \Eprint{http://arxiv.org/abs/arXiv:0808.0341}{arXiv:arXiv:0808.0341}%
  \bibAnnoteFile{NoStop}{Bunn_Bourdon}%
\bibitem{Schwarz2004}%
  \BibitemOpen
  \bibfield{author}{%
  \bibinfo {author} {\bibfnamefont{D.~J.}\ \bibnamefont{Schwarz}}, \bibinfo
  {author} {\bibfnamefont{G.~D.}\ \bibnamefont{Starkman}}, \bibinfo {author}
  {\bibfnamefont{D.}~\bibnamefont{Huterer}},\ and\ \bibinfo {author}
  {\bibfnamefont{C.~J.}\ \bibnamefont{Copi}},\ }%
  \bibfield{journal}{%
  \bibinfo {journal} {\physrevlett}\ }%
  \textbf{\bibinfo {volume} {93}},\ \bibinfo {pages} {221301} (\bibinfo {month}
  {Mar.}\ \bibinfo {year} {2004}),\
  \Eprint{http://arxiv.org/abs/astro-ph/0403353}{astro-ph/0403353}%
  \bibAnnoteFile{NoStop}{Schwarz2004}%
\bibitem{TOH}%
  \BibitemOpen
  \bibfield{author}{%
  \bibinfo {author} {\bibfnamefont{M.}~\bibnamefont{{Tegmark}}}, \bibinfo
  {author} {\bibfnamefont{A.}~\bibnamefont{{de Oliveira-Costa}}},\ and\
  \bibinfo {author} {\bibfnamefont{A.~J.}\ \bibnamefont{{Hamilton}}},\ }%
  \bibfield{journal}{%
  \bibinfo {journal} {\physrev}\ }%
  \textbf{\bibinfo {volume} {D68}},\ \bibinfo {pages} {123523} (\bibinfo
  {month} {Dec.}\ \bibinfo {year} {2003})%
  \bibAnnoteFile{NoStop}{TOH}%
\bibitem{Land2005a}%
  \BibitemOpen
  \bibfield{author}{%
  \bibinfo {author} {\bibfnamefont{K.}~\bibnamefont{{Land}}}\ and\ \bibinfo
  {author} {\bibfnamefont{J.}~\bibnamefont{{Magueijo}}},\ }%
  \bibfield{journal}{%
  \bibinfo {journal} {\physrevlett}\ }%
  \textbf{\bibinfo {volume} {95}},\ \bibinfo {pages} {071301} (\bibinfo {year}
  {2005})%
  \bibAnnoteFile{NoStop}{Land2005a}%
\bibitem{Yoho:2010pb}%
  \BibitemOpen
  \bibfield{author}{%
  \bibinfo {author} {\bibfnamefont{A.}~\bibnamefont{Yoho}}, \bibinfo {author}
  {\bibfnamefont{F.}~\bibnamefont{Ferrer}},\ and\ \bibinfo {author}
  {\bibfnamefont{G.~D.}\ \bibnamefont{Starkman}},\ }%
  \bibfield{journal}{%
  \Doi{10.1103/PhysRevD.83.083525}{\bibinfo {journal} {Phys. Rev.}}\ }%
  \textbf{\bibinfo {volume} {D83}},\ \bibinfo {pages} {083525} (\bibinfo {year}
  {2011}),\ \Eprint{http://arxiv.org/abs/1005.5389}{arXiv:1005.5389
  [astro-ph.CO]}%
  \bibAnnoteFile{NoStop}{Yoho:2010pb}%
\bibitem{Eriksen_asym}%
  \BibitemOpen
  \bibfield{author}{%
  \bibinfo {author} {\bibfnamefont{H.~K.}\ \bibnamefont{{Eriksen}}}, \bibinfo
  {author} {\bibfnamefont{F.~K.}\ \bibnamefont{{Hansen}}}, \bibinfo {author}
  {\bibfnamefont{A.~J.}\ \bibnamefont{{Banday}}}, \bibinfo {author}
  {\bibfnamefont{K.~M.}\ \bibnamefont{{G{\' o}rski}}},\ and\ \bibinfo {author}
  {\bibfnamefont{P.~B.}\ \bibnamefont{{Lilje}}},\ }%
  \bibfield{journal}{%
  \bibinfo {journal} {\apj}\ }%
  \textbf{\bibinfo {volume} {605}},\ \bibinfo {pages} {14} (\bibinfo {year}
  {2004}),\ \Eprint{http://arxiv.org/abs/astro-ph/0307507}{astro-ph/0307507}%
  \bibAnnoteFile{NoStop}{Eriksen_asym}%
\bibitem{Hansen_asym}%
  \BibitemOpen
  \bibfield{author}{%
  \bibinfo {author} {\bibfnamefont{F.~K.}\ \bibnamefont{{Hansen}}}, \bibinfo
  {author} {\bibfnamefont{P.}~\bibnamefont{{Cabella}}}, \bibinfo {author}
  {\bibfnamefont{D.}~\bibnamefont{{Marinucci}}},\ and\ \bibinfo {author}
  {\bibfnamefont{N.}~\bibnamefont{{Vittorio}}},\ }%
  \bibfield{journal}{%
  \bibinfo {journal} {\apj}\ }%
  \textbf{\bibinfo {volume} {607}},\ \bibinfo {pages} {L67} (\bibinfo {year}
  {2004}),\ \Eprint{http://arxiv.org/abs/astro-ph/0402396}{astro-ph/0402396}%
  \bibAnnoteFile{NoStop}{Hansen_asym}%
\bibitem{Eriksen:2007pc}%
  \BibitemOpen
  \bibfield{author}{%
  \bibinfo {author} {\bibfnamefont{H.~K.}\ \bibnamefont{Eriksen}}, \bibinfo
  {author} {\bibfnamefont{A.~J.}\ \bibnamefont{Banday}}, \bibinfo {author}
  {\bibfnamefont{K.~M.}\ \bibnamefont{Gorski}}, \bibinfo {author}
  {\bibfnamefont{F.~K.}\ \bibnamefont{Hansen}},\ and\ \bibinfo {author}
  {\bibfnamefont{P.~B.}\ \bibnamefont{Lilje}},\ }%
  \bibfield{journal}{%
  \bibinfo {journal} {\apj}\ }%
  \textbf{\bibinfo {volume} {660}},\ \bibinfo {pages} {L81} (\bibinfo {year}
  {2007}),\ \Eprint{http://arxiv.org/abs/astro-ph/0701089}{astro-ph/0701089}%
  \bibAnnoteFile{NoStop}{Eriksen:2007pc}%
\bibitem{Hansen:2008ym}%
  \BibitemOpen
  \bibfield{author}{%
  \bibinfo {author} {\bibfnamefont{F.~K.}\ \bibnamefont{Hansen}}, \bibinfo
  {author} {\bibfnamefont{A.~J.}\ \bibnamefont{Banday}}, \bibinfo {author}
  {\bibfnamefont{K.~M.}\ \bibnamefont{Gorski}}, \bibinfo {author}
  {\bibfnamefont{H.~K.}\ \bibnamefont{Eriksen}},\ and\ \bibinfo {author}
  {\bibfnamefont{P.~B.}\ \bibnamefont{Lilje}},\ }%
  \bibfield{journal}{%
  \Doi{10.1088/0004-637X/704/2/1448}{\bibinfo {journal} {Astrophys. J.}}\ }%
  \textbf{\bibinfo {volume} {704}},\ \bibinfo {pages} {1448} (\bibinfo {year}
  {2009}),\ \Eprint{http://arxiv.org/abs/arXiv:0812.3795}{arXiv:arXiv:0812.3795
  [astro-ph]}%
  \bibAnnoteFile{NoStop}{Hansen:2008ym}%
\bibitem{Hoftuft}%
  \BibitemOpen
  \bibfield{author}{%
  \bibinfo {author} {\bibfnamefont{J.}~\bibnamefont{Hoftuft}} \emph{et~al.},\
  }%
  \bibfield{journal}{%
  \Doi{10.1088/0004-637X/699/2/985}{\bibinfo {journal} {Astrophys. J.}}\ }%
  \textbf{\bibinfo {volume} {699}},\ \bibinfo {pages} {985} (\bibinfo {year}
  {2009}),\ \Eprint{http://arxiv.org/abs/arXiv:0903.1229}{arXiv:arXiv:0903.1229
  [astro-ph.CO]}%
  \bibAnnoteFile{NoStop}{Hoftuft}%
\bibitem{Raeth:2010kx}%
  \BibitemOpen
  \bibfield{author}{%
  \bibinfo {author} {\bibfnamefont{C.}~\bibnamefont{Raeth}} \emph{et~al.}}%
   (\bibinfo {year} {2010}),\
  \Eprint{http://arxiv.org/abs/1012.2985}{arXiv:1012.2985 [astro-ph.CO]}%
  \bibAnnoteFile{NoStop}{Raeth:2010kx}%
\bibitem{lowl2}%
  \BibitemOpen
  \bibfield{author}{%
  \bibinfo {author} {\bibfnamefont{C.~J.}\ \bibnamefont{Copi}}, \bibinfo
  {author} {\bibfnamefont{D.}~\bibnamefont{Huterer}}, \bibinfo {author}
  {\bibfnamefont{D.~J.}\ \bibnamefont{Schwarz}},\ and\ \bibinfo {author}
  {\bibfnamefont{G.~D.}\ \bibnamefont{Starkman}},\ }%
  \bibfield{journal}{%
  \bibinfo {journal} {Mon. Not. Roy. Astron. Soc.}\ }%
  \textbf{\bibinfo {volume} {367}},\ \bibinfo {pages} {79} (\bibinfo {year}
  {2006}),\ \Eprint{http://arxiv.org/abs/astro-ph/0508047}{astro-ph/0508047}%
  \bibAnnoteFile{NoStop}{lowl2}%
\bibitem{Bennett_anomalies}%
  \BibitemOpen
  \bibfield{author}{%
  \bibinfo {author} {\bibfnamefont{C.~L.}\ \bibnamefont{Bennett}}
  \emph{et~al.},\ }%
  \bibfield{journal}{%
  \Doi{10.1088/0067-0049/192/2/17}{\bibinfo {journal} {Astrophys. J. Suppl.}}\
  }%
  \textbf{\bibinfo {volume} {192}},\ \bibinfo {pages} {17} (\bibinfo {year}
  {2011}),\ \Eprint{http://arxiv.org/abs/1001.4758}{arXiv:1001.4758
  [astro-ph.CO]}%
  \bibAnnoteFile{NoStop}{Bennett_anomalies}%
\bibitem{Huterer_NewAst_review}%
  \BibitemOpen
  \bibfield{author}{%
  \bibinfo {author} {\bibfnamefont{D.}~\bibnamefont{Huterer}},\ }%
  \bibfield{journal}{%
  \bibinfo {journal} {New Astron. Rev.}\ }%
  \textbf{\bibinfo {volume} {50}},\ \bibinfo {pages} {868} (\bibinfo {year}
  {2006}),\
  \Eprint{http://arxiv.org/abs/astro-ph/0608318}{arXiv:astro-ph/0608318}%
  \bibAnnoteFile{NoStop}{Huterer_NewAst_review}%
\bibitem{CHSS_review}%
  \BibitemOpen
  \bibfield{author}{%
  \bibinfo {author} {\bibfnamefont{C.~J.}\ \bibnamefont{Copi}}, \bibinfo
  {author} {\bibfnamefont{D.}~\bibnamefont{Huterer}}, \bibinfo {author}
  {\bibfnamefont{D.~J.}\ \bibnamefont{Schwarz}},\ and\ \bibinfo {author}
  {\bibfnamefont{G.~D.}\ \bibnamefont{Starkman}},\ }%
  \bibfield{journal}{%
  \bibinfo {journal} {Adv.Astron.}\ }%
  \textbf{\bibinfo {volume} {2010}},\ \bibinfo {pages} {847541} (\bibinfo
  {year} {2010}),\ \Eprint{http://arxiv.org/abs/1004.5602}{arXiv:1004.5602
  [astro-ph.CO]}%
  \bibAnnoteFile{NoStop}{CHSS_review}%
\bibitem{Hirata}%
  \BibitemOpen
  \bibfield{author}{%
  \bibinfo {author} {\bibfnamefont{C.~M.}\ \bibnamefont{Hirata}},\ }%
  \bibfield{journal}{%
  \Doi{10.1088/1475-7516/2009/09/011}{\bibinfo {journal} {JCAP}}\ }%
  \textbf{\bibinfo {volume} {0909}},\ \bibinfo {pages} {011} (\bibinfo {year}
  {2009}),\ \Eprint{http://arxiv.org/abs/arXiv:0907.0703}{arXiv:arXiv:0907.0703
  [astro-ph.CO]}%
  \bibAnnoteFile{NoStop}{Hirata}%
\bibitem{Pullen_Hirata}%
  \BibitemOpen
  \bibfield{author}{%
  \bibinfo {author} {\bibfnamefont{A.~R.}\ \bibnamefont{Pullen}}\ and\ \bibinfo
  {author} {\bibfnamefont{C.~M.}\ \bibnamefont{Hirata}},\ }%
  \bibfield{journal}{%
  \Doi{10.1088/1475-7516/2010/05/027}{\bibinfo {journal} {JCAP}}\ }%
  \textbf{\bibinfo {volume} {1005}},\ \bibinfo {pages} {027} (\bibinfo {year}
  {2010}),\ \Eprint{http://arxiv.org/abs/1003.0673}{arXiv:1003.0673
  [astro-ph.CO]}%
  \bibAnnoteFile{NoStop}{Pullen_Hirata}%
\bibitem{Ando:2008zza}%
  \BibitemOpen
  \bibfield{author}{%
  \bibinfo {author} {\bibfnamefont{S.}~\bibnamefont{Ando}}\ and\ \bibinfo
  {author} {\bibfnamefont{M.}~\bibnamefont{Kamionkowski}},\ }%
  \bibfield{journal}{%
  \Doi{10.1103/PhysRevLett.100.071301}{\bibinfo {journal} {Phys. Rev. Lett.}}\
  }%
  \textbf{\bibinfo {volume} {100}},\ \bibinfo {pages} {071301} (\bibinfo {year}
  {2008}),\ \Eprint{http://arxiv.org/abs/arXiv:0711.0779}{arXiv:arXiv:0711.0779
  [astro-ph]}%
  \bibAnnoteFile{NoStop}{Ando:2008zza}%
\bibitem{Gordon2005}%
  \BibitemOpen
  \bibfield{author}{%
  \bibinfo {author} {\bibfnamefont{C.}~\bibnamefont{Gordon}}, \bibinfo {author}
  {\bibfnamefont{W.}~\bibnamefont{Hu}}, \bibinfo {author}
  {\bibfnamefont{D.}~\bibnamefont{Huterer}},\ and\ \bibinfo {author}
  {\bibfnamefont{T.~M.}\ \bibnamefont{Crawford}},\ }%
  \bibfield{journal}{%
  \Doi{10.1103/PhysRevD.72.103002}{\bibinfo {journal} {Phys. Rev.}}\ }%
  \textbf{\bibinfo {volume} {D72}},\ \bibinfo {pages} {103002} (\bibinfo {year}
  {2005}),\
  \Eprint{http://arxiv.org/abs/astro-ph/0509301}{arXiv:astro-ph/0509301}%
  \bibAnnoteFile{NoStop}{Gordon2005}%
\bibitem{ACW}%
  \BibitemOpen
  \bibfield{author}{%
  \bibinfo {author} {\bibfnamefont{L.}~\bibnamefont{Ackerman}}, \bibinfo
  {author} {\bibfnamefont{S.~M.}\ \bibnamefont{Carroll}},\ and\ \bibinfo
  {author} {\bibfnamefont{M.~B.}\ \bibnamefont{Wise}},\ }%
  \bibfield{journal}{%
  \Doi{10.1103/PhysRevD.75.083502}{\bibinfo {journal} {Phys. Rev.}}\ }%
  \textbf{\bibinfo {volume} {D75}},\ \bibinfo {pages} {083502} (\bibinfo {year}
  {2007}),\
  \Eprint{http://arxiv.org/abs/astro-ph/0701357}{arXiv:astro-ph/0701357}%
  \bibAnnoteFile{NoStop}{ACW}%
\bibitem{Donoghue:2007ze}%
  \BibitemOpen
  \bibfield{author}{%
  \bibinfo {author} {\bibfnamefont{J.~F.}\ \bibnamefont{Donoghue}}, \bibinfo
  {author} {\bibfnamefont{K.}~\bibnamefont{Dutta}},\ and\ \bibinfo {author}
  {\bibfnamefont{A.}~\bibnamefont{Ross}},\ }%
  \bibfield{journal}{%
  \Doi{10.1103/PhysRevD.80.023526}{\bibinfo {journal} {Phys. Rev.}}\ }%
  \textbf{\bibinfo {volume} {D80}},\ \bibinfo {pages} {023526} (\bibinfo {year}
  {2009}),\
  \Eprint{http://arxiv.org/abs/astro-ph/0703455}{arXiv:astro-ph/0703455}%
  \bibAnnoteFile{NoStop}{Donoghue:2007ze}%
\bibitem{ArmendarizPicon07}%
  \BibitemOpen
  \bibfield{author}{%
  \bibinfo {author} {\bibfnamefont{C.}~\bibnamefont{Armendariz-Picon}},\ }%
  \bibfield{journal}{%
  \Doi{10.1088/1475-7516/2007/09/014}{\bibinfo {journal} {JCAP}}\ }%
  \textbf{\bibinfo {volume} {0709}},\ \bibinfo {pages} {014} (\bibinfo {year}
  {2007}),\ \Eprint{http://arxiv.org/abs/arXiv:0705.1167}{arXiv:arXiv:0705.1167
  [astro-ph]}%
  \bibAnnoteFile{NoStop}{ArmendarizPicon07}%
\bibitem{Gumrukcuoglu}%
  \BibitemOpen
  \bibfield{author}{%
  \bibinfo {author} {\bibfnamefont{A.~E.}\ \bibnamefont{Gumrukcuoglu}},
  \bibinfo {author} {\bibfnamefont{C.~R.}\ \bibnamefont{Contaldi}},\ and\
  \bibinfo {author} {\bibfnamefont{M.}~\bibnamefont{Peloso}},\ }%
  \bibfield{journal}{%
  \Doi{10.1088/1475-7516/2007/11/005}{\bibinfo {journal} {JCAP}}\ }%
  \textbf{\bibinfo {volume} {0711}},\ \bibinfo {pages} {005} (\bibinfo {year}
  {2007}),\ \Eprint{http://arxiv.org/abs/arXiv:0707.4179}{arXiv:arXiv:0707.4179
  [astro-ph]}%
  \bibAnnoteFile{NoStop}{Gumrukcuoglu}%
\bibitem{Rodrigues_Bianchi}%
  \BibitemOpen
  \bibfield{author}{%
  \bibinfo {author} {\bibfnamefont{D.~C.}\ \bibnamefont{Rodrigues}},\ }%
  \bibfield{journal}{%
  \Doi{10.1103/PhysRevD.77.023534}{\bibinfo {journal} {Phys. Rev.}}\ }%
  \textbf{\bibinfo {volume} {D77}},\ \bibinfo {pages} {023534} (\bibinfo {year}
  {2008}),\ \Eprint{http://arxiv.org/abs/arXiv:0708.1168}{arXiv:arXiv:0708.1168
  [astro-ph]}%
  \bibAnnoteFile{NoStop}{Rodrigues_Bianchi}%
\bibitem{Pullen_Kam}%
  \BibitemOpen
  \bibfield{author}{%
  \bibinfo {author} {\bibfnamefont{A.~R.}\ \bibnamefont{Pullen}}\ and\ \bibinfo
  {author} {\bibfnamefont{M.}~\bibnamefont{Kamionkowski}},\ }%
  \bibfield{journal}{%
  \Doi{10.1103/PhysRevD.76.103529}{\bibinfo {journal} {Phys. Rev.}}\ }%
  \textbf{\bibinfo {volume} {D76}},\ \bibinfo {pages} {103529} (\bibinfo {year}
  {2007}),\ \Eprint{http://arxiv.org/abs/0709.1144}{arXiv:0709.1144
  [astro-ph]}%
  \bibAnnoteFile{NoStop}{Pullen_Kam}%
\bibitem{Pitrou08}%
  \BibitemOpen
  \bibfield{author}{%
  \bibinfo {author} {\bibfnamefont{C.}~\bibnamefont{Pitrou}}, \bibinfo {author}
  {\bibfnamefont{T.~S.}\ \bibnamefont{Pereira}},\ and\ \bibinfo {author}
  {\bibfnamefont{J.-P.}\ \bibnamefont{Uzan}},\ }%
  \bibfield{journal}{%
  \Doi{10.1088/1475-7516/2008/04/004}{\bibinfo {journal} {JCAP}}\ }%
  \textbf{\bibinfo {volume} {0804}},\ \bibinfo {pages} {004} (\bibinfo {year}
  {2008}),\ \Eprint{http://arxiv.org/abs/arXiv:0801.3596}{arXiv:arXiv:0801.3596
  [astro-ph]}%
  \bibAnnoteFile{NoStop}{Pitrou08}%
\bibitem{Erickcek_hem}%
  \BibitemOpen
  \bibfield{author}{%
  \bibinfo {author} {\bibfnamefont{A.~L.}\ \bibnamefont{Erickcek}}, \bibinfo
  {author} {\bibfnamefont{M.}~\bibnamefont{Kamionkowski}},\ and\ \bibinfo
  {author} {\bibfnamefont{S.~M.}\ \bibnamefont{Carroll}},\ }%
  \bibfield{journal}{%
  \Doi{10.1103/PhysRevD.78.123520}{\bibinfo {journal} {Phys. Rev.}}\ }%
  \textbf{\bibinfo {volume} {D78}},\ \bibinfo {pages} {123520} (\bibinfo {year}
  {2008}),\ \Eprint{http://arxiv.org/abs/arXiv:0806.0377}{arXiv:arXiv:0806.0377
  [astro-ph]}%
  \bibAnnoteFile{NoStop}{Erickcek_hem}%
\bibitem{Erickcek_iso}%
  \BibitemOpen
  \bibfield{author}{%
  \bibinfo {author} {\bibfnamefont{A.~L.}\ \bibnamefont{Erickcek}}, \bibinfo
  {author} {\bibfnamefont{C.~M.}\ \bibnamefont{Hirata}},\ and\ \bibinfo
  {author} {\bibfnamefont{M.}~\bibnamefont{Kamionkowski}},\ }%
  \bibfield{journal}{%
  \Doi{10.1103/PhysRevD.80.083507}{\bibinfo {journal} {Phys. Rev.}}\ }%
  \textbf{\bibinfo {volume} {D80}},\ \bibinfo {pages} {083507} (\bibinfo {year}
  {2009}),\ \Eprint{http://arxiv.org/abs/arXiv:0907.0705}{arXiv:arXiv:0907.0705
  [astro-ph.CO]}%
  \bibAnnoteFile{NoStop}{Erickcek_iso}%
\bibitem{Battye09}%
  \BibitemOpen
  \bibfield{author}{%
  \bibinfo {author} {\bibfnamefont{R.}~\bibnamefont{Battye}}\ and\ \bibinfo
  {author} {\bibfnamefont{A.}~\bibnamefont{Moss}},\ }%
  \bibfield{journal}{%
  \Doi{10.1103/PhysRevD.80.023531}{\bibinfo {journal} {Phys. Rev.}}\ }%
  \textbf{\bibinfo {volume} {D80}},\ \bibinfo {pages} {023531} (\bibinfo {year}
  {2009}),\ \Eprint{http://arxiv.org/abs/arXiv:0905.3403}{arXiv:arXiv:0905.3403
  [astro-ph.CO]}%
  \bibAnnoteFile{NoStop}{Battye09}%
\bibitem{Groeneboom:2008fz}%
  \BibitemOpen
  \bibfield{author}{%
  \bibinfo {author} {\bibfnamefont{N.~E.}\ \bibnamefont{Groeneboom}}\ and\
  \bibinfo {author} {\bibfnamefont{H.~K.}\ \bibnamefont{Eriksen}},\ }%
  \bibfield{journal}{%
  \Doi{10.1088/0004-637X/690/2/1807}{\bibinfo {journal} {Astrophys. J.}}\ }%
  \textbf{\bibinfo {volume} {690}},\ \bibinfo {pages} {1807} (\bibinfo {year}
  {2009}),\ \Eprint{http://arxiv.org/abs/arXiv:0807.2242}{arXiv:arXiv:0807.2242
  [astro-ph]}%
  \bibAnnoteFile{NoStop}{Groeneboom:2008fz}%
\bibitem{Hanson_Lewis}%
  \BibitemOpen
  \bibfield{author}{%
  \bibinfo {author} {\bibfnamefont{D.}~\bibnamefont{Hanson}}\ and\ \bibinfo
  {author} {\bibfnamefont{A.}~\bibnamefont{Lewis}},\ }%
  \bibfield{journal}{%
  \Doi{10.1103/PhysRevD.80.063004}{\bibinfo {journal} {Phys. Rev.}}\ }%
  \textbf{\bibinfo {volume} {D80}},\ \bibinfo {pages} {063004} (\bibinfo {year}
  {2009}),\ \Eprint{http://arxiv.org/abs/arXiv:0908.0963}{arXiv:arXiv:0908.0963
  [astro-ph.CO]}%
  \bibAnnoteFile{NoStop}{Hanson_Lewis}%
\bibitem{Groeneboom_sys}%
  \BibitemOpen
  \bibfield{author}{%
  \bibinfo {author} {\bibfnamefont{N.~E.}\ \bibnamefont{Groeneboom}}, \bibinfo
  {author} {\bibfnamefont{L.}~\bibnamefont{Ackerman}}, \bibinfo {author}
  {\bibfnamefont{I.~K.}\ \bibnamefont{Wehus}},\ and\ \bibinfo {author}
  {\bibfnamefont{H.~K.}\ \bibnamefont{Eriksen}},\ }%
  \bibfield{journal}{%
  \Doi{10.1088/0004-637X/722/1/452}{\bibinfo {journal} {Astrophys.J.}}\ }%
  \textbf{\bibinfo {volume} {722}},\ \bibinfo {pages} {452} (\bibinfo {year}
  {2010}),\ \Eprint{http://arxiv.org/abs/0911.0150}{arXiv:0911.0150
  [astro-ph.CO]}%
  \bibAnnoteFile{NoStop}{Groeneboom_sys}%
\bibitem{Hanson_beams}%
  \BibitemOpen
  \bibfield{author}{%
  \bibinfo {author} {\bibfnamefont{D.}~\bibnamefont{Hanson}}, \bibinfo {author}
  {\bibfnamefont{A.}~\bibnamefont{Lewis}},\ and\ \bibinfo {author}
  {\bibfnamefont{A.}~\bibnamefont{Challinor}},\ }%
  \bibfield{journal}{%
  \Doi{10.1103/PhysRevD.81.103003}{\bibinfo {journal} {Phys. Rev.}}\ }%
  \textbf{\bibinfo {volume} {D81}},\ \bibinfo {pages} {103003} (\bibinfo {year}
  {2010}),\ \Eprint{http://arxiv.org/abs/1003.0198}{arXiv:1003.0198
  [astro-ph.CO]}%
  \bibAnnoteFile{NoStop}{Hanson_beams}%
\bibitem{deOliveira2004}%
  \BibitemOpen
  \bibfield{author}{%
  \bibinfo {author} {\bibfnamefont{A.}~\bibnamefont{de~Oliveira-Costa}},
  \bibinfo {author} {\bibfnamefont{M.}~\bibnamefont{Tegmark}}, \bibinfo
  {author} {\bibfnamefont{M.}~\bibnamefont{Zaldarriaga}},\ and\ \bibinfo
  {author} {\bibfnamefont{A.}~\bibnamefont{Hamilton}},\ }%
  \bibfield{journal}{%
  \bibinfo {journal} {\physrev}\ }%
  \textbf{\bibinfo {volume} {D69}},\ \bibinfo {pages} {063516} (\bibinfo {year}
  {2004}),\ \Eprint{http://arxiv.org/abs/astro-ph/0307282}{astro-ph/0307282}%
  \bibAnnoteFile{NoStop}{deOliveira2004}%
\bibitem{Copi:2003kt}%
  \BibitemOpen
  \bibfield{author}{%
  \bibinfo {author} {\bibfnamefont{C.~J.}\ \bibnamefont{Copi}}, \bibinfo
  {author} {\bibfnamefont{D.}~\bibnamefont{Huterer}},\ and\ \bibinfo {author}
  {\bibfnamefont{G.~D.}\ \bibnamefont{Starkman}},\ }%
  \bibfield{journal}{%
  \Doi{10.1103/PhysRevD.70.043515}{\bibinfo {journal} {Phys. Rev.}}\ }%
  \textbf{\bibinfo {volume} {D70}},\ \bibinfo {pages} {043515} (\bibinfo {year}
  {2004}),\
  \Eprint{http://arxiv.org/abs/astro-ph/0310511}{arXiv:astro-ph/0310511}%
  \bibAnnoteFile{NoStop}{Copi:2003kt}%
\bibitem{Maxwell}%
  \BibitemOpen
  \bibfield{author}{%
  \bibinfo {author} {\bibfnamefont{J.~C.}\ \bibnamefont{{Maxwell}}},\ }%
  \emph{\bibinfo {title} {{A Treatise on Electricity and Magnetism}}},\
  \bibinfo {edition} {3rd}\ ed.,\ Vol.~\bibinfo {volume} {I}\ (\bibinfo
  {publisher} {Clarendon Press},\ \bibinfo {address} {London},\ \bibinfo {year}
  {1891})%
  \bibAnnoteFile{NoStop}{Maxwell}%
\bibitem{Katz2004}%
  \BibitemOpen
  \bibfield{author}{%
  \bibinfo {author} {\bibfnamefont{G.}~\bibnamefont{Katz}}\ and\ \bibinfo
  {author} {\bibfnamefont{J.}~\bibnamefont{Weeks}},\ }%
  \bibfield{journal}{%
  \bibinfo {journal} {\physrev}\ }%
  \textbf{\bibinfo {volume} {D70}},\ \bibinfo {pages} {063527} (\bibinfo {year}
  {2004}),\ \Eprint{http://arxiv.org/abs/astro-ph/0405631}{astro-ph/0405631}%
  \bibAnnoteFile{NoStop}{Katz2004}%
\bibitem{Dennis2005}%
  \BibitemOpen
  \bibfield{author}{%
  \bibinfo {author} {\bibfnamefont{M.~R.}\ \bibnamefont{{Dennis}}},\ }%
  \bibfield{journal}{%
  \bibinfo {journal} {J. Phys. A: Math. Gen.}\ }%
  \textbf{\bibinfo {volume} {38}},\ \bibinfo {pages} {1653} (\bibinfo {month}
  {Feb.}\ \bibinfo {year} {2005})%
  \bibAnnoteFile{NoStop}{Dennis2005}%
\bibitem{MV_code}%
  \BibitemOpen
  \bibfield{author}{%
  \bibinfo {author} {\bibfnamefont{C.~J.}\ \bibnamefont{{Copi}}},\ }%
  \enquote{\bibinfo {title} {Freely downloadable multipole vector code},}\
  \bibinfo {note}
  {\texttt{http://www.phys.cwru.edu/projects/mpvectors/\#code}}%
  \bibAnnoteFile{NoStop}{MV_code}%
\bibitem{Weeks04}%
  \BibitemOpen
  \bibfield{author}{%
  \bibinfo {author} {\bibfnamefont{J.~R.}\ \bibnamefont{{Weeks}}},\ }%
  \bibfield{journal}{%
  \bibinfo {journal} {astro-ph/0412231}}%
   (\bibinfo {month} {Dec.}\ \bibinfo {year} {2004})%
  \bibAnnoteFile{NoStop}{Weeks04}%
\bibitem{Helling}%
  \BibitemOpen
  \bibfield{author}{%
  \bibinfo {author} {\bibfnamefont{R.~C.}\ \bibnamefont{{Helling}}}, \bibinfo
  {author} {\bibfnamefont{P.}~\bibnamefont{{Schupp}}},\ and\ \bibinfo {author}
  {\bibfnamefont{T.}~\bibnamefont{{Tesileanu}}},\ }%
  \bibfield{journal}{%
  \Doi{10.1103/PhysRevD.74.063004}{\bibinfo {journal} {\prd}}\ }%
  \textbf{\bibinfo {volume} {74}},\ \bibinfo {pages} {063004} (\bibinfo {month}
  {Sep.}\ \bibinfo {year} {2006}),\
  \Eprint{http://arxiv.org/abs/astro-ph/0603594}{astro-ph/0603594}%
  \bibAnnoteFile{NoStop}{Helling}%
\bibitem{Hinshaw:2006ia}%
  \BibitemOpen
  \bibfield{author}{%
  \bibinfo {author} {\bibfnamefont{G.}~\bibnamefont{Hinshaw}} \emph{et~al.}
  (\bibinfo {collaboration} {WMAP}),\ }%
  \bibfield{journal}{%
  \bibinfo {journal} {\apjs}\ }%
  \textbf{\bibinfo {volume} {170}},\ \bibinfo {pages} {288} (\bibinfo {year}
  {2007}),\ \Eprint{http://arxiv.org/abs/astro-ph/0603451}{astro-ph/0603451}%
  \bibAnnoteFile{NoStop}{Hinshaw:2006ia}%
\bibitem{Dennis2004}%
  \BibitemOpen
  \bibfield{author}{%
  \bibinfo {author} {\bibfnamefont{M.~R.}\ \bibnamefont{{Dennis}}},\ }%
  \bibfield{journal}{%
  \bibinfo {journal} {J. Phys. A: Math. Gen.}\ }%
  \textbf{\bibinfo {volume} {37}},\ \bibinfo {pages} {9487} (\bibinfo {month}
  {Oct.}\ \bibinfo {year} {2004})%
  \bibAnnoteFile{NoStop}{Dennis2004}%
\bibitem{Bennett_foregr}%
  \BibitemOpen
  \bibfield{author}{%
  \bibinfo {author} {\bibfnamefont{C.~L.}\ \bibnamefont{{Bennett}}}
  \emph{et~al.},\ }%
  \bibfield{journal}{%
  \bibinfo {journal} {\apjs}\ }%
  \textbf{\bibinfo {volume} {148}},\ \bibinfo {pages} {97} (\bibinfo {year}
  {2003}),\ \Eprint{http://arxiv.org/abs/astro-ph/0302208}{astro-ph/0302208}%
  \bibAnnoteFile{NoStop}{Bennett_foregr}%
\bibitem{LILC}%
  \BibitemOpen
  \bibfield{author}{%
  \bibinfo {author} {\bibfnamefont{H.~K.}\ \bibnamefont{{Eriksen}}}, \bibinfo
  {author} {\bibfnamefont{A.~J.}\ \bibnamefont{{Banday}}}, \bibinfo {author}
  {\bibfnamefont{K.~M.}\ \bibnamefont{{G{\'o}rski}}},\ and\ \bibinfo {author}
  {\bibfnamefont{P.~B.}\ \bibnamefont{{Lilje}}},\ }%
  \bibfield{journal}{%
  \bibinfo {journal} {\apj}\ }%
  \textbf{\bibinfo {volume} {612}},\ \bibinfo {pages} {633} (\bibinfo {year}
  {2004}),\ \Eprint{http://arxiv.org/abs/astro-ph/0403098}{astro-ph/0403098}%
  \bibAnnoteFile{NoStop}{LILC}%
\bibitem{Tegmark:1995pn}%
  \BibitemOpen
  \bibfield{author}{%
  \bibinfo {author} {\bibfnamefont{M.}~\bibnamefont{Tegmark}}\ and\ \bibinfo
  {author} {\bibfnamefont{G.}~\bibnamefont{Efstathiou}},\ }%
  \bibfield{journal}{%
  \bibinfo {journal} {\mnras}\ }%
  \textbf{\bibinfo {volume} {281}},\ \bibinfo {pages} {1297} (\bibinfo {month}
  {aug}\ \bibinfo {year} {1996}),\
  \Eprint{http://arxiv.org/abs/astro-ph/9507009}{arXiv:astro-ph/9507009}%
  \bibAnnoteFile{NoStop}{Tegmark:1995pn}%
\bibitem{Hansen:2006rj}%
  \BibitemOpen
  \bibfield{author}{%
  \bibinfo {author} {\bibfnamefont{F.~K.}\ \bibnamefont{Hansen}}, \bibinfo
  {author} {\bibfnamefont{A.~J.}\ \bibnamefont{Banday}}, \bibinfo {author}
  {\bibfnamefont{H.~K.}\ \bibnamefont{Eriksen}}, \bibinfo {author}
  {\bibfnamefont{K.~M.}\ \bibnamefont{Gorski}},\ and\ \bibinfo {author}
  {\bibfnamefont{P.~B.}\ \bibnamefont{Lilje}},\ }%
  \bibfield{journal}{%
  \Doi{10.1086/506015}{\bibinfo {journal} {Astrophys. J.}}\ }%
  \textbf{\bibinfo {volume} {648}},\ \bibinfo {pages} {784} (\bibinfo {year}
  {2006}),\
  \Eprint{http://arxiv.org/abs/astro-ph/0603308}{arXiv:astro-ph/0603308}%
  \bibAnnoteFile{NoStop}{Hansen:2006rj}%
\bibitem{deOliveiraCosta:2006zj}%
  \BibitemOpen
  \bibfield{author}{%
  \bibinfo {author} {\bibfnamefont{A.}~\bibnamefont{de~Oliveira-Costa}}\ and\
  \bibinfo {author} {\bibfnamefont{M.}~\bibnamefont{Tegmark}},\ }%
  \bibfield{journal}{%
  \bibinfo {journal} {Phys. Rev.}\ }%
  \textbf{\bibinfo {volume} {D74}},\ \bibinfo {pages} {023005} (\bibinfo {year}
  {2006}),\ \Eprint{http://arxiv.org/abs/astro-ph/0603369}{astro-ph/0603369}%
  \bibAnnoteFile{NoStop}{deOliveiraCosta:2006zj}%
\bibitem{Tegmark:1997vs}%
  \BibitemOpen
  \bibfield{author}{%
  \bibinfo {author} {\bibfnamefont{M.}~\bibnamefont{Tegmark}},\ }%
  \bibfield{journal}{%
  \Doi{10.1103/PhysRevD.56.4514}{\bibinfo {journal} {Phys. Rev.}}\ }%
  \textbf{\bibinfo {volume} {D56}},\ \bibinfo {pages} {4514} (\bibinfo {year}
  {1997}),\
  \Eprint{http://arxiv.org/abs/astro-ph/9705188}{arXiv:astro-ph/9705188}%
  \bibAnnoteFile{NoStop}{Tegmark:1997vs}%
\bibitem{HTBJ}%
  \BibitemOpen
  \bibfield{author}{%
  \bibinfo {author} {\bibfnamefont{D.}~\bibnamefont{Huterer}}, \bibinfo
  {author} {\bibfnamefont{M.}~\bibnamefont{Takada}}, \bibinfo {author}
  {\bibfnamefont{G.}~\bibnamefont{Bernstein}},\ and\ \bibinfo {author}
  {\bibfnamefont{B.}~\bibnamefont{Jain}},\ }%
  \bibfield{journal}{%
  \Doi{10.1111/j.1365-2966.2005.09782.x}{\bibinfo {journal} {Mon. Not. Roy.
  Astron. Soc.}}\ }%
  \textbf{\bibinfo {volume} {366}},\ \bibinfo {pages} {101} (\bibinfo {year}
  {2006}),\
  \Eprint{http://arxiv.org/abs/astro-ph/0506030}{arXiv:astro-ph/0506030}%
  \bibAnnoteFile{NoStop}{HTBJ}%
\bibitem{healpix}%
  \BibitemOpen
  \bibfield{author}{%
  \bibinfo {author} {\bibfnamefont{K.~M.}\ \bibnamefont{{G{\' o}rski}}},
  \bibinfo {author} {\bibfnamefont{E.}~\bibnamefont{{Hivon}}}, \bibinfo
  {author} {\bibfnamefont{A.~J.}\ \bibnamefont{{Banday}}}, \bibinfo {author}
  {\bibfnamefont{B.~D.}\ \bibnamefont{{Wandelt}}}, \bibinfo {author}
  {\bibfnamefont{F.~K.}\ \bibnamefont{{Hansen}}}, \bibinfo {author}
  {\bibfnamefont{M.}~\bibnamefont{{Reinecke}}},\ and\ \bibinfo {author}
  {\bibfnamefont{M.}~\bibnamefont{{Bartelmann}}},\ }%
  \bibfield{journal}{%
  \Doi{10.1086/427976}{\bibinfo {journal} {\apj}}\ }%
  \textbf{\bibinfo {volume} {622}},\ \bibinfo {pages} {759} (\bibinfo {month}
  {Apr.}\ \bibinfo {year} {2005})%
  \bibAnnoteFile{NoStop}{healpix}%
\bibitem{Efstathiou2009}%
  \BibitemOpen
  \bibfield{author}{%
  \bibinfo {author} {\bibfnamefont{G.}~\bibnamefont{{Efstathiou}}}, \bibinfo
  {author} {\bibfnamefont{Y.}~\bibnamefont{{Ma}}},\ and\ \bibinfo {author}
  {\bibfnamefont{D.}~\bibnamefont{{Hanson}}}}%
   (\bibinfo {month} {Nov.}\ \bibinfo {year} {2009}),\
  \Eprint{http://arxiv.org/abs/arXiv:0911.5399}{arXiv:arXiv:0911.5399}%
  \bibAnnoteFile{NoStop}{Efstathiou2009}%
\bibitem{Pontzen_Peiris}%
  \BibitemOpen
  \bibfield{author}{%
  \bibinfo {author} {\bibfnamefont{A.}~\bibnamefont{Pontzen}}\ and\ \bibinfo
  {author} {\bibfnamefont{H.~V.}\ \bibnamefont{Peiris}},\ }%
  \bibfield{journal}{%
  \Doi{10.1103/PhysRevD.81.103008}{\bibinfo {journal} {Phys.Rev.}}\ }%
  \textbf{\bibinfo {volume} {D81}},\ \bibinfo {pages} {103008} (\bibinfo {year}
  {2010}),\ \Eprint{http://arxiv.org/abs/1004.2706}{arXiv:1004.2706
  [astro-ph.CO]}%
  \bibAnnoteFile{NoStop}{Pontzen_Peiris}%
\bibitem{Aurich_Lustig}%
  \BibitemOpen
  \bibfield{author}{%
  \bibinfo {author} {\bibfnamefont{R.}~\bibnamefont{Aurich}}\ and\ \bibinfo
  {author} {\bibfnamefont{S.}~\bibnamefont{Lustig}},\ }%
  \bibfield{journal}{%
  \Doi{10.1111/j.1365-2966.2010.17667.x}{\bibinfo {journal}
  {Mon.Not.Roy.Astron.Soc.}}\ }%
  \textbf{\bibinfo {volume} {411}},\ \bibinfo {pages} {124} (\bibinfo {year}
  {2011}),\ \Eprint{http://arxiv.org/abs/1005.5069}{arXiv:1005.5069
  [astro-ph.CO]}%
  \bibAnnoteFile{NoStop}{Aurich_Lustig}%
\bibitem{Gibelyou:2010qe}%
  \BibitemOpen
  \bibfield{author}{%
  \bibinfo {author} {\bibfnamefont{C.}~\bibnamefont{Gibelyou}}, \bibinfo
  {author} {\bibfnamefont{D.}~\bibnamefont{Huterer}},\ and\ \bibinfo {author}
  {\bibfnamefont{W.}~\bibnamefont{Fang}},\ }%
  \bibfield{journal}{%
  \Doi{10.1103/PhysRevD.82.123009}{\bibinfo {journal} {Phys.Rev.}}\ }%
  \textbf{\bibinfo {volume} {D82}},\ \bibinfo {pages} {123009} (\bibinfo {year}
  {2010}),\ \Eprint{http://arxiv.org/abs/1007.0757}{arXiv:1007.0757
  [astro-ph.CO]}%
  \bibAnnoteFile{NoStop}{Gibelyou:2010qe}%
\bibitem{Tegmark:1997cx}%
  \BibitemOpen
  \bibfield{author}{%
  \bibinfo {author} {\bibfnamefont{M.}~\bibnamefont{Tegmark}}}%
   (\bibinfo {year} {1997}),\
  \Eprint{http://arxiv.org/abs/astro-ph/9708021}{arXiv:astro-ph/9708021}%
  \bibAnnoteFile{NoStop}{Tegmark:1997cx}%
\bibitem{EDSGC}%
  \BibitemOpen
  \bibfield{author}{%
  \bibinfo {author} {\bibfnamefont{D.}~\bibnamefont{{Huterer}}}, \bibinfo
  {author} {\bibfnamefont{L.}~\bibnamefont{{Knox}}},\ and\ \bibinfo {author}
  {\bibfnamefont{R.~C.}\ \bibnamefont{{Nichol}}},\ }%
  \bibfield{journal}{%
  \Doi{10.1086/323328}{\bibinfo {journal} {\apj}}\ }%
  \textbf{\bibinfo {volume} {555}},\ \bibinfo {pages} {547} (\bibinfo {month}
  {Jul.}\ \bibinfo {year} {2001}),\
  \Eprint{http://arxiv.org/abs/astro-ph/0011069}{astro-ph/0011069}%
  \bibAnnoteFile{NoStop}{EDSGC}%
\bibitem{Baugh_Efstathiou}%
  \BibitemOpen
  \bibfield{author}{%
  \bibinfo {author} {\bibfnamefont{C.~M.}\ \bibnamefont{{Baugh}}}\ and\
  \bibinfo {author} {\bibfnamefont{G.}~\bibnamefont{{Efstathiou}}},\ }%
  \bibfield{journal}{%
  \bibinfo {journal} {\mnras}\ }%
  \textbf{\bibinfo {volume} {267}},\ \bibinfo {pages} {323} (\bibinfo {month}
  {Mar.}\ \bibinfo {year} {1994})%
  \bibAnnoteFile{NoStop}{Baugh_Efstathiou}%
\bibitem{Efstathiou_Moody}%
  \BibitemOpen
  \bibfield{author}{%
  \bibinfo {author} {\bibfnamefont{G.}~\bibnamefont{{Efstathiou}}}\ and\
  \bibinfo {author} {\bibfnamefont{S.~J.}\ \bibnamefont{{Moody}}},\ }%
  \bibfield{journal}{%
  \Doi{10.1046/j.1365-8711.2001.04576.x}{\bibinfo {journal} {\mnras}}\ }%
  \textbf{\bibinfo {volume} {325}},\ \bibinfo {pages} {1603} (\bibinfo {month}
  {Aug.}\ \bibinfo {year} {2001}),\
  \Eprint{http://arxiv.org/abs/astro-ph/0010478}{astro-ph/0010478}%
  \bibAnnoteFile{NoStop}{Efstathiou_Moody}%
\bibitem{Cole:2005sx}%
  \BibitemOpen
  \bibfield{author}{%
  \bibinfo {author} {\bibfnamefont{S.}~\bibnamefont{Cole}} \emph{et~al.}
  (\bibinfo {collaboration} {The 2dFGRS}),\ }%
  \bibfield{journal}{%
  \Doi{10.1111/j.1365-2966.2005.09318.x}{\bibinfo {journal} {Mon. Not. Roy.
  Astron. Soc.}}\ }%
  \textbf{\bibinfo {volume} {362}},\ \bibinfo {pages} {505} (\bibinfo {year}
  {2005}),\
  \Eprint{http://arxiv.org/abs/astro-ph/0501174}{arXiv:astro-ph/0501174}%
  \bibAnnoteFile{NoStop}{Cole:2005sx}%
\bibitem{Tegmark:2006az}%
  \BibitemOpen
  \bibfield{author}{%
  \bibinfo {author} {\bibfnamefont{M.}~\bibnamefont{Tegmark}} \emph{et~al.}
  (\bibinfo {collaboration} {SDSS}),\ }%
  \bibfield{journal}{%
  \Doi{10.1103/PhysRevD.74.123507}{\bibinfo {journal} {Phys. Rev.}}\ }%
  \textbf{\bibinfo {volume} {D74}},\ \bibinfo {pages} {123507} (\bibinfo {year}
  {2006}),\
  \Eprint{http://arxiv.org/abs/astro-ph/0608632}{arXiv:astro-ph/0608632}%
  \bibAnnoteFile{NoStop}{Tegmark:2006az}%
\end{thebibliography}%

\end{document}